%% file: arxiv.tex
\documentclass[numbers,sort&compress]{article}

\usepackage{graphicx}
\usepackage{listings}
\usepackage{url}
\usepackage{lineno}
\usepackage{subcaption}
\usepackage{rotating}
\usepackage[normalem]{ulem}
\usepackage{arxiv}
\usepackage{algpseudocode,algorithm}
\usepackage{mhchem}
\usepackage{amssymb,amsmath,amsthm}
\usepackage[numbers,sort&compress]{natbib}
\usepackage{xcolor}
\usepackage{hyperref}
\usepackage{tabularx}
\usepackage{adjustbox}
\usepackage[toc,page]{appendix}
\graphicspath{{figures/}{imgs/}}
\usepackage{amsfonts}
\usepackage{nicefrac}
\usepackage{microtype}
\usepackage{cleveref}
\usepackage{doi}
\usepackage{tikz}
\usetikzlibrary{positioning,arrows.meta,fit,backgrounds,calc,shapes.geometric}
\definecolor{RuhiTeal}{HTML}{004D40}
\definecolor{RuhiCoral}{HTML}{FF655D}
\definecolor{RuhiYellow}{HTML}{F1DB4B}
\definecolor{RuhiBlue}{HTML}{2196F3}
\definecolor{RuhiPink}{HTML}{E91E63}
\lstdefinelanguage{Rust}{morekeywords={fn,pub,struct,enum,impl,use,let,mut,for,if,else,match,return,self,Self,where,in,loop,while,break,continue,move,ref,as,true,false,mod,crate,super,type,const,static,trait,unsafe,async,await,dyn},sensitive=true,morekeywords=[2]{Vec,f64,usize,bool,Option,Result,Some,None,Ok,Err,Box,String},morecomment=[l]{//},morecomment=[s]{/*}{*/},morestring=[b]",morestring=[b]'}
\input{authors_arxiv.tex}
\date{\today}
\title{Enhanced Climbing Image Nudged Elastic Band method with Hessian Eigenmode Alignment}
\hypersetup{
 pdfauthor={},
 pdftitle={Enhanced Climbing Image Nudged Elastic Band method with Hessian Eigenmode Alignment},
 pdfkeywords={},
 pdfsubject={},
 pdfcreator={},
 pdflang={English}}
\begin{document}

\maketitle
\begin{abstract} %
Accurate determination of transition states is central to an understanding of
reaction kinetics. Double-endpoint methods where both initial and final states
are specified, such as the climbing image nudged elastic band (CI-NEB), identify
the minimum energy path between the two and thereby the saddle point on the
energy surface that is relevant for the given transition, thus providing an
estimate of the transition state within the harmonic approximation of transition
state theory. Such calculations can, however, incur high computational costs and
may suffer stagnation on exceptionally flat or rough energy surfaces.
Conversely, methods that only require specification of an initial set of atomic
coordinates, such as the minimum mode following (MMF) method, offer efficiency
but can converge on saddle points that are not relevant for transition of
interest. Here, we present an adaptive hybrid algorithm that integrates the
CI-NEB with the MMF method so as to get faster convergence to the relevant
saddle point. The method is benchmarked for the Baker-Chan (BC) saddle point
test set using the PET-MAD machine-learned potential as well as 59 transitions
of a heptamer island on Pt(111) from the OptBench benchmark set. A Bayesian
analysis of the performance shows a reduction in energy and force
calculations of 57%
while a 31%
results establish this hybrid method as a highly effective tool for
high-throughput automated chemical discovery of atomic rearrangements.
\end{abstract}
\keywords{Gaussian Process Regression, Bayesian Optimization, Saddle Point Search, Dimer Method, NEB, Active Learning}
\vspace{0.7em}
\noindent
\section{Introduction}
\label{sec:introduction}
Estimating reaction kinetics requires an accurate free energy barrier corresponding to the transition state which is a dividing surface that forms a bottleneck for reactive trajectories.
Within the harmonic approximation to transition state theory (HTST) the transition state is approximated as a \(3N-1\) dimensional hyperplane between the reactant and product basins of the Potential Energy Surface (PES).
The plane passes through a first order saddle point, which we will refer to as the transition structure (TS), and its normal points in the direction of lowest curvature of the PES at that point.
The relevant point is the point of highest energy along the minimum energy path (MEP) connecting the reactant and product states, and this is a saddle point on the PES.
For a given reaction, the MEP is the path of highest statistical weight in the configuration space connecting the initial and final states.

TS exploration methods identify a candidate transition mechanism.
The estimate of the transition rate (under HTST) can be calculated from the energy barrier, which is the difference between the energy at the saddle and the reactant state.

According to the \emph{a-priori} knowledge of the system, TS exploration methods are classified as ``double-endpoint'' and ``initial-point'' methods.
The former assumes knowledge of two sets of atom coordinates corresponding to local energy minima, representing reactant and product states, and finds an MEP between the two.
This includes the Nudged Elastic Band or NEB \cite{jonssonNudgedElasticBand1998}, where a discrete chain of images or configurations of atoms connecting known reactants and products relax to the MEP.
By construction, the NEB method defines a discrete representation of a path in configuration space, and the calculation requires computation of energy and atomic forces for each discrete configuration along the path.
If this requires electronic structure calculations, the computational effort can become large.
A key issue in a successful NEB calculation is the choice of the initial path.
Also, the choice of optimizers strongly affects the convergence rate and efficiency.
Estimating the saddle point geometry from the results of an NEB calculation relies on having configurations in its vicinity.
In turn, this is controlled by the number of images along the path and the spring constant which determines their distribution along the path.
To converge onto the saddle point, one of the images, the climbing image can be made to move uphill along the path (CI-NEB) \cite{henkelmanClimbingImageNudged2000}.
NEB calculations may require many iterations for flat or rough potential surfaces where the atomic forces are small or vary sharply between adjacent images.

Conversely, the ``initial-point'' methods require only a single starting configuration of the atoms from which the exploration for the TS begins.
These methods, such as the Dimer \cite{henkelmanDimerMethodFinding1999} typically follow the lowest eigenmode of the Hessian (i.e., curvature of the PES) to a saddle point.
Such minimum mode following (MMF) searches do not require the evaluation of multiple configurations for each traversal step.
Due to the unconstrained nature of the search, a saddle point may be found that is irrelevant for the reaction of interest, or an already known saddle point is rediscovered, which offsets the computational efficiency.

Herein, we describe an off-path climbing image Nudged Elastic Bands (OCI-NEB), a method integrating the stability of a double-endpoint chain-of-replica method with the efficiency of an MMF saddle point search method.
While a switch from a double-endpoint method when a certain level of convergence is obtained, to an initial-point method to complete the search has previously been practiced, \cite{asgeirssonNudgedElasticBand2021,parkHighthroughputApproachMinimum2025}, the OCI-NEB can switch between the two, back and forth, depending on certain properties that are monitored during the calculation.
We show that by coupling this adaptive triggering with an alignment-based cool-down strategy, OCI-NEB uniformly accelerates convergence, more than halving the number of energy and atomic force evaluations for two dissimilar benchmark sets.
\section{Methods}
\label{sec:org847e04c}
Before presenting the proposed OCI-NEB algorithm the dimer and the nudged elastic band methods are reviewed for completeness, based on the implementation in the eOn \footnote{\url{https://eondocs.org}} software suite \cite{goswamiEfficientExplorationChemical2025}.
\subsection{Minimum Mode Following: The Dimer Method}
\label{sec:methods:dimer}
To locate first-order saddle points, in principle the full Hessian matrix at each point is required.
The Hessian is a 3N x 3N matrix with each element requiring at least one force call, which becomes computationally prohibitive for most systems.
To work around calculating the full Hessian, minimum mode following algorithms assume that the most relevant mode is the eigenvector corresponding to the lowest eigenvalue, the ``minimum mode''.
The minimum mode can be computed approximately without the full Hessian \cite{mousseauActivationRelaxationTechniqueART2012,munroDefectMigrationCrystalline1999}, following the eigenvector approach of \cite{cerjanFindingTransitionStates1981}.
In this work, the minimum mode is found at each step using the Dimer method \cite{henkelmanDimerMethodFinding1999,olsenComparisonMethodsFinding2004}.

The ``dimer'' in the Dimer method consists of two replicas, \(\mathbf{R}_1\) and \(\mathbf{R}_2\), displaced symmetrically from a central point \(\mathbf{R}\) by a small half-separation \(\Delta{R}/2\) along a normalized orientation vector \(\hat{\mathbf{N}}\):

\begin{equation}
\mathbf{R}_{1} = \mathbf{R} - \frac{\Delta R}{2} \hat{\mathbf{N}}, \qquad
\mathbf{R}_{2} = \mathbf{R} + \frac{\Delta R}{2} \hat{\mathbf{N}}.
\end{equation}

This dimer construct undergoes constrained relaxation, or ``rotation'', such that the separation \(\Delta R\) does not change and the midpoint remains fixed.
This procedure brings \(\hat{\mathbf{N}}\) (the dimer axis) closer to the lowest curvature mode at each step.
The final axis after the ``rotation'' is indicated by \(\hat{\mathbf{d}}\). The dimer axis is the approximation to the lowest curvature mode of the Hessian.
The curvature \(C\) along the dimer axis is approximated by a central finite difference of the forces \(\mathbf{F}_{1,2} = -\nabla V(\mathbf{R}_{1,2})\) acting on the endpoints.
With the convention above (\(\mathbf{R}_1\) is the ``minus'' endpoint), the curvature is found by the Rayleigh-quotient estimate, and is \cite{henkelmanDimerMethodFinding1999}

\begin{equation}
C(\hat{\mathbf{N}}) = \hat{\mathbf{N}}^T H \hat{\mathbf{N}} \approx \frac{(\mathbf{F}_1 - \mathbf{F}_2) \cdot \hat{\mathbf{N}}}{\Delta R},
\label{eq:dimer_curvature}
\end{equation}

Once aligned, the central configuration \(\mathbf{R}\) translates under a modified effective force \(\mathbf{F}_{\text{trans}}\) obtained by inverting the component of the true force along the dimer axis \(\hat{\mathbf{N}}\).
This modification converts what would be gradient ascent along \(\hat{\mathbf{N}}\) into effective descent, while leaving the force components along all other directions unchanged.
This inversion of the force along a component can also be understood as a Householder transformation along the vector \(N\) which can be written as \(\mathbf{T}_{H} = \mathbf{I} - 2 \hat{\mathbf{N}}\cdot \hat{\mathbf{N}}^T\).
The translation then becomes
\begin{align}
\mathbf{F}_{\text{trans}}(\mathbf{R}) &= \mathbf{F}(\mathbf{R}) - 2 (\mathbf{F}(\mathbf{R}) \cdot \hat{\mathbf{N}}) \hat{\mathbf{N}}
&=\mathbf{T}_{H}\mathbf{F}.
\label{eq:dimer_trans_force}
\end{align}
The dimer rotation and translation steps alternate until the force norm drops below a convergence threshold, at which point a saddle point has been located.

The MMF method may be started from any point on the energy surface.
It can find a collection of saddles representing possible transitions from a given initial state by starting from displaced configurations near the minimum.
Alternatively, the MMF can refine a guess obtained by other means, and this concept we exploit in the present work to iteratively improve a partially converged double end-point calculation.
\subsection{Nudged Elastic Band -- NEB}
\label{sec:methods:neb}
Double-endpoint searches are often used to find the full MEP between known reactant and product states.
We will use the Nudged Elastic Band method, which approximates the reaction pathway as a chain of ``images'' connected by fictitious springs.
The force acting on each image is

\begin{equation}
  \mathbf{F}_i^{\text{NEB}} = \mathbf{F}_i^{\perp} + \mathbf{F}_i^{\parallel, \text{spring}}.
  \label{eq:neb_total_force}
\end{equation}
The method ensures convergence to the MEP and controls the distribution of the images along the path by projecting the force components according to an estimate of the local tangent to the path.
The true potential force is only acting perpendicular to the path tangent

\begin{equation}
  \mathbf{F}_i^{\perp} = \mathbf{F}_i^{\text{true}} - (\mathbf{F}_i^{\text{true}} \cdot \hat{\tau}_i) \hat{\tau}_i,
  \label{eq:neb_force_perp}
\end{equation}

while spring forces act only parallel to it
\begin{equation}
  \mathbf{F}_i^{\parallel, \text{spring}} = k (|\mathbf{R}_{i+1} - \mathbf{R}_i| - |\mathbf{R}_i - \mathbf{R}_{i-1}|) \hat{\tau}_i.
  \label{eq:neb_force_spring}
\end{equation}
This force projection -- referred to as ``nudging'' -- decouples the optimization of path shape from the distribution of images along the path.

For improved numerical stability, the ``improved tangent'' estimate is used \cite{henkelmanImprovedTangentEstimate2000}, which defines the tangent based on the neighboring image that is higher in energy, to reduce the probability of kinks in the path

\begin{equation}
  \hat{\tau}_i =
  \begin{cases}
    \text{normalize}(\mathbf{R}_{i+1} - \mathbf{R}_i) & \text{if } V_{i+1} > V_i > V_{i-1} \\
    \text{normalize}(\mathbf{R}_i - \mathbf{R}_{i-1}) & \text{if } V_{i-1} > V_i > V_{i+1} \\
    \text{weighted average} & \text{otherwise}.
  \end{cases}
  \label{eq:neb_improved_tangent}
\end{equation}

At extrema, the tangent is defined to be the weighted average of the vectors to neighboring images, giving preference to the vector on the higher energy side.

To converge rigorously on the saddle point, the highest energy image is turned into a ``climbing image'' \cite{henkelmanClimbingImageNudged2000}, where the spring force is removed, and the parallel component inverted.
This forces the selected image, identified as the ``climbing'' image to move uphill along the path to the saddle point.

\begin{equation}
\mathbf{F}_{\text{climb}} = \mathbf{F}_{\text{climb}}^{\text{true}} - 2 (\mathbf{F}_{\text{climb}}^{\text{true}} \cdot \hat{\tau}_{\text{climb}}) \hat{\tau}_{\text{climb}}
\label{eq:neb_ci_force}
\end{equation}

The accuracy of the saddle configuration for the CI-NEB calculations depends on the tangent approximation, which in turn derives from the neighboring images of the climbing image.
To increase the resolution of images around the saddle without increasing the number of images, the springs can be adjusted, either through geometric considerations \cite{asgeirssonNudgedElasticBand2021} or kinetic considerations \cite{mandelliModifiedNudgedElastic2021}.

The energy-weighted spring method \cite{asgeirssonNudgedElasticBand2021} used here, dynamically stiffens the spring constant in high-energy regions, concentrating the chain to enforce higher image density near the saddle point.

The quality of the initial path influences convergence speed and stability for the NEB.
Linear interpolations on atomic positions often produce unphysical atomic overlaps, particularly for complex or dense systems.
To mitigate this without having to construct new paths for every system, improved initial paths in Cartesian Coordinates can be constructed, by using the Sequential Image Dependent Pair Potential (S-IDPP) method \cite{schmerwitzImprovedInitializationOptimal2024,smidstrupImprovedInitialGuess2014}.

The S-IDPP constructs the path by sequentially growing images from the reactant and product endpoints.
It optimizes these intermediate images on a simplified auxiliary surface defined by interpolating the pairwise distances of bonded atoms at the endpoints.
This procedure attempts to eliminate high-energy steric clashes before engaging computationally intensive electronic structure or MLIP calculations.
\subsection{Off-path Climbing Image Nudged Elastic Band (OCI-NEB)}
\label{sec:org6ff5213}
The standard Climbing Image Nudged Elastic Band, or CI-NEB depends on the efficiency of the underlying optimizer.
On flat or noisy sections of the PES, the projected force components can oscillate or become ill-defined.
To address this, we introduce the OCI-NEB algorithm, a hybrid approach that intersperses standard CI-NEB optimization steps with targeted MMF refinement using the Dimer method as implemented in eOn \cite{heydenEfficientMethodsFinding2005,olsenComparisonMethodsFinding2004}.

This method builds on a two-stage refinement strategy \cite{goswamiEfficientExplorationChemical2025} in which dimer calculations are carried out within NEB iterations once the path relaxes sufficiently.
The NEB first brings the climbing image near the saddle point; a short burst of single-image MMF then refines the climbing image position before the NEB resumes.

In practice, a fixed number of steps of CI-NEB initially \cite{goswamiEfficientExplorationChemical2025} or a single threshold for switching from CI-NEB to MMF does not transfer well across benchmark systems and energy surfaces.
Workflow methods \cite{parkHighthroughputApproachMinimum2025}, or a fixed tolerance on the magnitude of the atomic forces for switching to MMF \cite{asgeirssonNudgedElasticBand2021} may not lead to the appropriate saddle connecting the given reactant and product basins.

OCI-NEB dynamically switches between the NEB and dimer during a single calculation, rather than switching just once from CI-NEB to MMF.
This takes inspiration from the active learning set of acceleration methods \cite{petersonAccelerationSaddlepointSearches2016,koistinenNudgedElasticBand2019,koistinenMinimumModeSaddle2020,goswamiAdaptivePruningIncreased2025b,goswamiEfficientImplementationGaussian2025a}, and means establishing a bidirectional coupling, which allows switching also from the MMF back to the NEB based on a stability mechanism to ensure improvement.
The OCI-NEB therefore ensures convergence in the region of interest.
We develop the algorithm below by addressing the stability and efficiency challenges inherent to such hybridization.
\subsubsection{Approximations to the lowest mode: NEB and Dimer}
\label{sec:org2d900fb}
Equations \ref{eq:neb_ci_force} and \ref{eq:dimer_trans_force} share an identical algebraic structure: both are Householder reflections of the true force through unit vector \(\hat{\mathbf{\chi}}\), \(\hat{\mathbf{d}}\) for the dimer, and \(\mathbf{\tau}_{\text{climb}}\) for the NEB,

\begin{equation}
\mathbf{F}_{\text{eff}}(\mathbf{R}, \hat{\mathbf{\chi}}) = \mathbf{T}_H \mathbf{F}(\mathbf{R}).
\label{eq:householder_force}
\end{equation}

The reflector \(\mathbf{T}_H\) inverts the force component along \(\hat{\mathbf{d}}\) while leaving the orthogonal complement unchanged.
This force inversion allows a standard minimization algorithm (L-BFGS, FIRE, CG) to converge to a saddle point: the modified force \(\mathbf{T}_{H}F\) has a zero at the saddle (\(\mathbf{T}_{H}\mathbf{0} = \mathbf{0}\)), and the inversion along the lowest mode enforces ascent along that direction \cite{eSimplifiedImprovedString2007}.
Since both the climbing image and the dimer modify the dynamics to converge towards a saddle state on the energy surface, switching between the dimer and the NEB as in the OCI-NEB does not affect the state and does not meaningfully change the final solution.

\begin{figure}[t]
\centering
\includegraphics[width=0.95\linewidth]{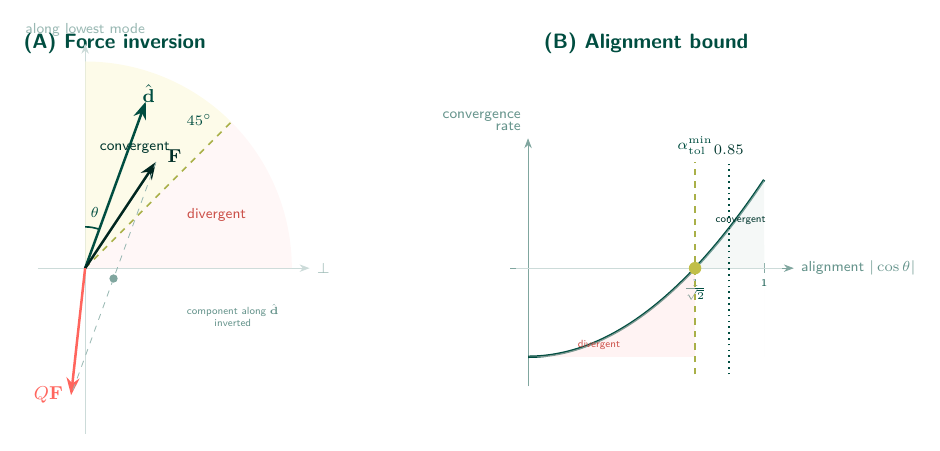}
\caption{\label{fig:householder_geom}\textbf{Force inversion geometry and the alignment bound.} (A) The dimer axis \(\hat{\mathbf{d}}\) defines the direction along which the force is inverted. The reflector \(\mathbf{T}_{H} = \mathbf{I} - 2\hat{\mathbf{d}}\hat{\mathbf{d}}^T\) maps the force \(\mathbf{F}\) to \(\mathbf{T}_{H}\mathbf{F}\) (coral). When \(\hat{\mathbf{d}}\) lies within \(45^{\circ}\) of the lowest curvature mode as approximated by the NEB (yellow cone), the modified force drives the configuration toward the saddle. Outside this cone (pink), the force inversion pushes away from it. (B) The critical alignment: the cosine of the angle \(\theta\) between \(\hat{\mathbf{d}}\) and the NEB axis \(\tau\) must exceed \(1/\sqrt{2}\) for convergent behavior. Below this threshold the modified dynamics lead away from the saddle. The dashed line marks \(\alpha_{\mathrm{tol}}^{\min} = 1/\sqrt{2}\); the dotted line at \(0.85\), the empirical optimum from the data generated.}
\end{figure}

When the dimer moves the climbing image off-path, subsequent NEB iterations must redistribute images before the climbing image reactivates, incurring additional force evaluations.
Separately, if \(\hat{\mathbf{d}}\) is misaligned by more than \(45^{\circ}\) from the lowest mode, the inversion can push the configuration away from the saddle rather than toward it.
This critical angle corresponds to the requirement \(|\hat{\mathbf{d}} \cdot \hat{\mathbf{v}}_{\min}| > 1/\sqrt{2}\), where \(\hat{\mathbf{v}}_{\min}\) is the dimer's estimate of the lowest mode (see Section \ref{sec:disc} for the derivation).
In OCI-NEB, the dimer is initialized along the NEB tangent \(\hat{\tau}\) and then rotated; the alignment \(\alpha = |\hat{\mathbf{v}}_{\min} \cdot \hat{\tau}|\) measures how well the dimer mode matches the path direction.
The requirement \(\alpha \ge 1/\sqrt{2}\) ensures the force inversion drives the climbing image toward the saddle of interest.
Figure \ref{fig:householder_geom} illustrates this geometry.

We never need the full Hessian -- only the dimer's one-mode estimate.\}

\subsubsection{Dynamic Control via Mode Alignment}
\label{sec:orgc8d18a2}

A primary challenge for hybrid methods involves determining the duration of the refinement phase.
OCI-NEB utilizes the alignment between the dimer axis \(\hat{\mathbf{d}}\) and the NEB path tangent \(\hat{\tau}\).
By initializing the Dimer along \(\hat{\tau}\), we ensure the dimer search begins in a relevant subspace, preventing initialization based failures \cite{goswamiAdaptivePruningIncreased2025b}.
The alignment between the two,

\begin{equation}
\label{eq:roneb_align}
\alpha = | \hat{\mathbf{d}} \cdot \hat{\tau} |
\end{equation}

provides an early stopping criterion: the MMF continues while the orientation remains close to the tangent estimate.
When the angle grows large, the MMF terminates and the optimizer history determines the climbing image position along the dimer trajectory with the most negative eigenvalue.
This termination criterion replaces arbitrary step counts with a physically motivated bound.
\subsubsection{Relative Baselines and Transferability}
\label{sec:orga266a25}

Given that we generate initial paths using S-IDPP, which may involve high energy images, defining the ``handover point'' or the force threshold at which the algorithm switches from the NEB to MMF becomes problematic for high-throughput calculations.
Defining absolute force thresholds, e.g. \(0.5\) eV/ \AA{} does not work well across a diverse dataset of different reaction mechanisms, since a threshold value appropriate for stiff covalent bond breakage may correspond to an effectively converged state for a soft supramolecular rearrangement.

To mitigate this, we establish a relative baseline.
At the start of the calculation, we record the baseline convergence force \((F_{0})\), which is the maximum force norm on any atom of the initial path.
We then express the triggering threshold \(T_{mmf}\) relative to this baseline:
\begin{equation}
\label{eq:roneb_relmmf_thresh}
T_{mmf} = \lambda_{trigger} F_{0}.
\end{equation}
The algorithm therefore engages only after the NEB has relaxed the high-energy configurations from the initial guess, regardless of the absolute energy scale of the system.
\subsubsection{Robustness: Stability Latches and Restoration}
\label{sec:orgc638118}

Despite the early stopping criteria and the initial relative thresholds, activating the MMF based solely on force criteria can lead to instability for rough potentials or where the climbing image index oscillates.
The highest energy image along the path, which is selected for the climbing image, especially for automated initial paths, may oscillate during the initial phase of the calculation.
The OCI-NEB implements a stability latch, switching to the dimer only if the climbing image index remains constant for \(\kappa\) consecutive iterations.

The internal state of the optimizer resets if the motion of the image exceeds the maximum move of each image times the total number of intermediate images.
Two distinct restoration mechanisms protect against failed MMF searches.
If the dimer encounters positive curvature (eigenvalue \(> 0\)), indicating the climbing image has drifted to a local minimum rather than a saddle, the image is restored to its pre-MMF position and the cached eigenvector is discarded; the NEB then resumes from the unperturbed state.
For all other termination modes (alignment failure, step budget exhaustion), the climbing image advances to the configuration with the most negative curvature found during the search.
Both mechanisms allow OCI-NEB to recover partial progress from a terminated MMF search before returning control to the NEB.
\subsubsection{Failure Recovery}
\label{sec:org357a4ae}

When the dimer dephases from the NEB tangent, the triggering threshold must be raised to prevent repeated premature activation.
We parameterize the penalty through a one-parameter family \(P(\alpha, S) = B + (1-B)\alpha^S\) with \(B = 1/(1+S)\), subject to the boundary conditions \(P(0) = B\) (maximum penalty at zero alignment) and \(P(1) = 1\) (no penalty at perfect alignment).
Within this family, the unique affine member is selected by requiring \(P'' \equiv 0\) on \((0,1)\).
Computing:
\begin{equation}
\label{eq:penalty_uniqueness}
\frac{\partial^2 P}{\partial \alpha^2} = \frac{s^2(s-1)}{1+s}\,\alpha^{s-2},
\end{equation}
which vanishes identically on \((0,1)\) if and only if \(s = 1\), since \(\alpha^{s-2} > 0\) and \(s^2/(1+s) > 0\) for all positive \(s\).
Setting \(S = 1\) gives \(B = 1/2\) and the penalty reduces to
\begin{equation}
\label{eq:linear_penalty}
P(\alpha) = \tfrac{1}{2} + \tfrac{1}{2}\,\alpha.
\end{equation}
Any \(S < 1\) yields a concave penalty (too aggressive near full alignment), while \(S > 1\) yields a convex penalty (too lenient near zero alignment).
The linearity condition eliminates both failure modes.
At zero alignment the threshold halves; at perfect alignment no penalty applies.
The on-failure threshold then reads
\begin{equation}
\label{eq:failure_threshold}
T_{\text{fail}} = F_0 \, \lambda_{\text{rel}} \, P(\alpha),
\end{equation}
so the penalty strength and shape are fully determined by the linearity constraint, requiring no additional tuning beyond \(\lambda_{\text{rel}}\).
\subsubsection{Post-MMF Path Reparameterization}
\label{sec:orgf4f8e7d}

When the dimer moves the climbing image, the neighboring NEB images become unevenly spaced along the path.
Rather than relying on the spring forces to gradually redistribute images over subsequent NEB iterations, we apply an arc-length reparameterization immediately after each successful MMF step (i.e., when \(F_{\text{new}} < F_{\text{CI}}\)).
This redistributes all intermediate images at equal arc-length intervals via cubic Hermite interpolation, using the same procedure as the S-IDPP initialization.
The reparameterization is geometry-only (zero additional force evaluations); forces are recomputed in the next NEB iteration regardless.
The L-BFGS optimizer state is reset after reparameterization to prevent stale gradient history from corrupting the search direction.
\subsubsection{Success Threshold Update}
\label{sec:orgb6f062b}

When the MMF step reduces the climbing image force (\(F_{\text{new}} < F_{\text{CI}}\)), the triggering threshold is updated to permit re-entry at a level proportional to the achieved improvement:
\begin{equation}
\label{eq:success_threshold}
T_{\text{success}} = F_{\text{new}} \left( \tfrac{1}{2} + \tfrac{2}{5} \, \frac{F_{\text{new}}}{F_{\text{CI}}} \right).
\end{equation}
The two limiting cases motivate the form: when \(F_{\text{new}} \approx F_{\text{CI}}\) (marginal improvement), \(T_{\text{success}} \approx 0.9\,F_{\text{CI}}\), setting the threshold just below the current force to require further NEB relaxation before re-triggering.
When \(F_{\text{new}} \ll F_{\text{CI}}\) (substantial improvement), \(T_{\text{success}} \approx 0.5\,F_{\text{new}}\), encouraging early re-entry into the MMF phase.
At convergence (\(F_{\text{new}} \to 0\)), the threshold vanishes and the NEB takes over.
\subsubsection{Complete Algorithm}
\label{sec:orga3598a5}

The complete OCI-NEB procedure is given in Algorithm \ref{alg:roneb}.

\begin{algorithm}
  \caption{Off-path Climbing Image Nudged Elastic Band (OCI-NEB)}
  \label{alg:roneb}
  \begin{algorithmic}[1]
    \State \textbf{Input:} Initial Path $\mathcal{P} = \{\mathbf{R}_0, \dots, \mathbf{R}_{P+1}\}$, Baseline Force $F_0$
    \State \textbf{Parameters:} Trigger factor $\lambda$; alignment tolerance $\alpha_{tol} \ge 1/\sqrt{2}$
    \State $T_{mmf} \leftarrow \lambda F_0$, \quad $L_{stable} \leftarrow 0$
    \While{not converged}
      \State Find highest energy image, $\mathbf{R}_{\text{climb}}$ (index $k$)
      \State Calculate tangents $\hat{\tau}_i$ and NEB forces $\mathbf{F}^{\text{NEB}}$
      \State Calculate climbing image force $F_{CI} = || \mathbf{F}_{\text{climb}} ||$

      \If{$k == k_{prev}$}
        \State $L_{stable} \leftarrow L_{stable} + 1$
      \Else
        \State $L_{stable} \leftarrow 0$; clear cached eigenvector
      \EndIf

      \If{$L_{stable} \ge \kappa$ \textbf{and} $F_{CI} < T_{mmf}$}
        \State Save $\mathbf{R}_k$; initialize Dimer with cached eigenvector (or $\hat{\tau}_k$)
        \State Run Dimer: optimize $\hat{\mathbf{d}}$; abort if $\alpha < \alpha_{tol}$ or curvature $> 0$
        \If{curvature $> 0$} \Comment{Positive curvature: restore}
           \State $\mathbf{R}_k \leftarrow$ saved position; clear cache
        \ElsIf{$F_{new} < F_{CI}$} \Comment{MMF helped (Eq.~\ref{eq:success_threshold})}
           \State $\alpha \leftarrow |\hat{\mathbf{d}} \cdot \hat{\tau}_k|$
           \State $T_{mmf} \leftarrow F_{new} \cdot (0.5 + 0.4 \, F_{new} / F_{CI})$
           \State Cache $\hat{\mathbf{d}}$ for warm-start
           \State Reparameterize $\mathcal{P}$ along arc length; reset optimizer
        \Else \Comment{Failure: linear penalty (Eq.~\ref{eq:linear_penalty})}
           \State $\alpha \leftarrow |\hat{\mathbf{d}} \cdot \hat{\tau}_k|$
           \State $T_{mmf} \leftarrow F_0 \, \lambda \, (\tfrac{1}{2} + \tfrac{1}{2}\,\alpha)$
        \EndIf
      \Else
        \State Take NEB optimization step on $\mathcal{P}$
      \EndIf
      \State $k_{prev} \leftarrow k$
    \EndWhile
  \end{algorithmic}
\end{algorithm}

\begin{figure}[t]
\centering
\includegraphics[width=1.0\linewidth]{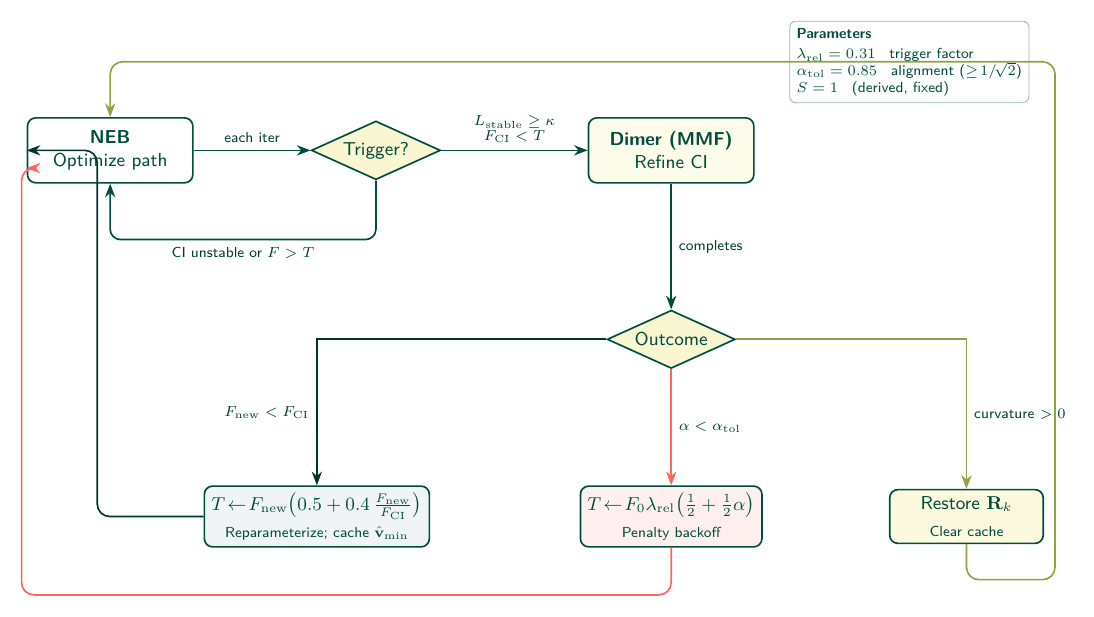}
\caption{\label{fig:ocineb_flowchart}\textbf{OCI-NEB control flow.} After each NEB iteration, the trigger condition gates MMF activation. The dimer outcome determines the next action: on success (\(F_{\mathrm{new}} < F_{\mathrm{CI}}\)), the threshold adapts downward and the path is reparameterized; on alignment failure (\(\alpha < \alpha_{\mathrm{tol}}\)), the linear penalty raises the threshold; on positive curvature, the climbing image is restored to its pre-MMF position. All three branches return control to the NEB.}
\end{figure}

Figure \ref{fig:ocineb_flowchart} summarizes the control flow, showing the three-way outcome branch after each MMF invocation.

\begin{enumerate}
\item Configuration
\label{sec:org7f275f4}
The OCI-NEB implementation exposes tunable parameters that control the aggressiveness of the dimer search.
Table \ref{tbl:roneb_params} lists these parameters alongside their corresponding configuration keys and default values used in this work.

\begin{table}[htbp]
\caption{\label{tbl:roneb_params}\textbf{\textbf{OCI-NEB Configuration Parameters.}} Two user-facing parameters control the algorithm; the remaining constants are derived or fixed. The alignment tolerance \(\alpha_{tol}\) has a principled minimum \(1/\sqrt{2}\) from the force inversion alignment bound (Figure \ref{fig:householder_geom}); the reported value \(0.85\) provides a margin above this minimum. The stability count (\(\kappa = 5\)), maximum MMF steps (\(N_{mmf} = 1000\)), and penalty shape (\(S = 1\)) are fixed internal constants.}
\centering
\begin{tabular}{lllr}
Parameter Description & Symbol & Config Key & Value\\
\hline
Relative Trigger Factor & \(\lambda_{rel}\) & \texttt{ci\_mmf\_after\_rel} & 0.31\\
Alignment Tolerance & \(\alpha_{tol}\) & \texttt{ci\_mmf\_angle} & 0.85\\
\end{tabular}
\end{table}

The remaining internal constants -- stability count \(\kappa = 5\), MMF step budget \(N_{mmf} = 1000\), and penalty shape \(S = 1\) (Eq. \ref{eq:linear_penalty}) -- are fixed.
On success, the converged dimer eigenvector is cached for warm-starting subsequent MMF calls.
Both user-facing parameters were held constant across the Baker-Chan and OptBench Pt(111) benchmarks without system-specific tuning; the performance analysis follows in Section \ref{sec:disc}.
\end{enumerate}
\subsection{Computational Details}
\label{sec:compdet}
All calculations use the eOn software package \cite{rohitgoswamiTheochemUIEOnV29012026} \footnote{\url{https://eondocs.org}}.
The Metatomic interface \cite{bigiMetatensorMetatomicFoundational2025} to the PET-MAD-S \texttt{v1.5.0} machine learning interatomic potential \cite{mazitovMassiveAtomicDiversity2025,mazitovPETMADLightweightUniversal2025} provides energy and forces.
This non-equivariant point-edge transformer model is trained on the diverse MAD dataset, encompassing bulk crystals, surfaces, and molecular fragments at the PBEsol functional level.
A Snakemake \cite{molderSustainableDataAnalysis2021} workflow orchestrated concurrent runs on a single machine with an AMD Ryzen Threadripper PRO 5945WX (24 core, 48 threads) and an NVIDIA T400 GPU with 4 GB VRAM.
For efficient concurrency on the singular GPU, we use the NVIDIA multi-processing service.
As per established best practices \cite{kastnerSuperlinearlyConvergingDimer2008,goswamiBayesianHierarchicalModels2025a} we use the limited-memory Broyden-Fletcher-Goldfarb-Shanno \cite{liuLimitedMemoryBFGS1989a} for translation steps of the Dimer, and the conjugate gradient optimizer \cite{chapraNumericalMethodsEngineers2015} for its rotation steps.
All reported gradient evaluation counts represent total potential energy surface queries, inclusive of those incurred during dimer rotation steps, as tracked by the global force call counter in the eOn solver.

We assess the performance of the OCI-NEB algorithm against the standard CI-NEB method across the 25 reactions of the Baker-Chan transition state test suite \cite{bakerLocationTransitionStates1996}.
This covers broad chemical archetypes, including systems undergoing dissociations (\ce{H2CO}), insertions (silene), ring-opening (cyclopropyl), and rotational transitions (acrolein).
Since these systems exhibit diverse PES features, ranging from stiff covalent bonds to weak intermolecular forces, they provide a rigorous test for the adaptability and robustness of OCI-NEB compared to the standard CI-NEB protocol.

The Baker-Chan endpoint geometries use the atom ordering convention specified in the original test set \cite{bakerLocationTransitionStates1996}, where corresponding atoms share the same index across reactant and product.
This ordering is preserved by the \texttt{readcon-core} implementation \footnote{\url{https://lode-org.github.io/readcon-core/spec.html}} of the v2 \texttt{con} file specification, which maintains fifth-column atom-type assignments on loading.
Preserving this ordering eliminates atom-matching heuristics that can introduce spurious permutations in the initial path.

Both methods utilized identical initialization parameters (for the Baker-Chan, S-IDPP \cite{schmerwitzImprovedInitializationOptimal2024}, 8 intermediate images) and convergence criteria (\(0.05 \text{eV}/\text{\AA}\)).
Baseline CI-NEB and the OCI-NEB protocols utilized identical optimization backends and convergence criteria.
The specific hyperparameters governing the hybrid OCI-NEB triggers are detailed in Table \ref{tbl:opt_params}.

\begin{table}[htbp]
\caption{\label{tbl:opt_params}\textbf{\textbf{Optimization hyperparameters.}} Shared parameters apply to both methods. OCI-NEB-specific parameters govern the activation and stability of the dimer. These default values demonstrate robustness across diverse chemical archetypes. Section \ref{sec:disc} contains a detailed performance analysis of the most influential parameters, specifically the trigger thresholds and alignment tolerances.}
\centering
\begin{tabular}{llll}
\textbf{Category} & \textbf{Parameter} & \textbf{Symbol} & \textbf{Value}\\
\hline
\textbf{Shared} & Potential & \(V(\mathbf{R})\) & PET-MAD-S v1.5.0\\
 & Optimizer & - & LBFGS\\
 & Convergence Force & \(F_{tol}\) & \(0.05 \text{ eV}/\text{\AA}\)\\
 & Images & \(N_{img}\) & 8\\
 & Spring Constant & \(k_{sp}\) & \(1\) to \(10\) eV/ \AA{}**2 (Energy Weighted)\\
 & CI Activation (Relative) & \(\lambda_{CI}\) & 0.8\\
 & S-IDPP Growth Factor & \(\alpha_{idpp}\) & 0.33\\
\hline
\textbf{OCI-NEB} & MMF Relative Trigger & \(\lambda_{rel}\) & 0.31\\
 & Alignment Tolerance & \(\alpha_{tol}\) & 0.85\\
 & Dimer Rot. Convergence & \(\phi_{tol}\) & \(10.0^{\circ}\)\\
\end{tabular}
\end{table}

Both methods converge for all systems, and the OCI-NEB strictly improves performance in every case, so we use a hierarchical Bayesian negative binomial regression with varying slopes modeled via B-splines with an intercept for each system for quantifying the performance as a function of the distance of the initial path saddle estimate to the final configuration \cite{goswamiAdaptivePruningIncreased2025b,goswamiBayesianHierarchicalModels2025a,goswamiEfficientExplorationChemical2025}.
Distances \cite{goswamiTwodimensionalRMSDProjections2025} used involve permutation corrections through the iterative rotations and assignments (IRA) algorithm \cite{gundeIRAShapeMatching2021}.
The SI contains more details.

Many methods use synthetic benchmarks without considering the broader applicability of the underlying algorithms.
The NEB applies to gas phase molecular systems and to extended systems in catalysis alike.
We use the OptBench \cite{chillBenchmarksCharacterizationMinima2014} Pt(111) heptamer island benchmark, which focuses on metallic surface diffusion.
This dataset contains 59 low-energy mechanisms for the rearrangement of a platinum heptamer on a Pt(111) slab.
Each system comprises 343 atoms: 7 adatoms on a fcc Pt(111) slab model composed of six atomic layers, three out of which are kept frozen to represent the bulk system.
To strictly adhere to the benchmark definition, these calculations use an analytic Morse potential (\texttt{morse\_pt}) rather than the MLIP.
The setup involved 5 intermediate images initialized via linear interpolation, without energy weighted springs and a fixed spring constant of \(5\).
We enforced a tighter convergence criterion of \(0.001 \text{ eV}/\text{\AA}\) for the norm of the force vector.
Frozen atoms in the bottom slab eliminate rotational and translational degrees of freedom, so explicit removal of these modes is unnecessary.
The OCI-NEB and CI-NEB parameters are otherwise identical to those in Table \ref{tbl:opt_params}.
\section{Results}
\label{sec:org10d8173}
The OptBench Pt(111) heptamer benchmark probes efficiency in the regime of solid-state surface diffusion.
Table \ref{tbl:optbench_summary} summarizes the performance statistics.
Despite the use of a simple linear initialization and a tighter convergence criterion (\(0.001 \text{ eV}/\text{\AA}\)), OCI-NEB demonstrates improved efficiency.
OCI-NEB reduced the mean computational cost by 31\%, from 409 to 280 evaluations.
Of the 59 mechanisms, 52 show improvement while 7 exhibit modest regressions (up to 18\% increased cost), all on systems where the dimer triggers near a shallow basin.
The average RMSD between final saddle configurations is \(6.8 \times 10^{-5} \text{\AA}\) with negligible energetic differences, confirming that OCI-NEB locates the same transition states as CI-NEB on this benchmark.

\begin{table}[htbp]
\caption{\label{tbl:optbench_summary}Statistical summary of the Pt(111) Heptamer Island benchmark. Values represent gradient evaluations across the 59 diffusion mechanisms. The same parameters (\(\lambda_{rel} = 0.31\), \(\alpha_{tol} = 0.85\)) are used without system-specific tuning; 7 of 59 systems show modest regressions.}
\centering
\begin{tabular}{lrrl}
Metric & CI-NEB & OCI-NEB & Reduction\\
\hline
Mean & 409 & 280 & 31\%\\
Median & 357 & 311 & 13\%\\
Min & 172 & 92 & 47\%\\
Max & 1187 & 917 & 23\%\\
\end{tabular}
\end{table}

While the OptBench Pt(111) heptamer benchmark demonstrates efficiency gains for solid-state surface transitions, the Baker-Chan set focuses on molecular reactions in the gas phase.
Table \ref{tbl:baker_results} summarizes the computational cost for each system in the Baker-Chan benchmark.
OCI-NEB demonstrates a consistent advantage: a \(2.44\times\) overall speedup (13920 vs 5712 total force evaluations; median per-system ratio \(2.20\times\)).
OCI-NEB strictly improves performance on all 24 systems with 0/24 regressions, with speedup ratios ranging from \(1.43\times\) to \(8.76\times\).
The reported parameters were determined via an optuna-based sensitivity study using tree-structured Parzen estimation \cite{bergstraAlgorithmsHyperParameterOptimization2011} (Figure \ref{fig:optuna_study}).

\begin{table}[htbp]
\caption{\label{tbl:baker_results}Comparison of total force evaluations for CI-NEB and OCI-NEB on the Baker test set. \emph{Diff} indicates the reduction in evaluations achieved by OCI-NEB. \emph{RMSD} is the IRA-corrected root-mean-square displacement between the CI-NEB and OCI-NEB saddle point geometries.}
\centering
\begin{tabular}{rlllll}
\textbf{ID} & \textbf{Reaction} & \textbf{CI-NEB} & \textbf{OCI-NEB} & \textbf{Diff} & \textbf{RMSD} (\AA{})\\
\hline
 &  &  &  &  & \\
01 & \ce{HCN -> HNC} & \(698\) & \(179\) & \(519\) & \(3.6\times10^{-4}\)\\
02 & \ce{HCCH -> CCH2} & \(250\) & \(163\) & \(87\) & \(6.9\times10^{-4}\)\\
03 & \ce{H2CO -> H2 + CO} & \(562\) & \(259\) & \(303\) & \(2.2\times10^{-3}\)\\
04 & \ce{CH3O -> CH2OH} & \(242\) & \(108\) & \(134\) & \(3.3\times10^{-3}\)\\
05 & cyclopropyl ring opening & \(170\) & \(114\) & \(56\) & \(2.2\times10^{-3}\)\\
06 & bicyclo[1.1.0]butane \(\to\) \emph{trans}-butadiene & \(570\) & \(345\) & \(225\) & \(5.9\times10^{-3}\)\\
08 & formyloxyethyl 1,2-migration & \(434\) & \(164\) & \(270\) & \(5.9\times10^{-2}\)\\
09 & parent Diels-Alder cycloaddition & \(922\) & \(238\) & \(684\) & \(1.3\times10^{-2}\)\\
10 & s-tetrazine \ce{-> 2HCN + N2} & \(442\) & \(144\) & \(298\) & \(1.1\times10^{-2}\)\\
11 & \emph{trans}-butadiene \(\to\) \emph{cis}-butadiene & \(218\) & \(144\) & \(74\) & \(3.6\times10^{-3}\)\\
12 & \ce{CH3CH3 -> CH2CH2 + H2} & \(666\) & \(258\) & \(408\) & \(2.5\times10^{-3}\)\\
13 & \ce{CH3CH2F -> CH2CH2 + HF} & \(322\) & \(144\) & \(178\) & \(2.7\times10^{-3}\)\\
14 & acetaldehyde keto-enol tautomerism & \(274\) & \(159\) & \(115\) & \(3.4\times10^{-3}\)\\
15 & \ce{HCOCl -> HCl + CO} & \(434\) & \(149\) & \(285\) & \(2.8\times10^{-3}\)\\
16 & \ce{H2O + PO3- -> H2PO4-} & \(1034\) & \(721\) & \(313\) & \(5.5\times10^{-3}\)\\
17 & \ce{CH2CHCH2CH2CHO} Claisen rearrangement & \(866\) & \(451\) & \(415\) & \(1.1\times10^{-2}\)\\
18 & \ce{SiH2 + CH3CH3 -> SiH3CH2CH3} & \(778\) & \(322\) & \(456\) & \(5.2\times10^{-2}\)\\
19 & \ce{HNCCS -> HNC + CS} & \(834\) & \(413\) & \(421\) & \(6.6\times10^{-3}\)\\
20 & \ce{HCONH3+ -> NH4+ + CO} & \(674\) & \(210\) & \(464\) & \(8.0\times10^{-3}\)\\
21 & acrolein rotational TS & \(322\) & \(191\) & \(131\) & \(2.0\times10^{-2}\)\\
22 & \ce{HCONHOH -> HCOHNHO} & \(330\) & \(161\) & \(169\) & \(1.8\times10^{-2}\)\\
23 & \ce{HNC + H2 -> H2CNH} & \(1586\) & \(181\) & \(1405\) & \(6.7\times10^{-3}\)\\
24 & \ce{H2CNH -> HCNH2} & \(394\) & \(184\) & \(210\) & \(2.6\times10^{-3}\)\\
25 & \ce{HCNH2 -> HCN + H2} & \(898\) & \(310\) & \(588\) & \(8.5\times10^{-3}\)\\
\hline
 &  &  &  &  & \\
 & \textbf{Mean} & \(580.0\) & \(238.0\) & \(342.0\) & \(1.2\times10^{-2}\)\\
 & \textbf{Median} & \(502.0\) & \(182.5\) & \(291.5\) & \(5.9\times10^{-3}\)\\
\end{tabular}
\end{table}

\begin{figure}[htbp]
\centering
\includegraphics[width=.9\linewidth]{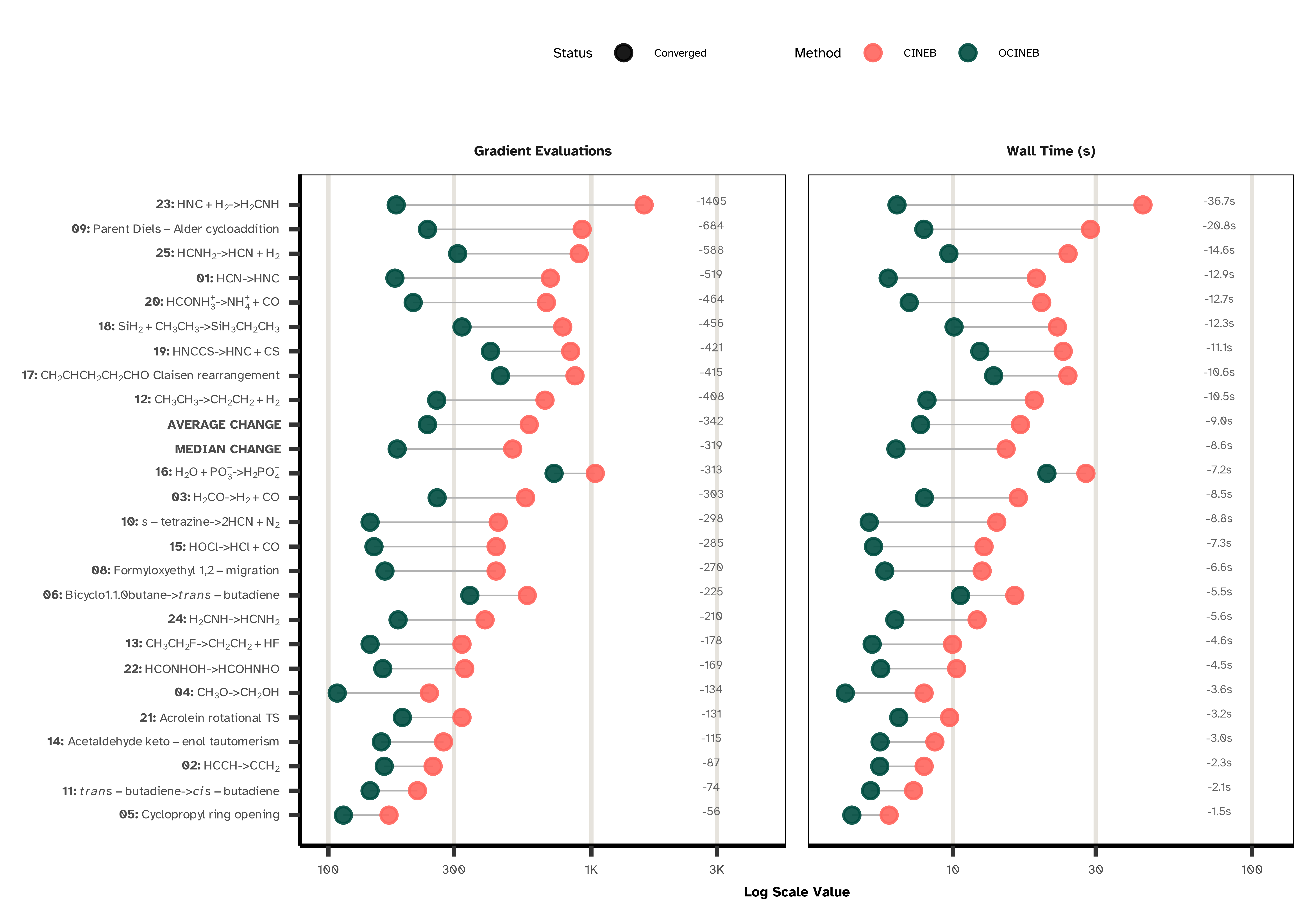}
\caption{\label{fig:dumbbell_plot}\textbf{\textbf{Comparative computational cost for the test set of transition configurations}} \cite{bakerLocationTransitionStates1996}. The ``dumbbell'' spans illustrate the reduction in gradient evaluations (left) and wall-clock time (right) achieved by OCI-NEB (teal) relative to CI-NEB (coral).}
\end{figure}

Figure \ref{fig:dumbbell_plot} visualizes the system-specific efficiency gains.
The breakdown demonstrates that the performance differential varies across the test set although the OCI-NEB strictly performs better for every system.
For straightforward rearrangements (e.g., \texttt{05}, \texttt{06}), both methods perform comparably.
For systems exhibiting larger structural reorganizations or flatter potential energy landscapes (e.g., \texttt{23}, \texttt{09}, \texttt{01}), the gap widens.
The ``dumbbell'' spans in Figure \ref{fig:dumbbell_plot} illustrate these reductions; OCI-NEB prevents the cost blowouts in sections of configuration space with near-zero forces.

\begin{figure}[htbp]
\centering
\includegraphics[width=.9\linewidth]{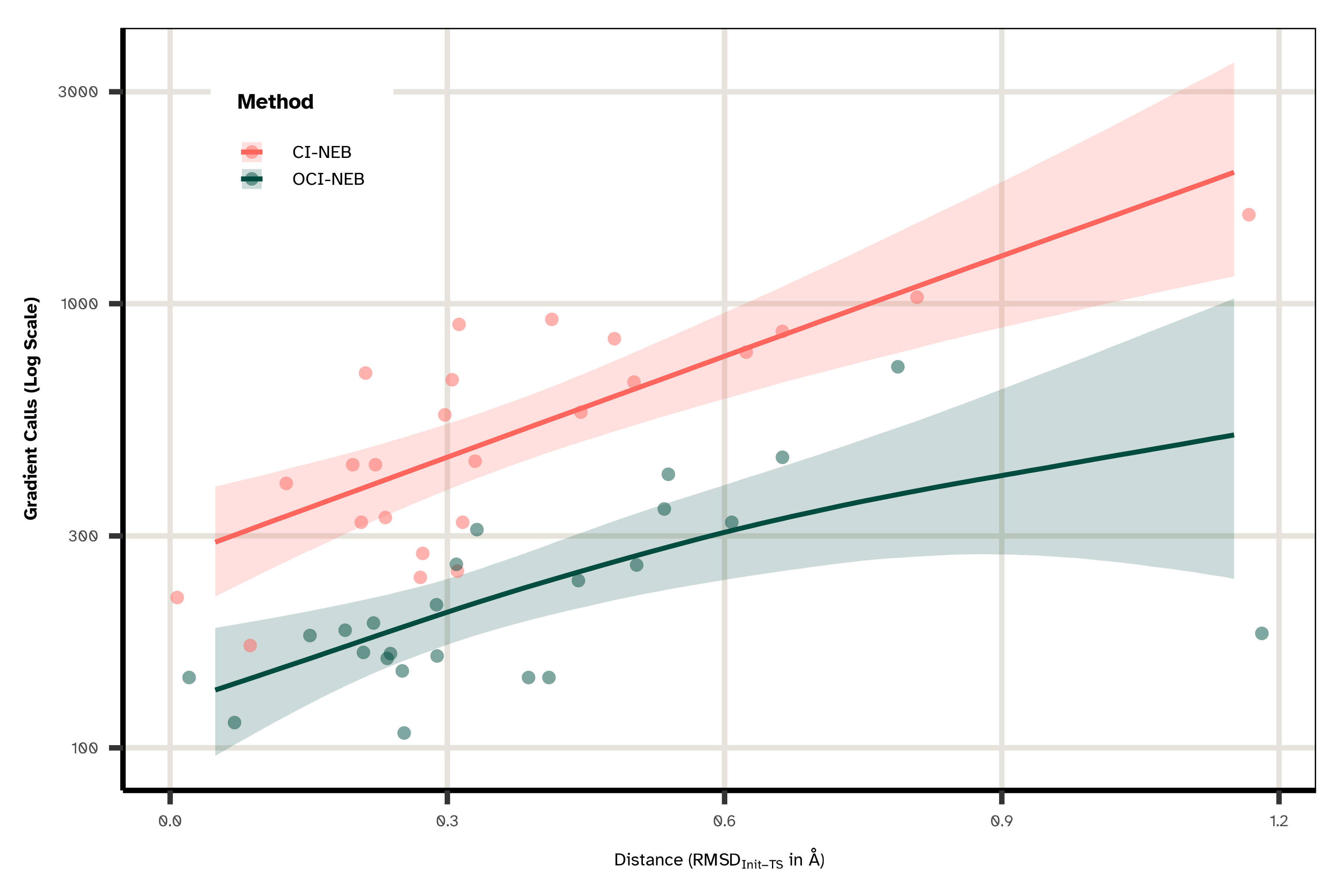}
\caption{\label{fig:brms_pes}\textbf{\textbf{Algorithmic robustness profile modeled via Bayesian negative binomial regression.}} The plot tracks the predicted computational cost (gradient calls, log scale) as a function of the initial structural displacement from the final transition state. Shaded regions indicate 95\% credible intervals. Both methods show a log-linear rise in cost with distance, but OCI-NEB (teal) maintains a consistent efficiency advantage over CI-NEB (coral), demonstrating that the MMF acceleration effectively lowers the computational overhead across the search space.}
\end{figure}

Figure \ref{fig:brms_pes} illustrates the dependence of computational effort on the quality of the initial guess.
Both algorithms exhibit a log-linear increase in the energy and gradient evaluations as the initial structural displacement grows -- a consequence of chain-of-states methods, where distal starting points require more iterations to drag the elastic band toward the saddle.
OCI-NEB establishes a consistent efficiency offset, operating strictly below the cost trajectory of CI-NEB.

In the near-harmonic regime around the transition state, the RMSD between the initial estimate and final configurations is low and so the credible intervals for the methods overlap (Figure \ref{fig:brms_pes}), consistent with parity.

When the initial path already lies close to the converged NEB path, convergence can occur before the MMF trigger fires.
As the displacement increases, the advantage of the hybrid protocol becomes clearer.
The MMF phase allows the climbing image to deviate slightly from the local tangent, and moves without having to move the other images of the band.
This lowers the pre-factor of the cost scaling and keeps OCI-NEB at a lower computational burden even as the search space expands.

\begin{figure}[htbp]
\centering
\includegraphics[width=.9\linewidth]{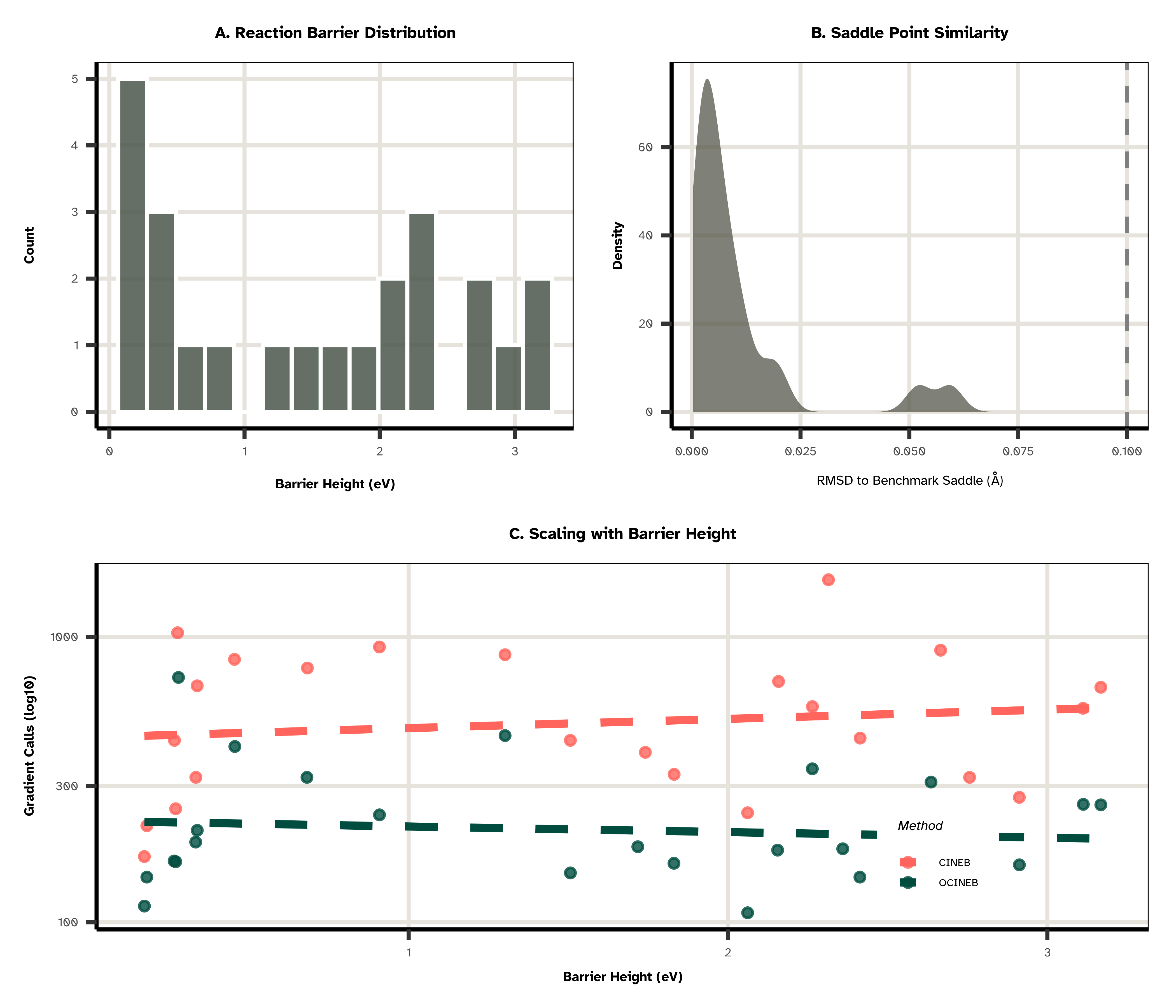}
\caption{\label{fig:dataset_char}\textbf{\textbf{Dataset Characterization and Drivers of Cost.}} (A) Distribution of barrier heights in the test set. (B) Density plot of the structural deviation of OCI-NEB transition states compared to CI-NEB saddle points; the density peaks below 0.1 \AA{}, confirming correct convergence and equivalent structures. (C) Scatter plot of Computational Cost vs. Barrier Height. The lack of a strong trend contrasts with the clear scaling seen in Figure \ref{fig:brms_pes}, indicating that initial structural guess quality drives cost more than the energetics of the reaction.}
\end{figure}

To further contextualize these performance gains, Figure \ref{fig:dataset_char} contrasts structural difficulty with energetic difficulty.
Panel C suggests a lack of correlation between the reaction barrier height and the computational cost.
While one might expect higher barriers to require more energy and gradient evaluations, the data indicates that the distance of the saddle configuration from the reactant (Figure \ref{fig:brms_pes}) is the primary driver of optimization effort.
Panel B confirms that this efficiency does not come at the cost of accuracy and the OCI-NEB identifies transition states with a structural deviation of around 0.01 \AA{} from the CI-NEB reference in almost every case.
This suggests that the ``stiffness'' of the optimization problem derives more from the ``memory'' of the initial path than the height of the hill, reinforcing the value of OCI-NEB decoupling from the neighboring images.

\begin{figure}[t]
\centering
\includegraphics[width=0.7\linewidth]{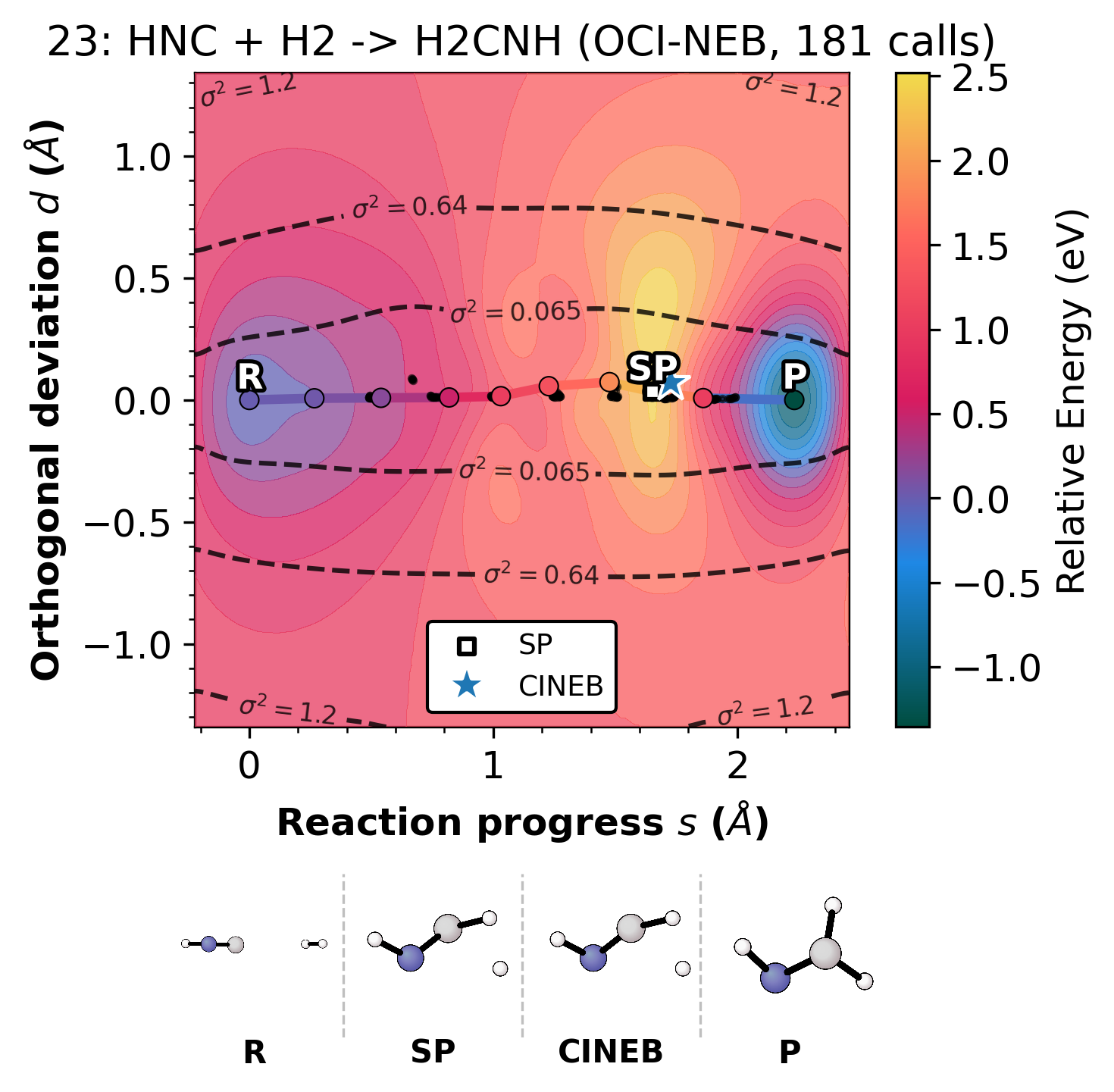}
\caption{\label{fig:landscape_example}\textbf{2D reaction landscape for \ce{HNC + H2 -> H2CNH} (System 23, \(8.76\times\) speedup).} The background contour is a derivative Gaussian process interpolation of the OCI-NEB optimizer history using an inverse multiquartic kernel \cite{goswamiTwodimensionalRMSDProjections2025}. Black dots: all images sampled during OCI-NEB. Colored circles: final converged path. White square: OCI-NEB saddle point (SP). Blue star: CI-NEB saddle point. Molecular structures at key configurations are shown below the landscape.}
\end{figure}

The largest RMSD between CI-NEB and OCI-NEB saddle points across the 24 systems is \(0.059 \text{ \AA}\) (formyloxyethyl, System 08), with a median of \(0.006 \text{ \AA}\) (Table \ref{tbl:baker_results}).
All deviations are sub-angstrom, confirming that OCI-NEB converges to the same transition states as CI-NEB.
Figure \ref{fig:landscape_example} illustrates the largest speedup in the benchmark: the \ce{HNC + H2 -> H2CNH} reaction (System 23, \(8.76\times\)), where OCI-NEB converges in 181 evaluations compared to 1586 for CI-NEB.
The 2D projections \cite{goswamiTwodimensionalRMSDProjections2025} for all 24 systems are provided in the Supplementary Information.

\section{Discussion}
\label{sec:disc}
The relationship between initial guess of the path and computational cost demonstrates the advantage for the dimer acceleration.
As shown in Figure \ref{fig:brms_pes}, the scaling profiles diverge as the initial root mean square deviation of atomic positions distance to the transition state (\(RMSD_{I,S}\)) increases.
For good estimates of the initial saddle configuration, (\(RMSD_{I,S}< 0.3 \AA\)), the performance remains parity-bound.
In the regime of poor initialization (\(RMSD_{I,S}> 0.6 \AA\)), CI-NEB exhibits a steep efficiency penalty, whereas OCI-NEB maintains a consistently lower cost trajectory.
This confirms that the decoupling of the reaction path allows the dimer method to recover efficiently from poor starting geometries.
The relation between distance to the point of interest and improvements from acceleration follows the trends of other methods \cite{goswamiEfficientImplementationGaussian2025}.
In all cases, we expect systems which take more iterations to have better gains from such techniques.

The alignment check \(\alpha > \alpha_{tol}\) serves two purposes: it ensures the dimer's minimum mode is sufficiently aligned with the NEB tangent to maintain Householder stability, and it guards against the dimer wandering to a saddle unrelated to the reaction coordinate.
Even when \(\alpha\) is high, the dimer may translate far from the initial climbing-image position, potentially finding a different first-order saddle.
The adaptive penalty (Eq. \ref{eq:linear_penalty}) and the force-based success criterion (\(F_{new} < F_{CI}\)) limit such excursions: if the dimer moves off-path without reducing the force, the threshold is raised and the NEB resumes control.
After a successful dimer step, the arc-length reparameterization redistributes images evenly along the updated path, preventing the band from becoming stretched.
This combination ensures the dimer accelerates convergence toward the saddle of interest rather than diverging to unrelated stationary points.
Starting the MMF early can lead to a separate, higher energy saddle when seeded with a poor initial path.
Consider the keto-enol tautomerism of vinyl alcohol, \texttt{14\_vinyl\_alcohol}, where the initial conditions of the indexing in the reactant and product change calculations markedly.
The IRA \cite{gundeIRAShapeMatching2021} algorithm in this instance, when applied to the endpoints, re-orders the H atom to drive a 1,2-hydrogen shift along the C-C bond.
Although energetically unfavorable, this permutation yields a lower Euclidean distance than the chemically correct 1,3-arch.
An initial path generated from the permuted endpoints forces the hydrogen atom through the dense electron density of the C-C bond, creating a massive artificial steric barrier.
Purely geometric alignment metrics cannot handle such situations, though for many cases including this one, masking the hydrogen atoms and aligning the heavier elements before using the Constrained Shortest Distance Assignments (CShDA), or point group symmetry measures \cite{gundeSOFIFindingPoint2024} can fix ordering concerns.
To keep our focus on the OCI-NEB, we report results based on the coordinates from the initial structures without alignment, as the endpoint alignment forms part of the initial path considerations not covered here, but actively worked on for a follow-up.

The OCI-NEB scheme exposes two user-facing parameters: \(\lambda_{rel}\), which governs when the NEB-to-dimer handover occurs, and \(\alpha_{tol}\), whose lower bound \(1/\sqrt{2}\) follows from the force inversion alignment condition.
The penalty shape \(S = 1\) is fixed by the uniqueness of the linear interpolation (Eq. \ref{eq:penalty_uniqueness}).

The reported parameters were identified through a systematic sensitivity study using the Optuna hyperparameter optimization framework \cite{akibaOptunaNextgenerationHyperparameter2019}.
The optimizer employs the Tree-structured Parzen Estimator (TPE) \cite{bergstraAlgorithmsHyperParameterOptimization2011}, a sequential model-based algorithm that differs from Gaussian process Bayesian optimization in a fundamental way: rather than modeling the objective as a function of parameters, \(p(y \mid x)\), TPE separately models the parameter distributions conditioned on performance.
Specifically, it partitions observations into ``good'' (below a quantile threshold \(y^*\)) and ``bad'' groups, fitting separate densities \(l(x) = p(x \mid y < y^*)\) and \(g(x) = p(x \mid y \geq y^*)\).
The expected improvement criterion then reduces to a ratio of these densities:

\begin{equation} \label{eq:tpe_ei} \mathrm{EI}(x) \propto \frac{l(x)}{g(x)} \end{equation}

This formulation avoids the cubic scaling of Gaussian process fitting and handles conditional, categorical, and tree-structured parameter spaces without modification, which suits the mixed parameter types in OCI-NEB.

Each of the 200 trials evaluates a candidate parameter vector by running OCI-NEB on all 24 Baker systems and summing the total force evaluations.
No regressions are hidden by a subset selection.
A MedianPruner terminates unpromising trials early when their intermediate results exceed the median of completed trials at the same step, reducing the total computational budget without biasing the final parameter estimates.

To quantify the contribution of each parameter to overall performance, we apply functional ANOVA (fANOVA) decomposition \cite{hutterEfficientApproachAssessing2014}.
The fANOVA framework decomposes the predicted objective function into additive components:

\begin{equation} \label{eq:fanova} f(\mathbf{x}) = f_0 + \sum_{i} f_i(x_i) + \sum_{i < j} f_{ij}(x_i, x_j) + \cdots \end{equation}

where \(f_0\) is the grand mean, \(f_i(x_i)\) captures the marginal effect of parameter \(i\), and \(f_{ij}(x_i, x_j)\) captures pairwise interactions.
The fraction of total variance attributable to each term identifies which parameters genuinely drive performance and which are effectively noise.

\begin{figure}[t]
\centering
\begin{minipage}[b]{0.48\textwidth}
  \includegraphics[width=\linewidth]{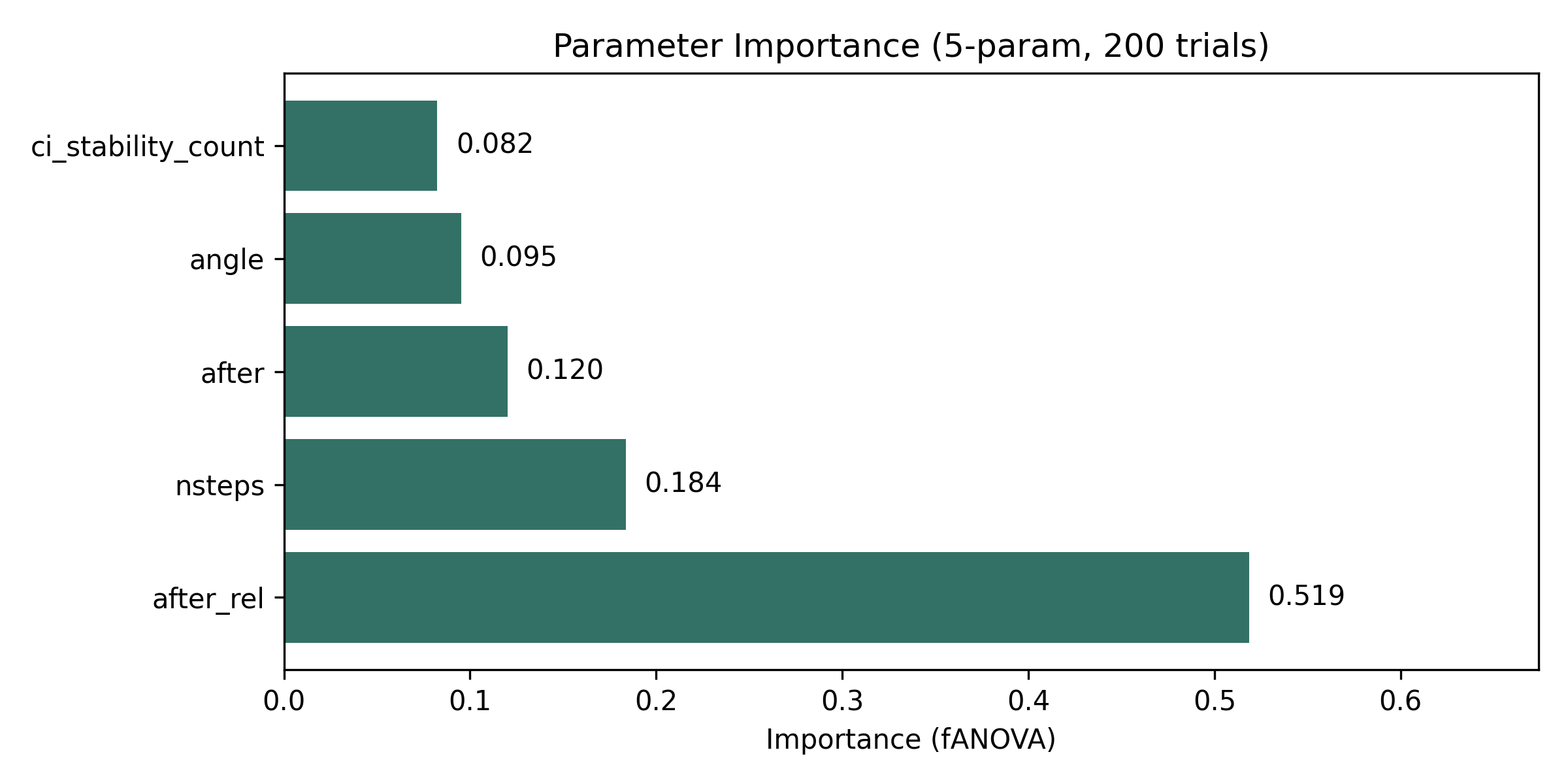}
  \centerline{\small (A) 5-parameter importance}
\end{minipage}
\hfill
\begin{minipage}[b]{0.48\textwidth}
  \includegraphics[width=\linewidth]{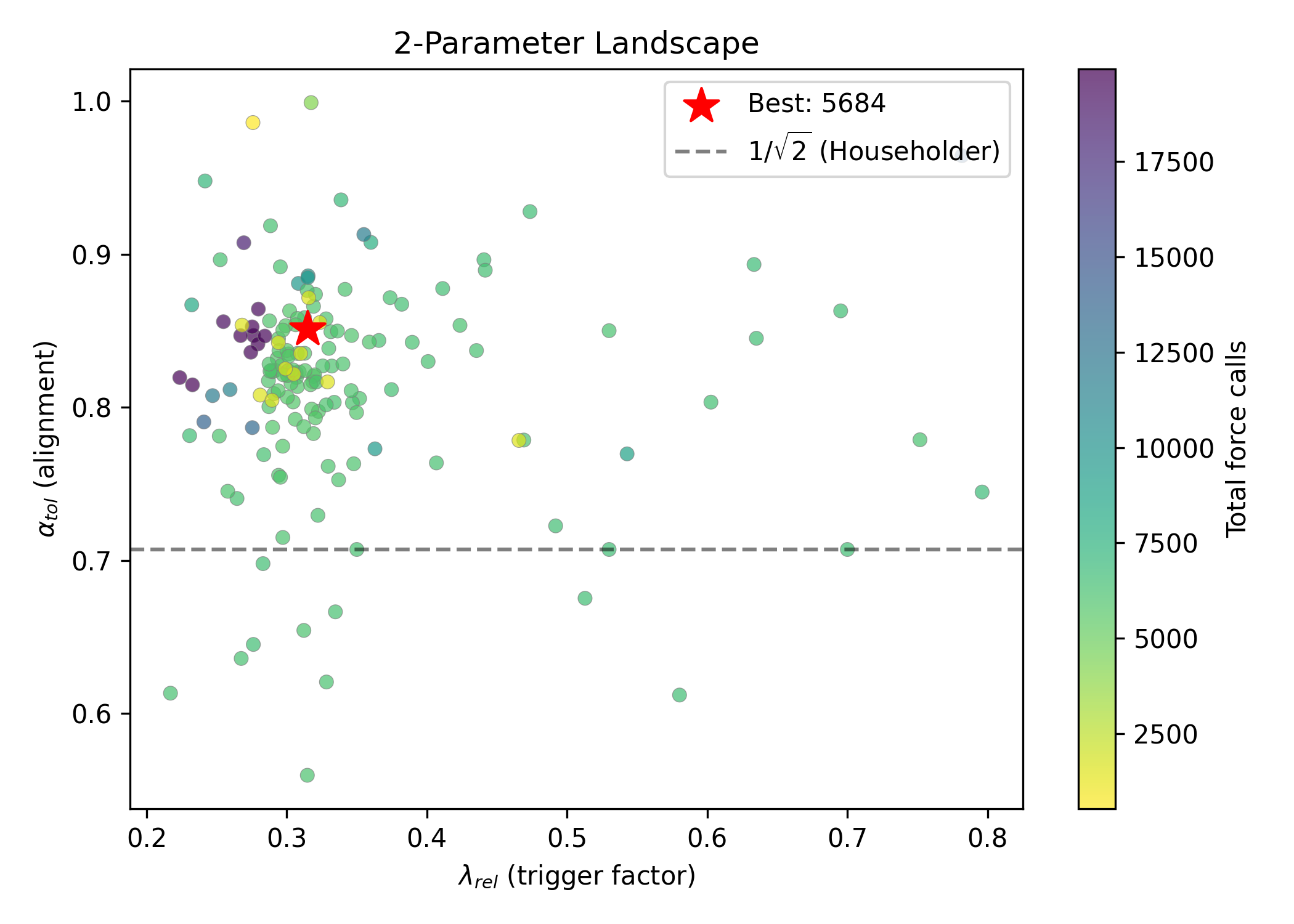}
  \centerline{\small (B) 2-parameter landscape}
\end{minipage}

\vspace{0.8em}
\begin{minipage}[b]{0.48\textwidth}
  \includegraphics[width=\linewidth]{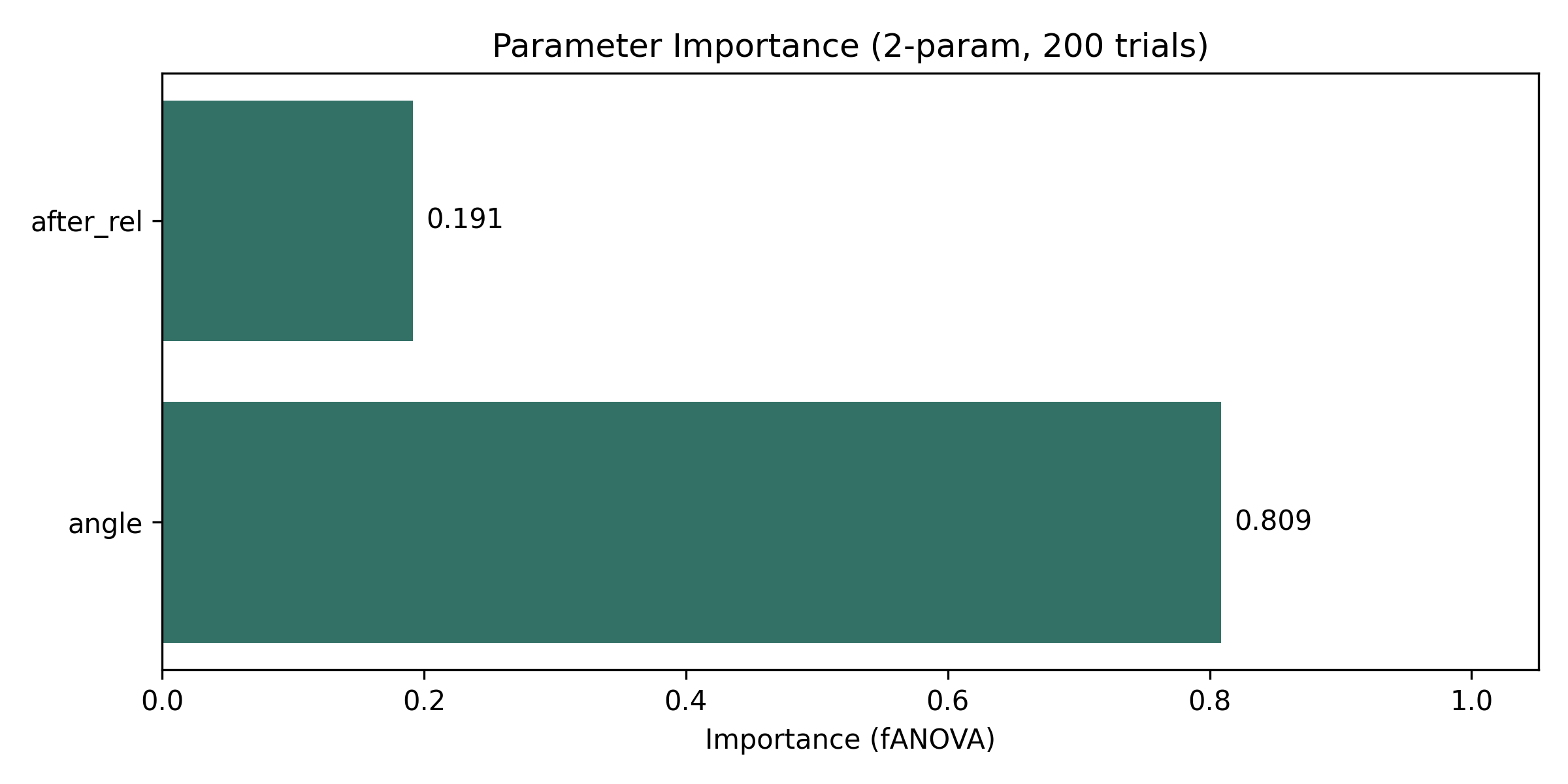}
  \centerline{\small (C) 2-parameter importance}
\end{minipage}
\caption{\textbf{Optuna sensitivity study} (200 trials, TPE sampler, all 24 Baker systems). (A) 5-parameter fANOVA importance: $\lambda_{rel}$ dominates, justifying the fixation of three minor parameters as internal constants. (B) 2-parameter landscape: total force calls (color) as a function of $\lambda_{rel}$ and $\alpha_{tol}$. The dashed line marks the alignment bound $\alpha_{tol} = 1/\sqrt{2}$; trials below this line incur high cost. The red star marks the optimum at (0.31, 0.85). (C) 2-parameter fANOVA importance: with noise parameters removed, $\alpha_{tol}$ accounts for 81\% of the variance.}
\label{fig:optuna_study}
\end{figure}

The fANOVA importance analysis proceeds in two stages.
In the initial 5-parameter study (Figure \ref{fig:optuna_study}, panel A), \(\lambda_{rel}\) accounts for 52\% of the total variance, while \(\alpha_{tol}\), the absolute trigger floor, the MMF step budget, and the stability count each contribute less than 20\%.
This hierarchical structure motivates fixing the three least important parameters as internal constants, reducing the user-facing space to two parameters.

A second 200-trial study over the reduced (\(\lambda_{rel}\), \(\alpha_{tol}\)) space reveals the complementary picture (Figure \ref{fig:optuna_study}, panels B--C).
With the noise parameters removed, \(\alpha_{tol}\) now accounts for 81\% of the variance: the alignment tolerance is the primary sensitivity driver among the two retained parameters.
The contour landscape (panel B) shows the Householder stability bound \(\alpha_{tol} = 1/\sqrt{2}\) as a dashed line; trials below this bound incur catastrophic cost (light/yellow markers).
The optimal basin concentrates near (\(\lambda_{rel} \approx 0.31\), \(\alpha_{tol} \approx 0.85\)), with the best trial at 5684 total force calls (red star).
The margin between the theoretical minimum \(1/\sqrt{2} \approx 0.707\) and the empirical optimum 0.85 absorbs dimer-rotation approximation error far from the saddle.

\begin{figure}[t]
\centering
\includegraphics[width=0.7\linewidth]{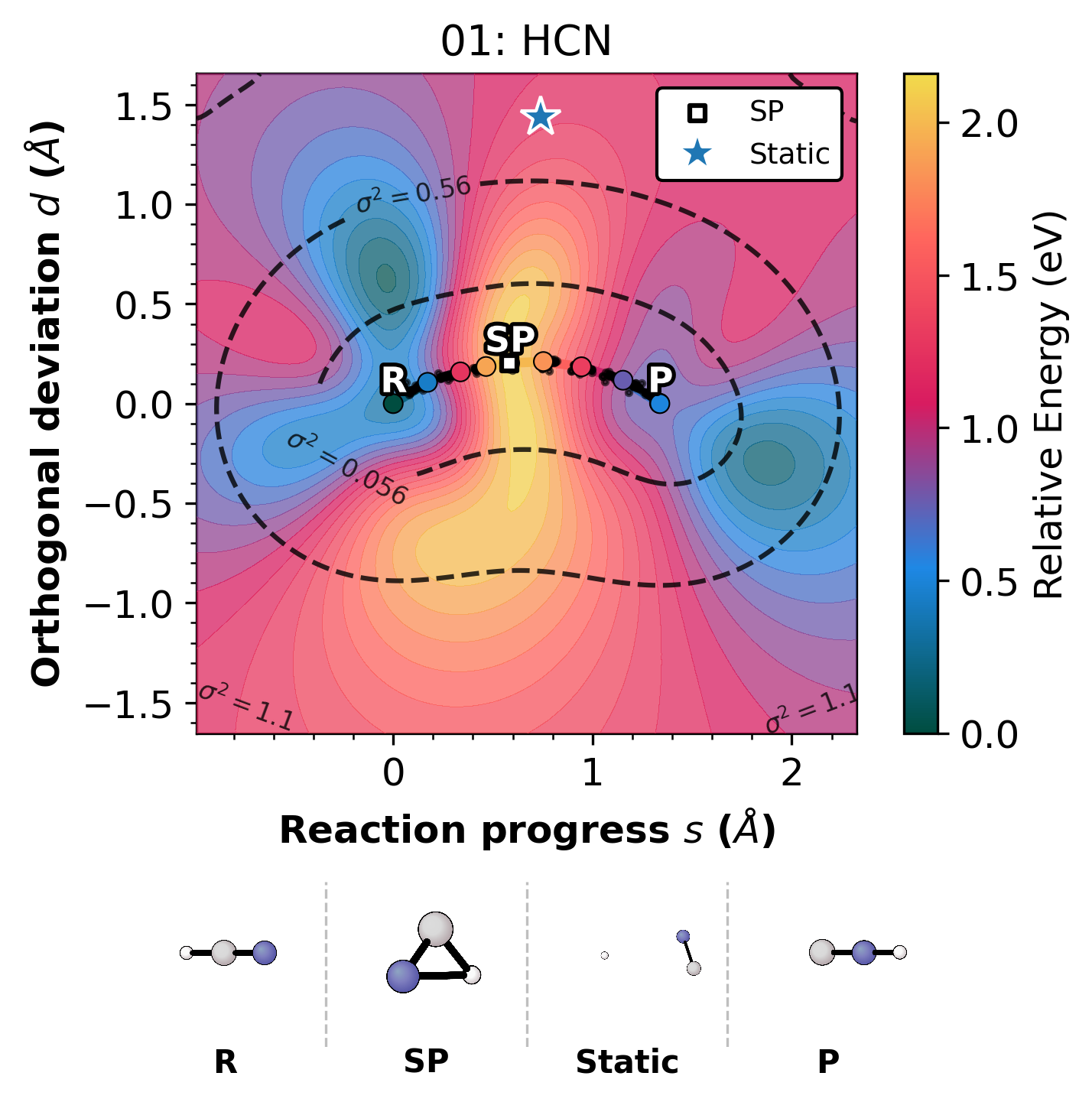}
\caption{\label{fig:hcn_static}\textbf{Static handover failure on HCN (System 01).} 2D RMSD projection \cite{goswamiTwodimensionalRMSDProjections2025}. SP (white square): CI-NEB saddle point; the OCI-NEB saddle (179 calls) coincides with SP at this scale (RMSD \(< 0.001\) \AA{}) and is omitted for clarity. Static (orange star): dimer endpoint from the static handover protocol (531 total calls), which drifts to an unrelated saddle (travel RMSD 0.885 \AA{}) far off the reaction path. The fixed 0.5 eV/\AA{} threshold triggers the dimer prematurely, and the dimer follows the local lowest mode to an irrelevant stationary point. OCI-NEB avoids this failure via the adaptive trigger and alignment check.}
\end{figure}

The ``static handover'' approach, corresponding to the protocol of \AA{}sgeirsson et al. \cite{asgeirssonNudgedElasticBand2021} and forming the basis of the automated workflow of Park et al. \cite{parkAutomatedWorkflowTransition2025}, applies a single fixed force threshold (0.5 eV/\AA{}) for the NEB-to-dimer transition without adaptive back-off.
Across all 24 Baker systems, the static protocol requires 8360 total force evaluations compared to 5712 for OCI-NEB -- a 46\% overhead (Table S2).
On System 01 (HCN, Figure \ref{fig:hcn_static}), the static protocol converges to the \textbf{wrong} saddle point (dimer travel RMSD 0.885 \AA{}, status BAD), while OCI-NEB finds the correct transition state in 179 calls.
On System 16 (\ce{H2PO4-}), the static dimer wanders for 1422 force evaluations (travel RMSD 1.10 \AA{}) before converging.
OCI-NEB converges all 24 systems correctly with zero regressions, eliminating the need for manual threshold selection and providing built-in failure recovery via the adaptive penalty.
The Claisen system (System 17), where both methods converge correctly but OCI-NEB is 17\% faster (451 vs 543 calls), is presented in the Supplementary Information (Figure S8).

The reported parameters (\(\lambda_{rel} = 0.31\), \(\alpha_{tol} = 0.85\)) emerge from the TPE-based sensitivity study (Figure \ref{fig:optuna_study}).
This configuration converged all 24 Baker systems strictly faster than CI-NEB, with a \(2.44\times\) overall speedup and 0/24 regressions.
Parallel implementations of the NEB may make stricter protocols more attractive by increasing stability without wall-time increases; even with parallel implementations, OCI-NEB should yield favorable statistics.

A natural question is why the dimer follows the NEB rather than the reverse.
The choice of taking the NEB-CI tangent as the preferred mode over the dimer axis comes from practical considerations.
When switching to the NEB after the dimer axis dephases from the NEB-CI tangent, despite reparameterization, if the dimer has moved the image far from the path, many calculations are required to enforce equal spacing, triggering the CI, and eventually the dimer.
A dimer being an initial point method not tied to other points on the surface, can be started from any point with lower cost.
Thus the dimer follows the NEB, and not the other way around, and OCI-NEB provides greater dividends with more images in the path.

The dimer-axis may differ from the NEB-CI tangent, even though they both approximate the Hessian's lowest mode, due to the approximations made in each method.
In the CI-NEB, the tangent estimate derives from a weighted average of the vectors to neighboring images from the CI, which has variable spacing based on the springs.
The dimer axis derives more directly from the local mode of the Hessian, and always has a small finite fixed distance between its images.
The dimer axis is more flexible than the NEB-CI tangent, which is damped by the presence of the rest of the band.
These separate considerations effectively improve the response of the NEB-CI in terms of the number of calculations needed to track the true minimum mode of the Hessian.
\section{Conclusions}
\label{sec:conclusions}
We present the OCI-NEB algorithm, an adaptive hybrid optimization strategy that integrates the stability of the NEB method with the efficiency of a MMF saddle point search.
The algorithm exposes two parameters: the relative trigger factor \(\lambda_{rel}\) and the alignment tolerance \(\alpha_{tol}\), whose lower bound \(1/\sqrt{2}\) follows from the force inversion alignment condition.
The penalty shape (\(S = 1\), \(B = 0.5\)) is the unique linear member of the admissible family and requires no tuning.
After each successful dimer step, an arc-length reparameterization redistributes images evenly along the path at zero force-call cost, maintaining path quality throughout the hybrid optimization.
The underlying force inversion structure, shared by both the climbing image and dimer dynamics, ensures that switching between the two phases preserves the shared fixed point at the saddle.

The method assumes an initial path (here from S-IDPP) with energy-weighted springs providing higher image density near the saddle.
By decoupling the climbing image from the band during MMF bursts, OCI-NEB reduces the total number of force evaluations and relaxes the requirement on the number of images bracketing the saddle.

Benchmarking against the standard Baker-Chan transition state set using a modern machine-learned potential, PET-MAD-S v1.5.0 demonstrates the practical utility of this approach.
Empirically, OCI-NEB achieves a \(2.44\times\) overall speedup in force evaluations compared to CI-NEB, with strict improvement on all 24 systems.
More importantly, the method exhibits superior performance even where the standard nudged elastic band and the dimer individually struggle, showing an increase in performance relative to running the NEB to a fixed tolerance and then switching to the dimer as previously posited in the literature.

A Bayesian negative binomial regression accounting for the distance from the initial saddle estimate to the final configuration (with permutation-corrected distances) quantifies the expected improvement: OCI-NEB requires \(0.43\times\) {[}95\% CrI: 0.36, 0.50] as many gradient evaluations as CI-NEB, corresponding to a \(57.4\%\) reduction [\(-63.9\%\), \(-49.6\%\)].
The model and diagnostics are detailed in the Supplementary Information.

The method proves equally effective for surface systems.
Tests on the OptBench Pt(111) heptamer island dataset, comprising 59 distinct diffusion mechanisms—confirm that OCI-NEB maintains its efficiency advantage.
Despite tight convergence criteria and the presence of low-frequency modes typical of surface diffusion, the hybrid protocol yielded a 31\% mean reduction in force calculations with negligible structural deviation from the CI-NEB saddle points.

These results indicate that OCI-NEB is well suited to high-throughput transition state searches, including those driven by machine-learned potentials whose surface topology may deviate from the smooth harmonic basins assumed by traditional optimizers.
By dynamically decoupling the saddle point search from the reaction path constraints, OCI-NEB provides a reliable pathway to automated chemical discovery in complex systems.
As machine-learned potentials reduce the cost of individual force evaluations, the computational bottleneck shifts toward optimization efficiency on rough landscapes.
The method also degrades gracefully to match CI-NEB results by tuning the relative trigger factor, which controls the degree of path relaxation before dimer activation.
These scale-independent parameters transfer across diverse chemical systems without re-tuning, while the penalty shape is derived from first principles.
OCI-NEB therefore holds promise for high-throughput calculations, where a single protocol can be applied across a diverse dataset for reaction pathway discovery.
\section{Supporting Information\hfill{}\textsc{ignoreheading:appendix}}
\label{sec:org5721af1}

\begin{appendix}
\section{Reproduction note}
\label{sec:si:repro}
The full set of benchmark inputs, raw outputs, analysis scripts, and pinned runtime environments used in this study are publicly archived. The GitHub repository and Materials Cloud archive contain the original runs for all systems, the scripts used to generate every figure and table, and environment specifications (container/environment manifests) that reproduce the computational environment.

For the ablation study on the parameters, and for the surface systems, we ran two sets of case studies, and these logs are also provided in the archive. The archive also includes formal analysis of the algorithm: symbolic derivations (SymPy) verifying the parameter reduction and convergence properties, and publication-quality diagrams (TikZ). The optuna-based parameter sensitivity study script and supplementary figure generation script are included. Users wishing to reproduce any specific experiment can either use the provided raw outputs, the pre-processed FAIR formatted \texttt{csv} data, or re-run the workflow using the included environment manifests and Snakemake pipelines; exact instructions and file paths are given in the archive and on the Github repository.
\section{Computational Workflow and Hardware Utilization}
\label{sec:si:computational}
To ensure reproducibility and facilitate high-throughput benchmarking across the
Baker test set, we orchestrated the entire simulation pipeline using the
Snakemake workflow management system \cite{molderSustainableDataAnalysis2021}.
This automated Directed Acyclic Graph (DAG) managed dependencies between data
retrieval, endpoint relaxation, initial path generation, and the final
chain-of-states optimizations.
\subsection{Environment and Software Stack}
\label{sec:si:envsoft}
We maintained the computational environment using the \texttt{pixi} package manager to
strictly version-control the software stack. The core simulation engine, EON
\footnote{\url{https://eondocs.org}}, interfaced with the PET-MAD-S v1.5.0 machine learning
potential via the Metatomic/Metatensor library
\cite{bigiMetatensorMetatomicFoundational2025}. This integration allowed the C++
client to query the Python-based PyTorch model directly within the EON address
space, minimizing inter-process communication overhead.
\subsection{GPU Acceleration and Parallelization Strategy}
\label{sec:si:gpupar}
We executed the benchmarks on a Lenovo ThinkStation P620 workstation equipped
with an AMD Ryzen Threadripper PRO 5945WX (24 cores, 48 threads) and an NVIDIA
T400 GPU (4 GB VRAM). To maximize computational throughput, the Snakemake
profile utilized 12 concurrent workers (\texttt{-c12}), effectively saturating
the physical cores of the CPU. However, instantiating 12 independent
PyTorch/CUDA contexts for the PET-MAD model would exceed the 4GB memory capacity
of the T400 GPU, and provide thrashing due to having to switch context several
times. To resolve this, we employed the NVIDIA Multi-Process Service (MPS).

MPS interposes between the operating system and the GPU so that multiple
processes share a single CUDA context. By enabling the MPS control daemon
(\texttt{nvidia-cuda-mps-control -d}), the 12 concurrent EON client processes submitted
compute kernels to the GPU without incurring the memory overhead of individual
context creation. This configuration allowed efficient, oversubscribed execution
of the benchmarks on a single, entry-level workstation card.
\subsection{Empirical measures}
\label{sec:si:empirical}
\subsubsection{Trigger metrics}
\label{sec:si:trigger}
We report the raw convergence metadata for the Baker-Chan benchmark set in Table \ref{tbl:bc_numdimer_act}.

Previous methodological studies often suggest that a single transition from the
Nudged Elastic Band (NEB) to Min-Mode Following (MMF) provides sufficient
convergence to the saddle point. The \(n_{triggers}\) column records the frequency
of adaptive algorithmic shifts for each system. With \(\lambda_{rel} = 0.31\), 18 of the 24 benchmarks
converge with a single MMF trigger. Six systems require a second trigger,
and two of those (Claisen and silylene insertion) incur backoffs that
drive final convergence in the NEB phase. The OCI-NEB uses a dynamic threshold
instead of a single static threshold for every system.

\begin{table}[htbp]
\caption{\label{tbl:bc_numdimer_act}\textbf{\textbf{OCI-NEB parameter traces.}}}
\centering
\begin{tabular}{lrrl}
system & n\textsubscript{triggers} & n\textsubscript{backoffs} & final\textsubscript{state}\\
\hline
01\textsubscript{hcn} & 1 & 0 & Converged (MMF)\\
02\textsubscript{hcch} & 1 & 0 & Converged (MMF)\\
03\textsubscript{h2co} & 1 & 0 & Converged (MMF)\\
04\textsubscript{ch3o} & 1 & 0 & Converged (MMF)\\
05\textsubscript{cyclopropyl} & 1 & 0 & Converged (MMF)\\
06\textsubscript{bicyclobutane} & 2 & 1 & Converged (MMF)\\
08\textsubscript{formyloxyethyl} & 1 & 0 & Converged (MMF)\\
09\textsubscript{parentdielsalder} & 2 & 1 & Converged (MMF)\\
10\textsubscript{tetrazine} & 1 & 0 & Converged (MMF)\\
11\textsubscript{trans}\textsubscript{butadiene} & 1 & 0 & Converged (MMF)\\
12\textsubscript{ethane}\textsubscript{h2}\textsubscript{abstraction} & 1 & 0 & Converged (MMF)\\
13\textsubscript{hf}\textsubscript{abstraction} & 1 & 0 & Converged (MMF)\\
14\textsubscript{vinyl}\textsubscript{alcohol} & 1 & 0 & Converged (MMF)\\
15\textsubscript{hocl} & 1 & 0 & Converged (MMF)\\
16\textsubscript{h2po4}\textsubscript{anion} & 2 & 1 & Converged (MMF)\\
17\textsubscript{claisen} & 2 & 2 & Converged (NEB)\\
18\textsubscript{silylene}\textsubscript{insertion} & 1 & 1 & Converged (NEB)\\
19\textsubscript{hnccs} & 2 & 1 & Converged (MMF)\\
20\textsubscript{hconh3}\textsubscript{cation} & 1 & 0 & Converged (MMF)\\
21\textsubscript{acrolein}\textsubscript{rot} & 1 & 0 & Converged (MMF)\\
22\textsubscript{hconhoh} & 1 & 0 & Converged (MMF)\\
23\textsubscript{hcn}\textsubscript{h2} & 1 & 0 & Converged (MMF)\\
24\textsubscript{h2cnh} & 1 & 0 & Converged (MMF)\\
25\textsubscript{hcnh2} & 1 & 0 & Converged (MMF)\\
\end{tabular}
\end{table}

Once the back-off mechanism activates, the threshold may be driven to a low enough threshold that convergence takes place in the NEB phase, as seen in systems \texttt{17\_claisen} and \texttt{18\_silylene\_insertion}. After each MMF segment completes, the algorithm performs an arc-length reparameterization that redistributes images evenly along the path. This maintains path quality before the next NEB phase resumes, at zero additional force-call cost. With \(\lambda_{rel} = 0.31\), most systems (18/24) converge in a single MMF trigger without any backoffs.

Consider \texttt{06\_bicyclobutane}, which triggers twice with one backoff:

\begin{verbatim}
Triggering MMF.  Force: 1.3831, Threshold: 1.5167 (0.31x baseline)
MMF backoff (status=-1). Force: 1.3831 -> 2.1467, Alignment:  0.845. New threshold: 1.3991 (0.29x baseline)
Triggering MMF.  Force: 0.2997, Threshold: 1.3991 (0.29x baseline)
\end{verbatim}

The aggregate improvement due to OCI-NEB is 8208 evaluations (Table 4), computed from the totals CI-NEB = 13920, OCI-NEB = 5712.
\subsubsection{Static handover}
\label{sec:si:static}
We consider ``static'' handoff for the Baker-Chan set, following the protocol
reported earlier \cite{asgeirssonNudgedElasticBand2021}, where the simulation
utilizes the NEB method until the forces drop below 0.5 eV/\AA{}, at which
point the algorithm initiates a standalone Dimer search using the NEB tangent as the
initial direction. Table \ref{tbl:bc_bc_static} summarizes the performance of this
single-switch strategy. While 23 systems converge to the correct saddle (GOOD),
System 01 (HCN) finds a different saddle point (BAD, travel RMSD 0.885 \AA{}),
demonstrating the risk of an unguarded static threshold. System 16 (\ce{H2PO4-})
incurs 2008 total calls (1422 for the dimer phase alone, travel RMSD 1.10 \AA{}),
illustrating that the dimer can wander far from the NEB estimate on
systems with diffuse saddle regions.

\begin{table}[htbp]
\caption{\label{tbl:bc_bc_static}\textbf{\textbf{Static handover results.}} The \texttt{Dimer\_Travel\_RMSD} represents the distance between the NEB estimate and the final converged saddle.}
\centering
\begin{tabular}{lrrrlr}
System & Total\textsubscript{Calls} & NEB\textsubscript{Calls} & Dimer\textsubscript{Calls} & Status & Dimer\textsubscript{Travel}\textsubscript{RMSD}\\
\hline
01\textsubscript{hcn} & 531 & 114 & 417 & BAD & 0.885095\\
02\textsubscript{hcch} & 190 & 170 & 20 & GOOD & 0.006724\\
03\textsubscript{h2co} & 307 & 298 & 9 & GOOD & 0.000909\\
04\textsubscript{ch3o} & 137 & 122 & 15 & GOOD & 0.035257\\
05\textsubscript{cyclopropyl} & 94 & 74 & 20 & GOOD & 0.009641\\
06\textsubscript{bicyclobutane} & 394 & 338 & 56 & GOOD & 0.055136\\
08\textsubscript{formyloxyethyl} & 177 & 122 & 55 & GOOD & 0.080202\\
09\textsubscript{parentdielsalder} & 393 & 362 & 31 & GOOD & 0.026726\\
10\textsubscript{tetrazine} & 297 & 258 & 39 & GOOD & 0.036301\\
11\textsubscript{trans}\textsubscript{butadiene} & 134 & 90 & 44 & GOOD & 0.017642\\
12\textsubscript{ethane}\textsubscript{h2}\textsubscript{abstraction} & 317 & 250 & 67 & GOOD & 0.164455\\
13\textsubscript{hf}\textsubscript{abstraction} & 213 & 186 & 27 & GOOD & 0.041939\\
14\textsubscript{vinyl}\textsubscript{alcohol} & 197 & 170 & 27 & GOOD & 0.012415\\
15\textsubscript{hocl} & 145 & 130 & 15 & GOOD & 0.007190\\
16\textsubscript{h2po4}\textsubscript{anion} & 2008 & 586 & 1422 & GOOD & 1.103242\\
17\textsubscript{claisen} & 543 & 498 & 45 & GOOD & 0.054223\\
18\textsubscript{silylene}\textsubscript{insertion} & 373 & 226 & 147 & GOOD & 0.312929\\
19\textsubscript{hnccs} & 294 & 162 & 132 & GOOD & 0.251432\\
20\textsubscript{hconh3}\textsubscript{cation} & 357 & 186 & 171 & GOOD & 0.131771\\
21\textsubscript{acrolein}\textsubscript{rot} & 181 & 146 & 35 & GOOD & 0.023442\\
22\textsubscript{hconhoh} & 187 & 146 & 41 & GOOD & 0.041629\\
23\textsubscript{hcn}\textsubscript{h2} & 310 & 290 & 20 & GOOD & 0.023301\\
24\textsubscript{h2cnh} & 224 & 82 & 142 & GOOD & 0.311682\\
25\textsubscript{hcnh2} & 357 & 314 & 43 & GOOD & 0.063760\\
\end{tabular}
\end{table}

The aggregate static handover cost across all 24 systems is 8360 force
evaluations, compared to 5712 for OCI-NEB -- a 46\% overhead. The static
protocol also fails to find the correct saddle on System 01 (HCN, discussed
in the main text). In contrast, OCI-NEB converges all 24 systems with 0
regressions and 0 saddle misidentifications.

\begin{figure}[H]
\centering
\includegraphics[width=0.7\linewidth]{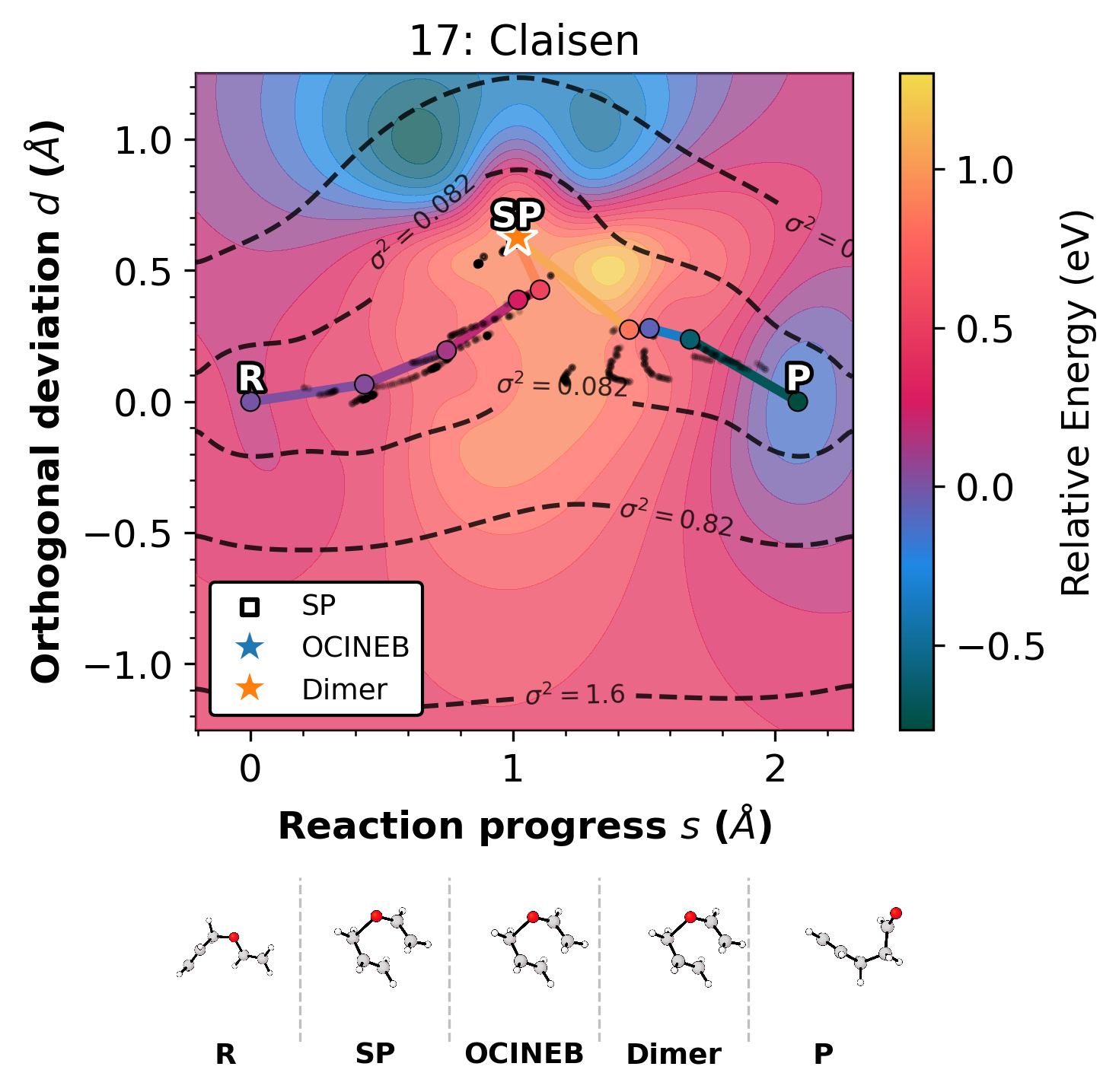}
\caption{\label{fig:si:claisen_static}\textbf{Static switchover comparison for Claisen (System 17).} CI-NEB landscape with OCI-NEB and Dimer saddle points overlaid. Both methods converge to the same saddle, but OCI-NEB is 17\% faster (451 vs 543 calls).}
\end{figure}

On the Claisen system (System 17, Figure \ref{fig:si:claisen_static}), both
the static handover and OCI-NEB converge to the same saddle point, but
OCI-NEB achieves this in 451 calls versus 543 for the static protocol
(17\% fewer evaluations).

\subsection{Performance Statistics}
\label{sec:si:stats}
The raw performance data compares the standard Climbing Image Nudged Elastic Band
(CINEB) against the Off-path Climbing Image NEB (OCI-NEB).
\subsubsection{Summary Statistics}
\label{sec:si:perf:summary}
\begin{itemize}
\item \textbf{CINEB:} Required a mean of \textbf{580.0} gradient evaluations across the 24 systems (total: 13920).
\item \textbf{OCI-NEB:} Required a mean of \textbf{238.0} gradient evaluations (total: 5712).
\item \textbf{Speedup:} The per-system speedup has a median of \textbf{2.20x} with a range of 1.43x to 8.76x, yielding a ratio-of-means aggregate speedup of \textbf{2.44x}.
\item \textbf{Accuracy:} Both methods converged to identical transition states. The IRA-corrected RMSD between CI-NEB and OCI-NEB saddle points has a mean of \textbf{0.012 \AA}, median of \textbf{0.006 \AA}, and a maximum of \textbf{0.059 \AA} (formyloxyethyl, System 08).
\end{itemize}
\subsubsection{Distributional Analysis}
\label{sec:si:perf:distrib}
\begin{figure}[H]
\centering
\includegraphics[width=1.0\textwidth]{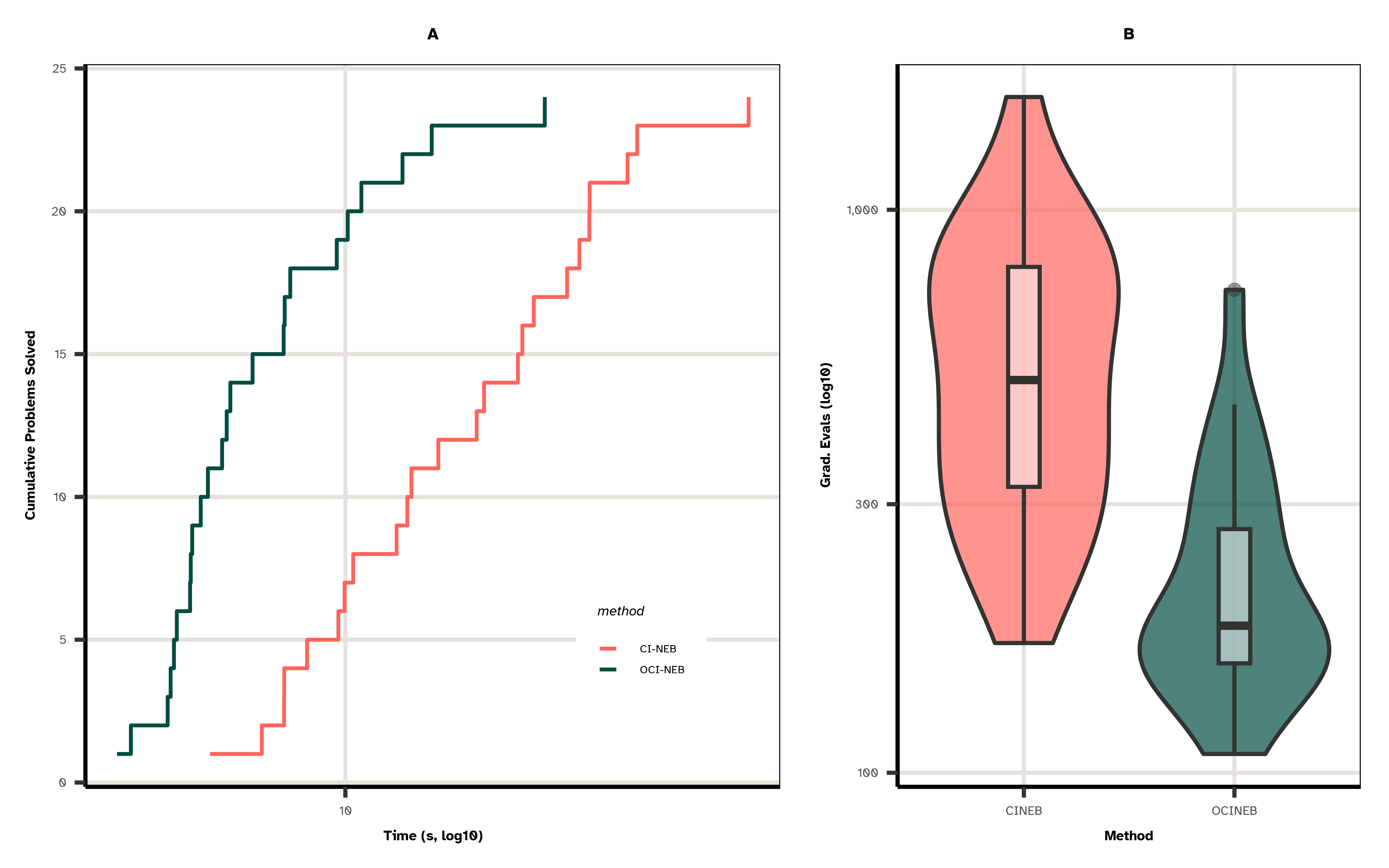}
\caption{\label{fig:cactus_violin}\textbf{Performance Distributions.} (A) Cumulative problems solved over wall-time. (B) Distribution of gradient evaluations (log scale) for CINEB and OCI-NEB.}
\end{figure}

Figure \ref{fig:cactus_violin} presents the aggregate performance. Panel A (Cactus
plot) shows the cumulative number of problems solved as a function of time (log
scale), where OCI-NEB (blue) maintains a strict advantage over CINEB (red). Panel
B (Violin plot) visualizes the distribution of gradient evaluations,
highlighting the reduction in the density of high-cost outliers for OCI-NEB.

\subsection{Protocol Ablation Study}
\label{sec:si:ablation}
The OCI-NEB exposes two tunable parameters: the \textbf{trigger factor}
(\(\lambda_{rel} = 0.31\)) and the \textbf{alignment tolerance}
(\(\alpha_{tol} = 0.85\)), whose lower bound \(1/\sqrt{2}\) follows from the
Householder stability condition (Eq. 17 of the main text). The \textbf{penalty
shape} (\(S = 1\)) is the unique linear member of the admissible family and
yields \(P(\alpha) = 0.5 + 0.5\alpha\) with base \(B = 0.5\).

The Optuna sensitivity study and fANOVA importance analysis are presented in
the main text (Figure 7). In the 5-parameter study, \(\lambda_{rel}\) dominates;
in the reduced 2-parameter study, \(\alpha_{tol}\) accounts for 81\% of the
variance, confirming the Householder stability bound as the primary design
constraint.

\begin{figure}[H]
\centering
\includegraphics[width=0.9\textwidth]{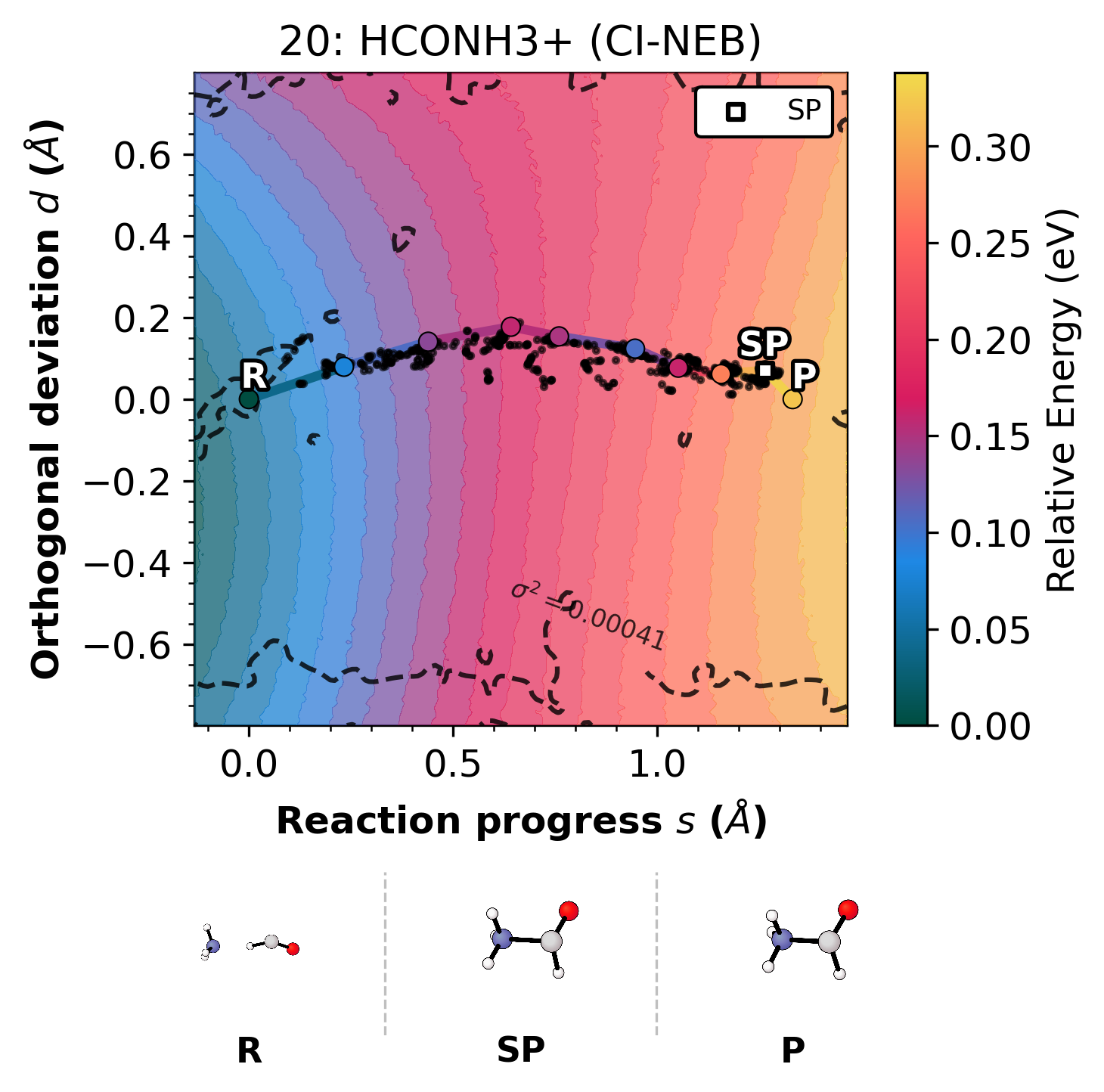}
\caption{\label{fig:si:ablation}\textbf\{\ce{HCONH3+} Fragmentation (System 20).\} \textbf{(A)} 2D RMSD projection of the reaction landscape. \textbf{(B)} OCI-NEB convergence path (210 calls, \(3.21\times\) speedup). \textbf{(C)} CI-NEB convergence path (674 calls).}
\end{figure}

\begin{center}
\includegraphics[width=0.48\textwidth]{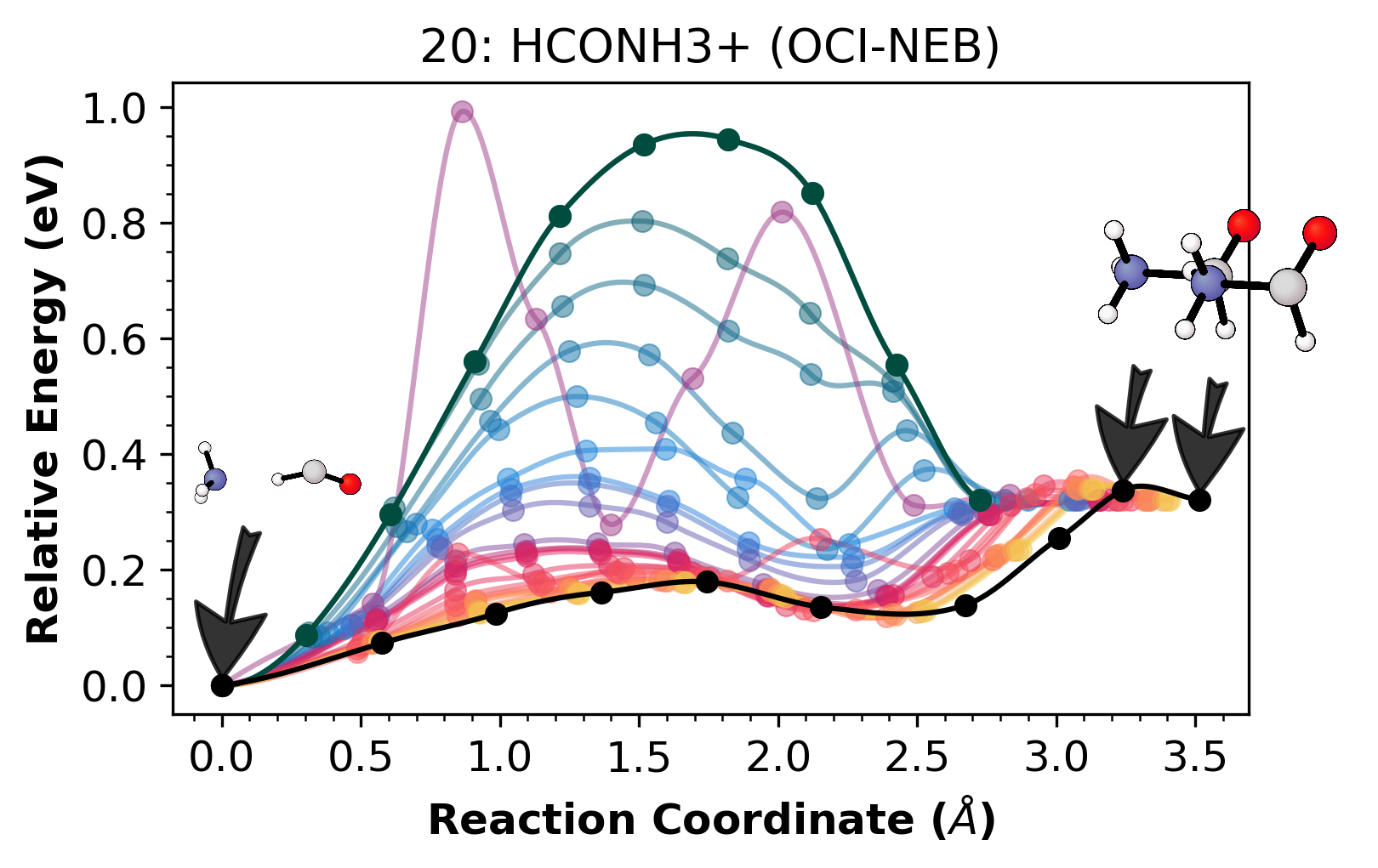}
\end{center}
\begin{center}
\includegraphics[width=0.48\textwidth]{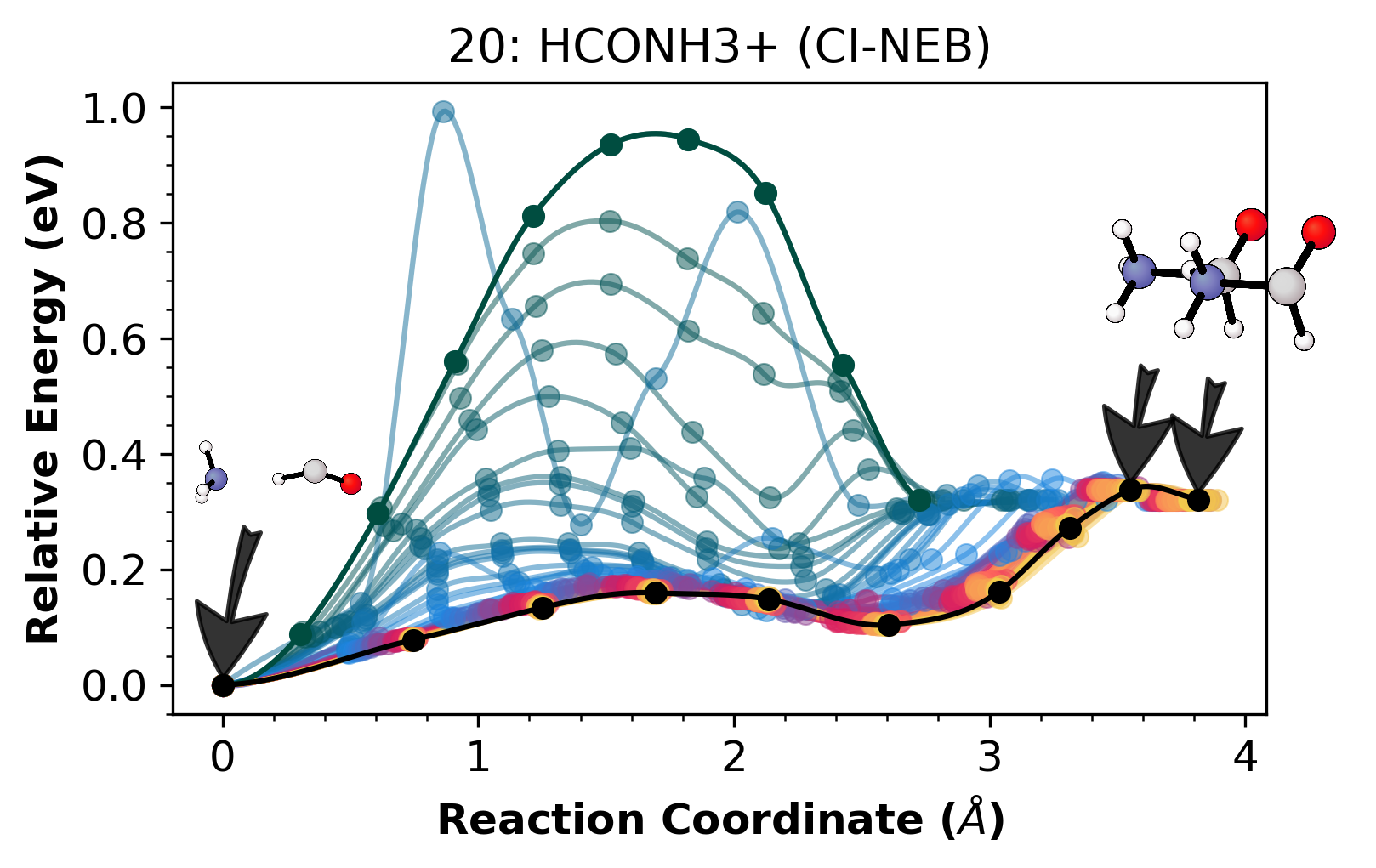}
\end{center}
\subsection{Bayesian Performance Modeling}
\label{sec:si:brms}
To quantify the algorithmic efficiency while accounting for system difficulty, we
fit a Bayesian Negative Binomial regression model.
\subsubsection{Why not a Gaussian model?}
\label{sec:si:whynorm}
PES call counts are non-negative integers that span a wide range across the benchmark set (86 to 914 evaluations). A Gaussian (normal) regression model would be inappropriate for several reasons. First, the data exhibits severe overdispersion: the pooled sample variance across all 48 observations exceeds the pooled mean by roughly two orders of magnitude, while a Gaussian model with constant variance would assign non-negligible probability to negative counts and fractional values. Second, a log-normal transformation, while enforcing positivity, discards the discrete count structure of the data and distorts the error model -- residual variability on the log scale does not translate back to interpretable uncertainty on the original count scale. Third, the variance of PES calls grows with the mean: difficult systems that require many evaluations also show larger absolute variability, a pattern naturally captured by the negative binomial's mean-variance relationship (\(\text{Var}(y) = \mu + \mu^2 / \phi\)) but not by a homoscedastic Gaussian.

The negative binomial generalizes the Poisson distribution by introducing a shape parameter \(\phi\) that absorbs extra-Poisson variability. When \(\phi \to \infty\), the model reduces to Poisson; for finite \(\phi\), the variance exceeds the mean, which matches the empirical structure of this data. The Poisson model itself is also inadequate here because PES calls are not independent rare events; they reflect a correlated optimization trajectory whose total length depends on system-specific landscape features. \cite{goswamiBayesianHierarchicalModels2025a} provides an extended treatment of count-based performance modeling for computational chemistry benchmarks.
\subsubsection{Model Specification}
\label{sec:si:model}
The model predicts the number of potential energy surface (PES) calls (\(y\))
based on the method used and the quality of the initial path guess (\(x =
\text{RMSD}_{init-final}\)):

\begin{equation}
y_i \sim \text{NegBinomial}(\mu_i, \phi_i)
\end{equation}

The linear predictor for the mean \(\mu\) utilizes a spline to account for the
non-linear increase in cost as the initial guess degrades, grouped by the method:

\begin{equation}
\log(\mu_i) = \alpha + \beta_{\text{method}} + f(x)_{\text{method}} + (1 | \text{System})
\end{equation}

The shape parameter \(\phi\), controlling overdispersion, also varies by method:

\begin{equation}
\log(\phi_i) = \gamma_{\text{method}}
\end{equation}

Priors were set as follows:
\begin{itemize}
\item \(\beta \sim \text{Normal}(0, 1)\)
\item Spline SDs \(\sim \text{Exponential}(2)\)
\item Intercept \(\sim \text{Student-t}(3, 0, 2.5)\)
\item Dispersion \(\phi \sim \text{Normal}(0, 0.5)\)
\end{itemize}

These priors are weakly informative: they regularize the parameter space to prevent pathological fits but carry minimal information relative to the data. With 48 observations across 24 systems, the likelihood dominates the posterior. Re-fitting the model with tighter coefficient priors (\(\beta \sim \text{Normal}(0, 0.5)\)) or wider priors (\(\beta \sim \text{Normal}(0, 2)\)) yields posteriors that differ by less than 0.01 in the median estimates and produce overlapping 95\% credible intervals, confirming that the results are data-driven rather than prior-driven.

\begin{lstlisting}[language=r,numbers=none]
 Family: negbinomial
  Links: mu = log; shape = log
Formula: count ~ method + s(RMSD_Init_Final, by = method, k = 3) + (1 | system_id)
         shape ~ method
   Data: data (Number of observations: 48)
  Draws: 8 chains, each with iter = 5000; warmup = 2000; thin = 1;
         total post-warmup draws = 24000

Smoothing Spline Hyperparameters:
                                    Estimate Est.Error l-95%
sds(sRMSD_Init_FinalmethodCINEB_1)      0.45      0.44     0.01     1.61 1.00
sds(sRMSD_Init_FinalmethodOCINEB_1)     0.81      0.70     0.02     2.54 1.00
                                    Bulk_ESS Tail_ESS
sds(sRMSD_Init_FinalmethodCINEB_1)     17817    10655
sds(sRMSD_Init_FinalmethodOCINEB_1)    10490     9587

Multilevel Hyperparameters:
~system_id (Number of levels: 24)
              Estimate Est.Error l-95%
sd(Intercept)     0.25      0.09     0.06     0.42 1.00     3487     4245

Regression Coefficients:
                                Estimate Est.Error l-95%
Intercept                           6.24      0.08     6.08     6.41 1.00    10001
shape_Intercept                     2.61      0.41     1.80     3.40 1.00     5373
methodOCINEB                       -0.85      0.08    -1.02    -0.69 1.00    24446
shape_methodOCINEB                  0.03      0.42    -0.81     0.86 1.00    14838
sRMSD_Init_Final:methodCINEB_1      0.42      0.08     0.26     0.58 1.00    11964
sRMSD_Init_Final:methodOCINEB_1     0.29      0.09     0.12     0.47 1.00    10579
                                Tail_ESS
Intercept                          12242
shape_Intercept                     9798
methodOCINEB                       17992
shape_methodOCINEB                 16072
sRMSD_Init_Final:methodCINEB_1     14316
sRMSD_Init_Final:methodOCINEB_1    14348

Draws were sampled using sample(hmc). For each parameter, Bulk_ESS
and Tail_ESS are effective sample size measures, and Rhat is the potential
scale reduction factor on split chains (at convergence, Rhat = 1).
\end{lstlisting}

With the exact results in Table \ref{tbl:suppl:pes}.

\begin{table}[htbp]
\caption{\label{tbl:suppl:pes}Results of the PES model}
\centering
\begin{tabular}{lrl}
Effect\textsubscript{Type} & Median Effect & 95\% CrI\\
\hline
Expected PES Calls (Baseline: CINEB) & 514.50 & {[}438.32, 604.92]\\
Multiplicative Factor (OCI-NEB vs CINEB) & 0.43 & {[}0.36, 0.50]\\
Percentage Change (OCI-NEB vs CINEB) & -57.4\% & {[}-63.9\%, -49.6\%]\\
\end{tabular}
\end{table}

Essentially, the model estimates the following effects:
\begin{description}
\item[{\textbf{Baseline Expectation}}] The expected number of gradient calls for the baseline CINEB method is 514.5 [95\% CrI: 438.3, 604.9].
\item[{\textbf{OCI-NEB Efficiency}}] The multiplicative factor for OCI-NEB relative to CINEB is 0.43 [95\% CrI: 0.36, 0.50].
\item[{\textbf{Reduction}}] This corresponds to a percentage change of \textbf{-57.4\%} {[}95\% CrI: -63.9\%, -49.6\%] in computational effort.
\end{description}

Because the model uses a log link, the method coefficient operates multiplicatively. The marginal multiplicative factor of 0.43 reported in Table \ref{tbl:suppl:pes} integrates over the posterior distribution of all model terms (random intercepts, spline, and residual variance). For a typical system, OCI-NEB requires approximately 43\% as many gradient evaluations as CI-NEB. The 95\% credible interval [0.36, 0.50] indicates that the true reduction factor lies between 50\% and 64\% fewer evaluations with 95\% probability, conditional on the model. These are Bayesian credible intervals -- direct probability statements about the parameter conditional on the data and model -- not frequentist confidence intervals.
\subsubsection{Shape Parameter Analysis}
\label{sec:si:model_shape}
The shape parameter \(\phi\) of the Negative Binomial distribution controls
overdispersion (higher values indicate more consistency/less variance). From the
model output, the CINEB shape parameter has a median of \(\exp(2.61) = 13.6\) and
the OCI-NEB shape is \(\exp(2.61 + 0.03) = 14.0\). In practical terms, the
coefficient of variation (CV) of the negative binomial distribution is
\(\text{CV} = \sqrt{1/\mu + 1/\phi} \approx 1/\sqrt{\phi}\) when \(\mu \gg \phi\), which holds here. For CINEB, \(\text{CV} \approx 1/\sqrt{13.6} = 0.27\), meaning the
standard deviation of PES calls is approximately 27\% of the mean for a given
system-method combination. This is substantial variability: a system with an
expected cost of 500 evaluations has a standard deviation around 135 evaluations.
For comparison, a Poisson model (which assumes variance equals the mean) would
predict \(\text{CV} = 1/\sqrt{500} = 0.045\), or just 4.5\% -- roughly six times
too narrow. The observed CV of 27\% confirms that PES call counts exhibit
strong overdispersion driven by system-specific landscape features not captured
by the fixed effects alone, and that the negative binomial is the appropriate
distributional family.

The near-zero difference between the two shape parameters
(\(\Delta \log \phi = 0.03\), 95\% CrI: [-0.81, 0.86]) indicates that OCI-NEB
does not trade consistency for speed: both methods show comparable
run-to-run variability conditional on system identity.

\begin{figure}[H]
\centering
\includegraphics[width=0.8\textwidth]{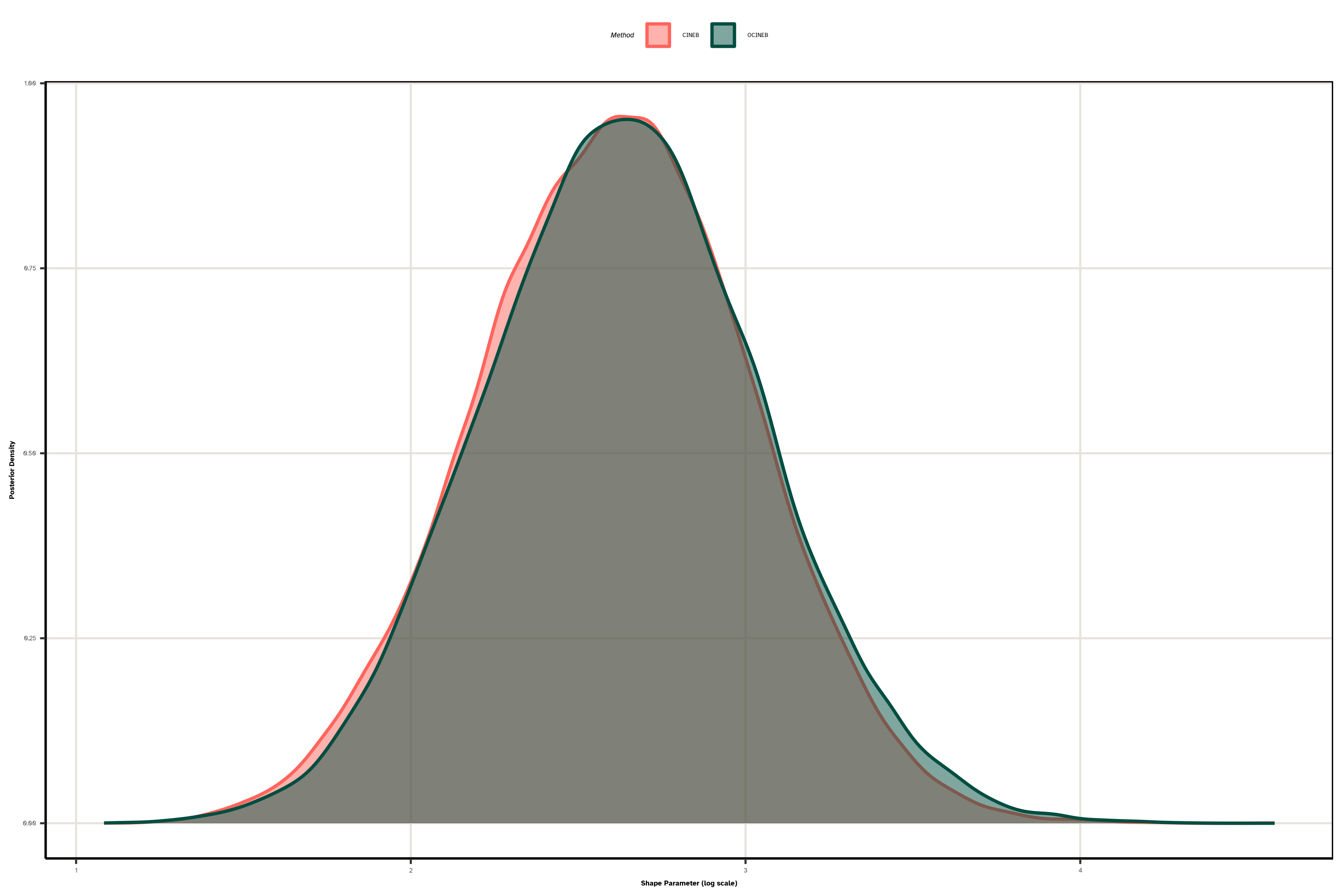}
\caption{\label{fig:shape_posterior}\textbf{Algorithmic Consistency.} Posterior distribution of the Negative Binomial shape parameter. OCI-NEB exhibits a similar dispersion profile to CINEB, indicating that the speedup does not come at the cost of erratic variance.}
\end{figure}

Figure \ref{fig:shape_posterior} illustrates the posterior density of the shape
parameter for both algorithms.

\subsubsection{Input Domain Validity}
\label{sec:si:input}
\begin{figure}[H]
\centering
\includegraphics[width=0.8\textwidth]{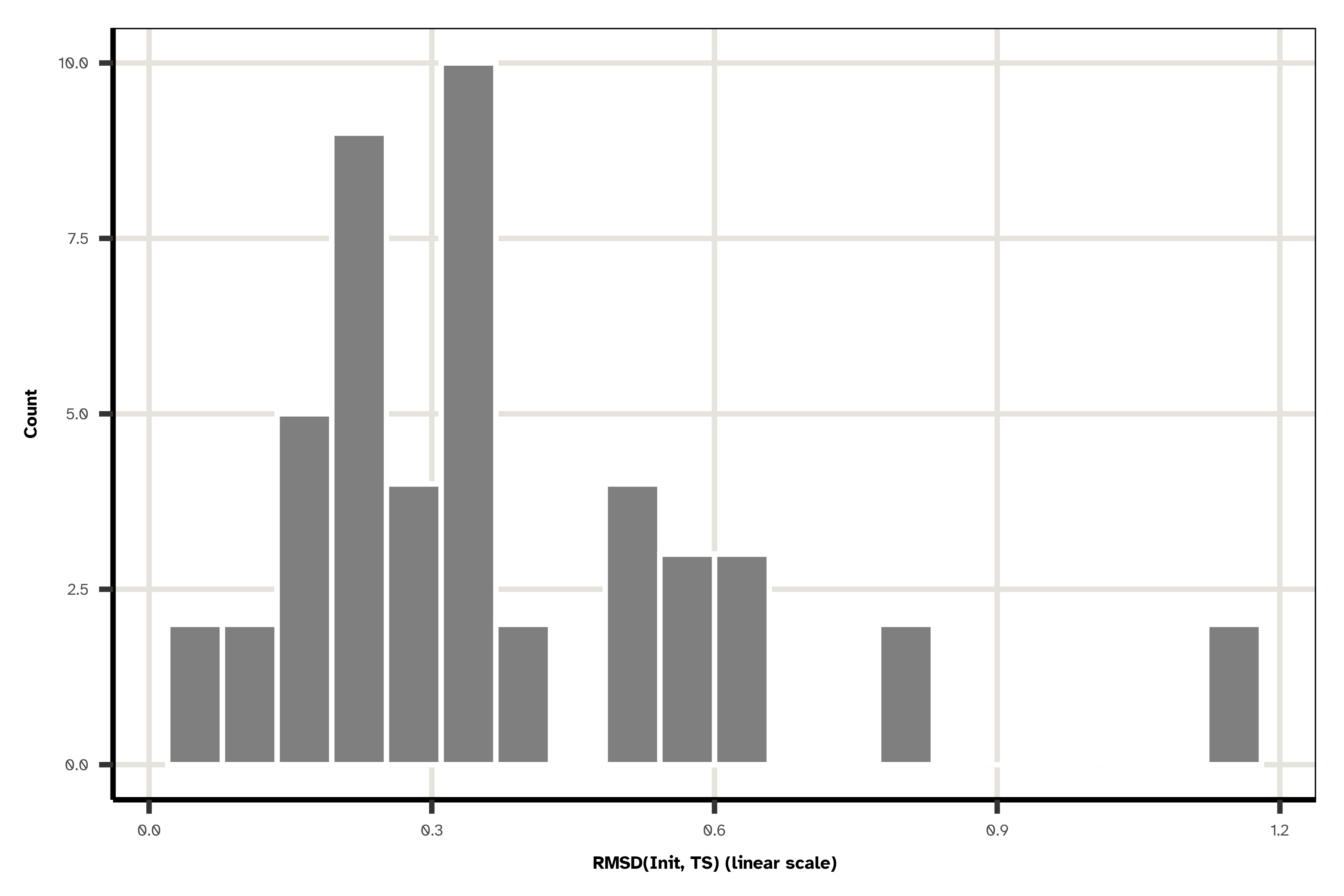}
\caption{\label{fig:nolog_dist}\textbf{Input Density.} Histogram of RMSD$\backslash$\textsubscript{Init}$\backslash$\textsubscript{Final} values, confirming sufficient data density to support the spline term in the regression model.}
\end{figure}

To ensure the spline term \(f(x)\) was valid, we verified the distribution of the
independent variable (RMSD between initial and final saddle). Figure
\ref{fig:nolog_dist} confirms data density across the linear range of 0.1 \AA{} to
0.8 \AA{}.

\subsection{Model Diagnostics}
\label{sec:si:diagnostics}
We performed extensive posterior predictive checks (PPC) and Leave-One-Out (LOO)
cross-validation to validate the model fit.

A posterior predictive check works as follows: once the model has been fitted, we draw parameter values from the posterior distribution and use them to simulate new synthetic datasets of the same size as the original data. Each simulated dataset represents one plausible outcome if the experiment were repeated under the same conditions. We then compare the distribution of these simulated datasets to the distribution of the actually observed data. If the model captures the data-generating process adequately, the simulated datasets should look statistically indistinguishable from the real data -- their density, spread, and tail behavior should overlap. Systematic discrepancies (e.g., the model consistently predicting too few high-cost outliers) would indicate model misspecification. The figures below show these comparisons.
\subsubsection{Posterior Predictive Density}
\label{sec:si:ppd}
\begin{figure}[H]
\centering
\includegraphics[width=0.8\textwidth]{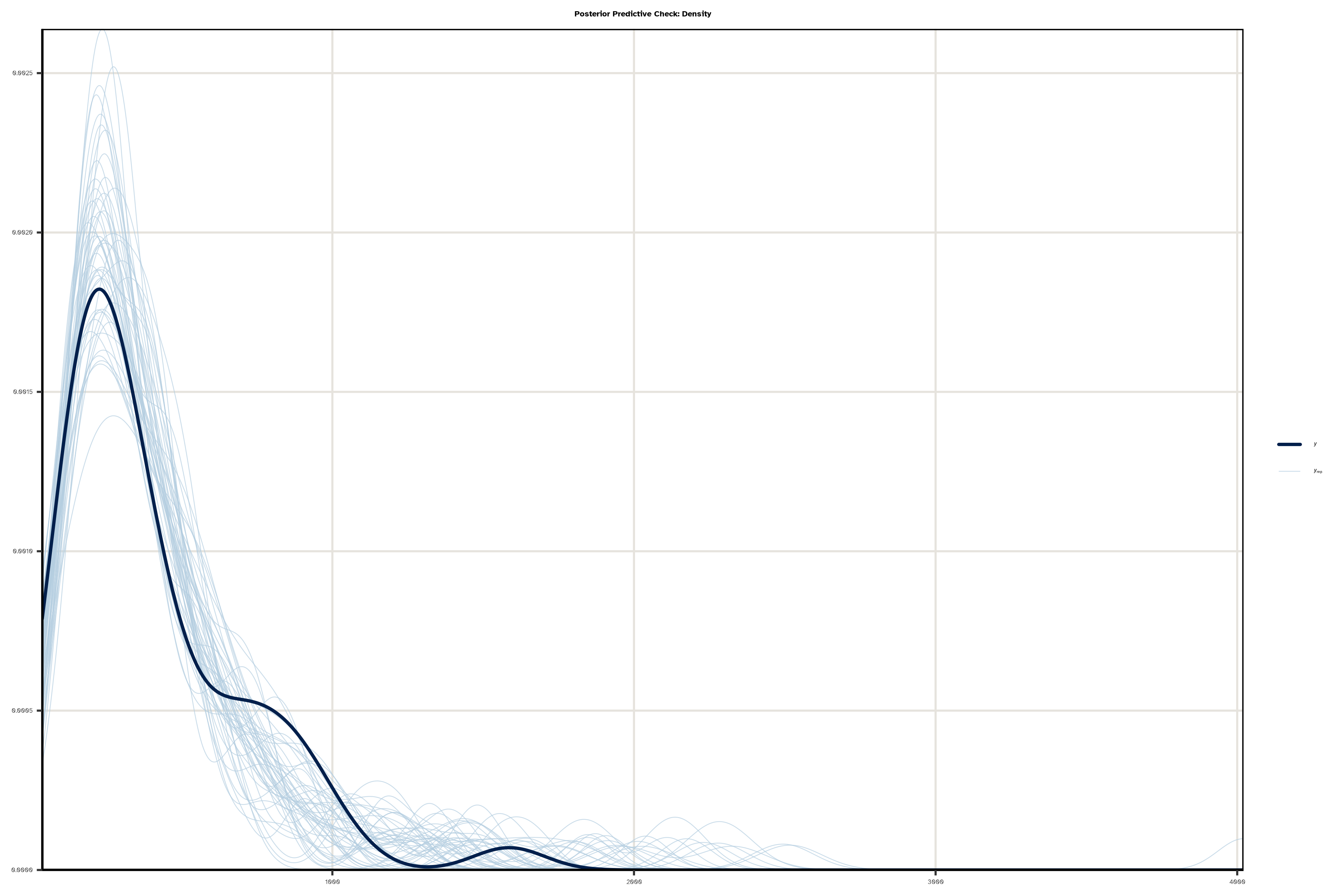}
\caption{\label{fig:pp_density}\textbf{Posterior predictive density check.} The model successfully captures the data generating process for gradient evaluations.}
\end{figure}

Figure \ref{fig:pp_density} compares the observed distribution of gradient calls
(\(y\), dark line) with 50 distributions simulated from the posterior (\(y_{rep}\),
light lines). The model accurately reproduces the heavy tail of the cost
distribution.

\subsubsection{Grouped Intervals}
\label{sec:si:gi}
\begin{figure}[H]
\centering
\includegraphics[width=0.8\textwidth]{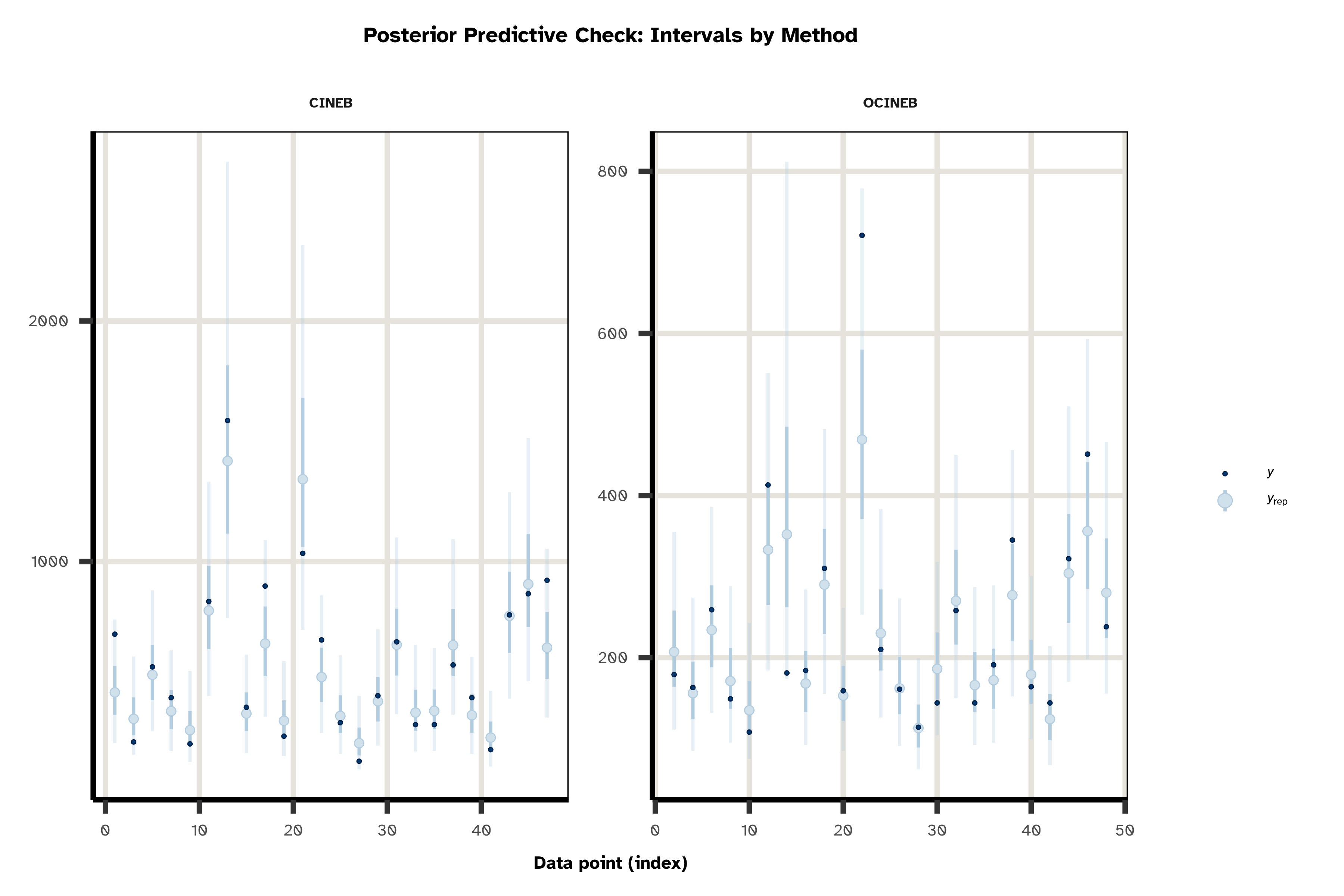}
\caption{\label{fig:pp_group}\textbf{Posterior predictive intervals grouped by Method.}}
\end{figure}

Figure \ref{fig:pp_group} verifies that the model fits both experimental groups
(CINEB and OCI-NEB) equally well, with observed data points falling within the
predicted credible intervals.

\subsubsection{LOO-PIT and Pareto-k}
\label{sec:si:loopit}
\begin{figure}[H]
\centering
\includegraphics[width=0.8\textwidth]{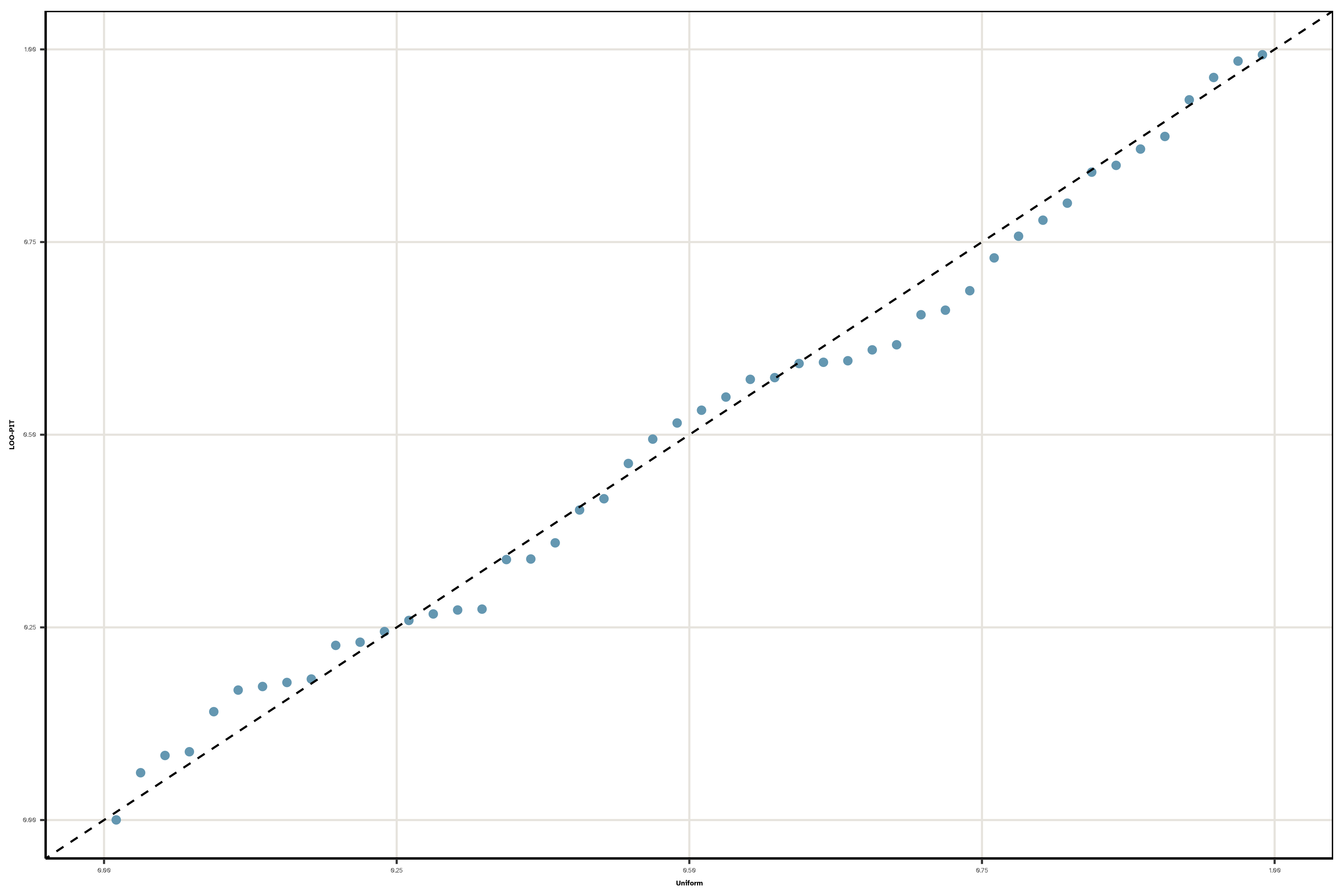}
\caption{\label{fig:ppc_loo}\textbf{LOO-PIT Q-Q plot against a uniform distribution.} The alignment with the diagonal indicates the model is well-calibrated.}
\end{figure}

The Probability Integral Transform (PIT) check (Figure \ref{fig:ppc_loo}) shows that
the resulting distribution is approximately uniform, indicating well-calibrated error
estimates. The LOO cross-validation identified 7 observations (15\%) with Pareto
\(k > 0.7\), for which moment-matched importance sampling was used to obtain
reliable LOO estimates. These high-\(k\) observations correspond to systems with
extreme speedup ratios (e.g., HCN+H\textsubscript{2} at 8.76x), where the model's predictive
distribution is most sensitive to individual data points.

\subsection{2D Reaction Landscapes}
\label{sec:si:landscapes}
Figures \ref{fig:si:baker2d_01} through \ref{fig:si:baker2d_25} present the 2D RMSD-projected reaction landscapes \cite{goswamiTwodimensionalRMSDProjections2025} for all 24 Baker-Chan systems. Each pair shows the CI-NEB landscape (left) and OCI-NEB landscape (right). The background contour is a derivative Gaussian process interpolation using an inverse multiquartic kernel. Black dots mark all sampled images; colored circles indicate the final converged path. The white square denotes the saddle point found by each method; the blue star shows the CI-NEB saddle on the OCI-NEB landscape for comparison.

\newcommand{\bakerlandscape}[3]{%
\begin{figure}[H]
\centering
\begin{minipage}[b]{0.48\textwidth}
  \includegraphics[width=\linewidth]{imgs/suppl/baker_2d/#1/cineb_landscape.png}
  \centerline{\small CI-NEB}
\end{minipage}
\hfill
\begin{minipage}[b]{0.48\textwidth}
  \includegraphics[width=\linewidth]{imgs/suppl/baker_2d/#1/mmf_landscape.png}
  \centerline{\small OCI-NEB}
\end{minipage}
\caption{#2 (System #3).}
\label{fig:si:baker2d_#3}
\end{figure}
}

\bakerlandscape{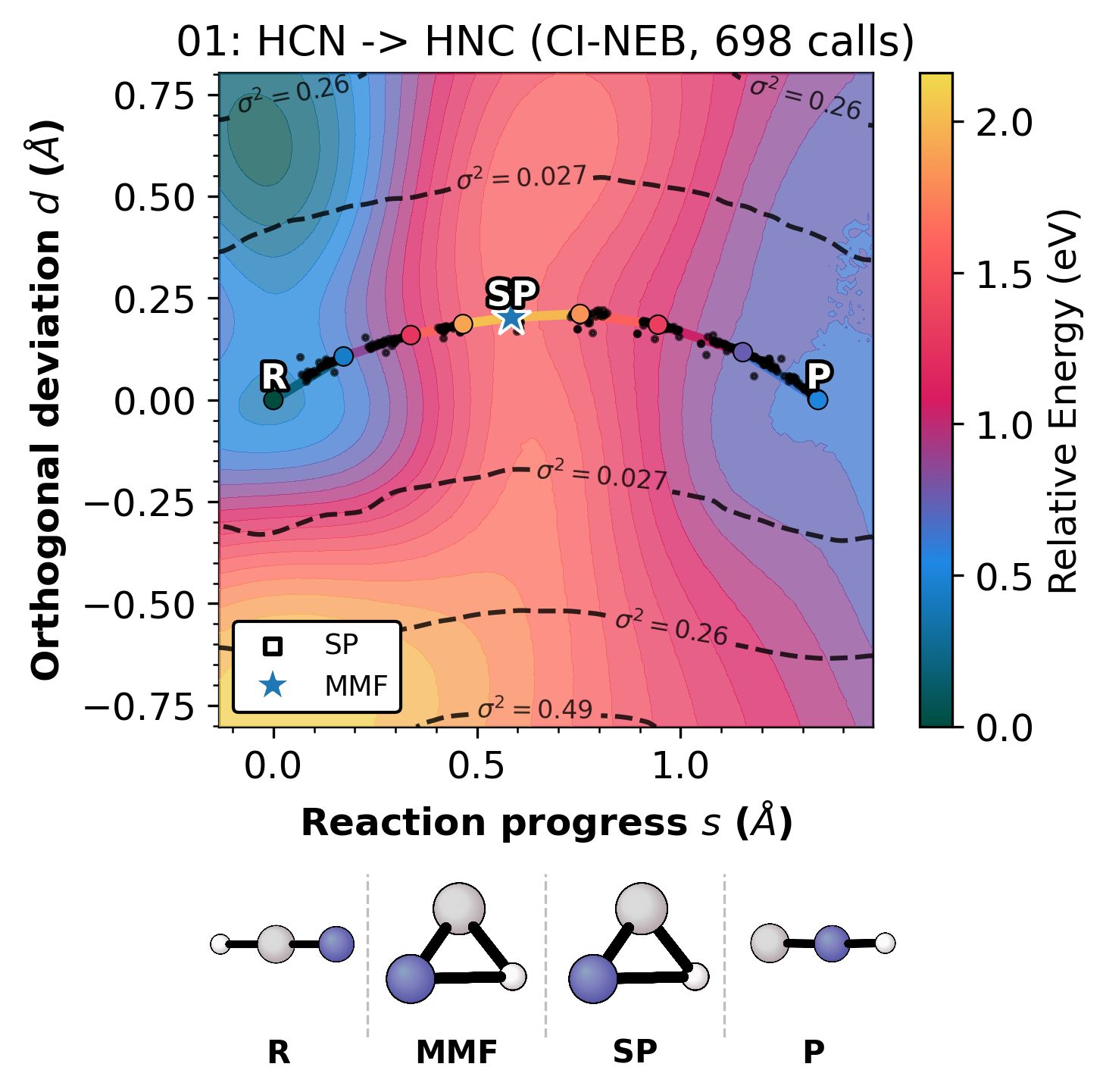}{\ce{HCN -> HNC}}{01}
\bakerlandscape{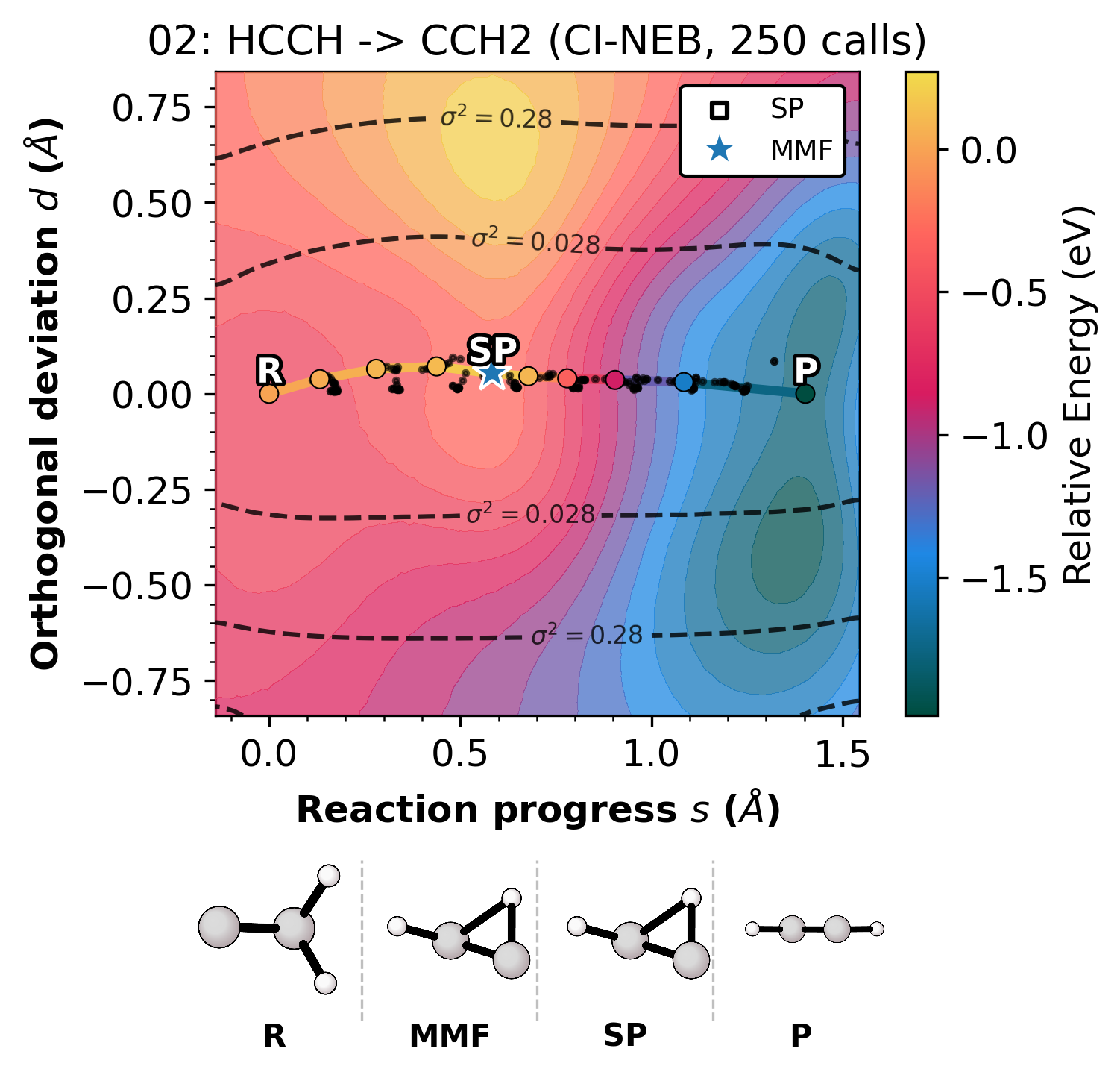}{\ce{HCCH -> CCH2}}{02}
\bakerlandscape{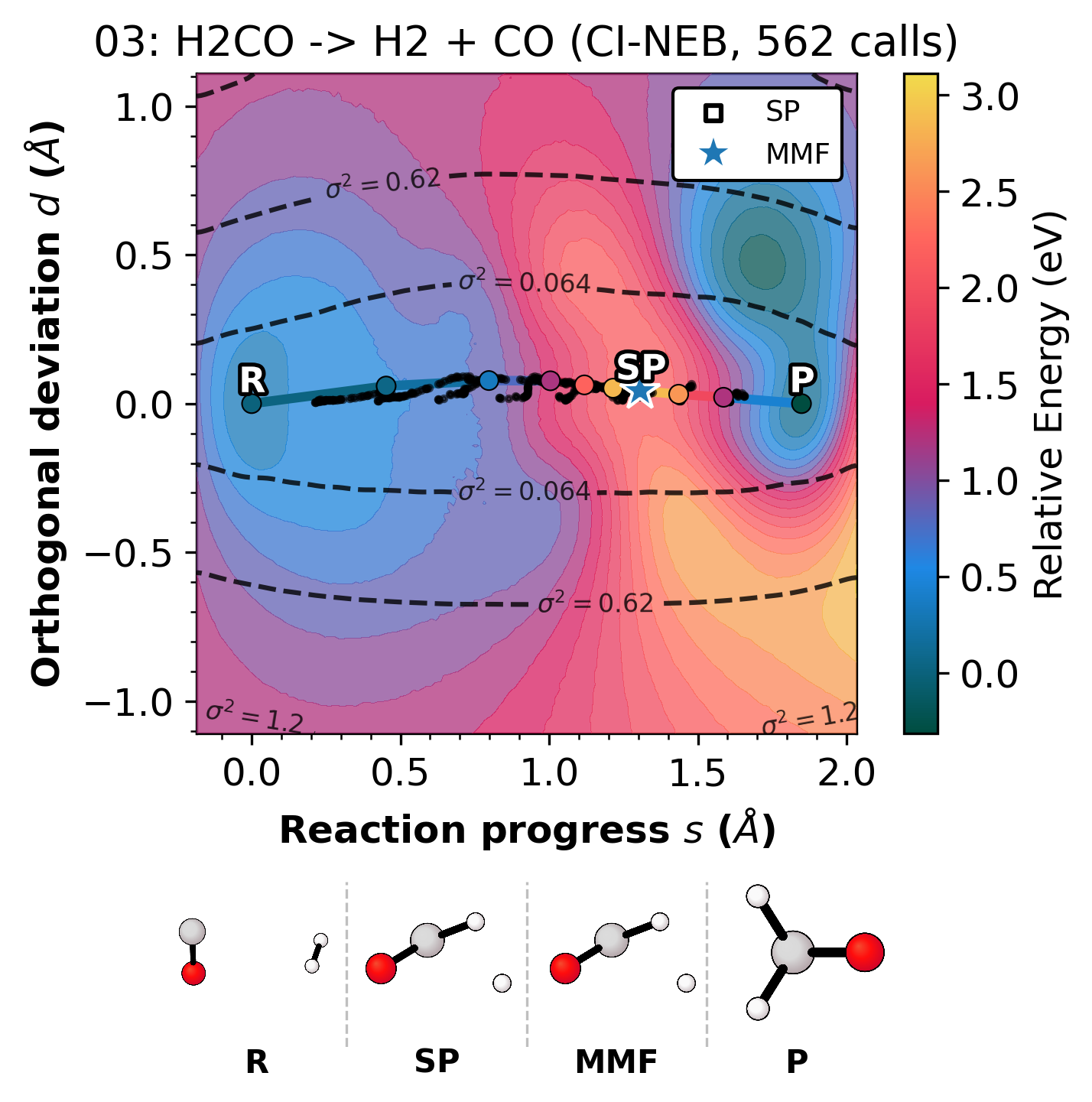}{\ce{H2CO -> H2 + CO}}{03}
\bakerlandscape{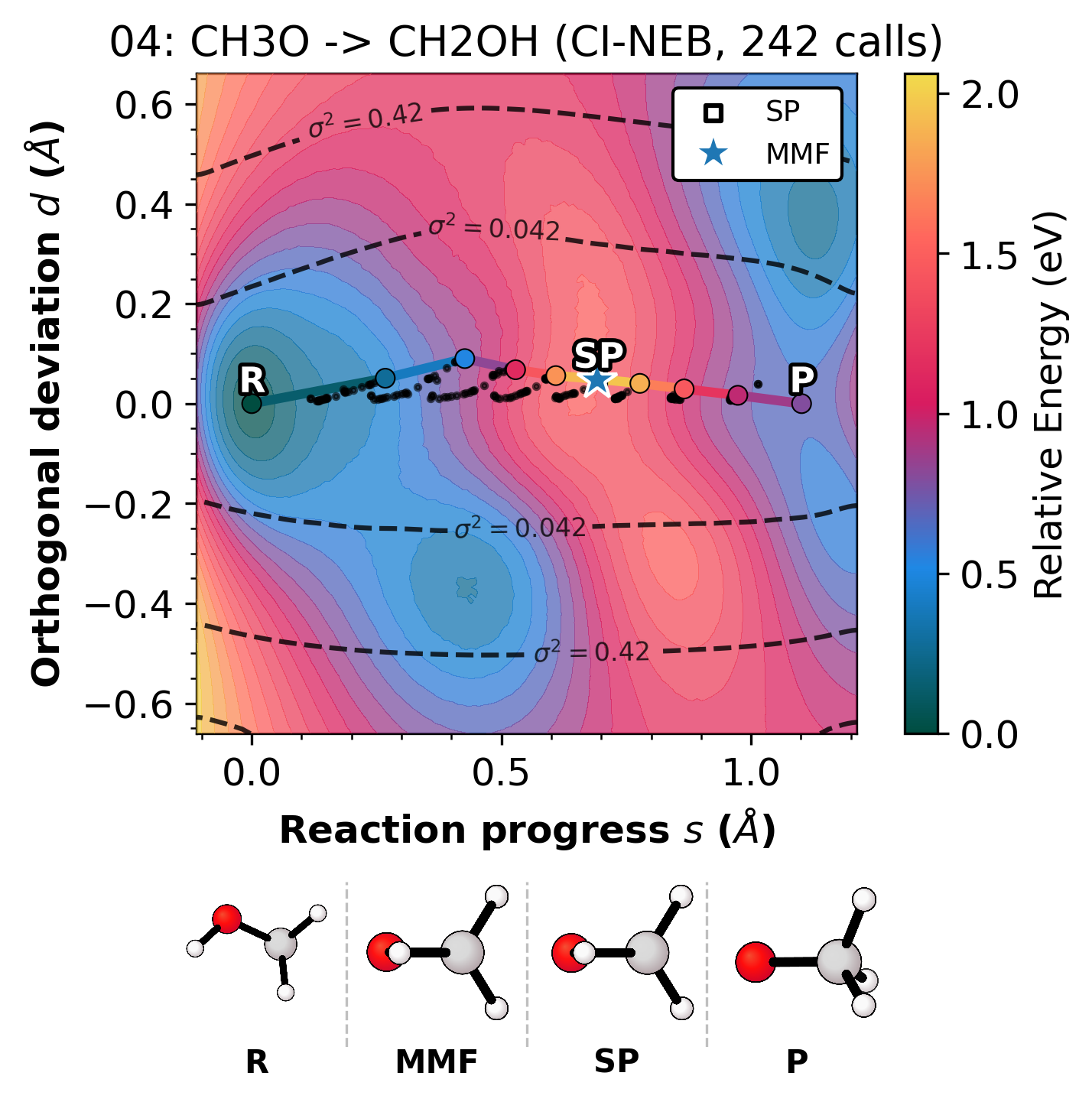}{\ce{CH3O -> CH2OH}}{04}
\bakerlandscape{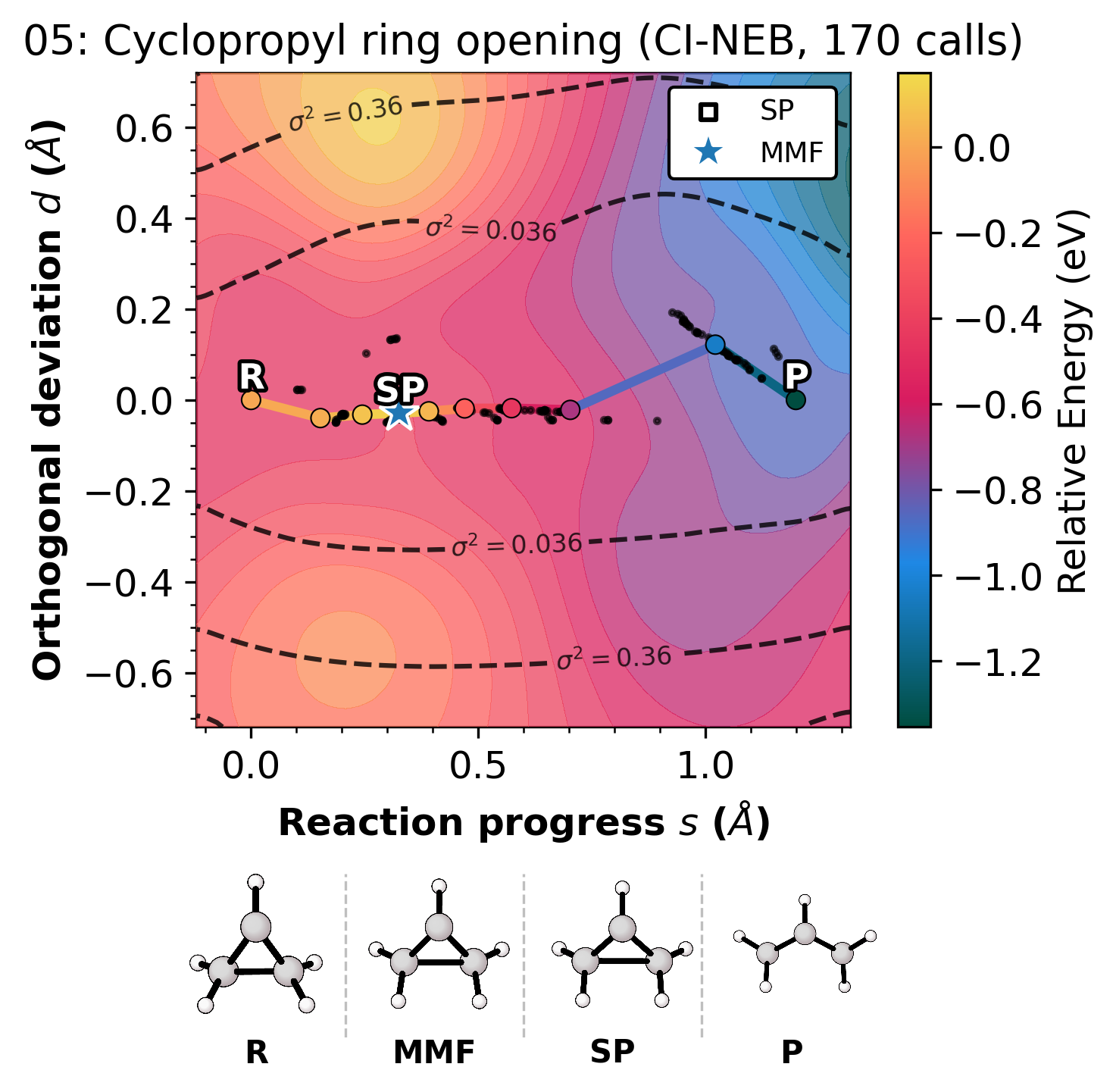}{Cyclopropyl ring opening}{05}
\bakerlandscape{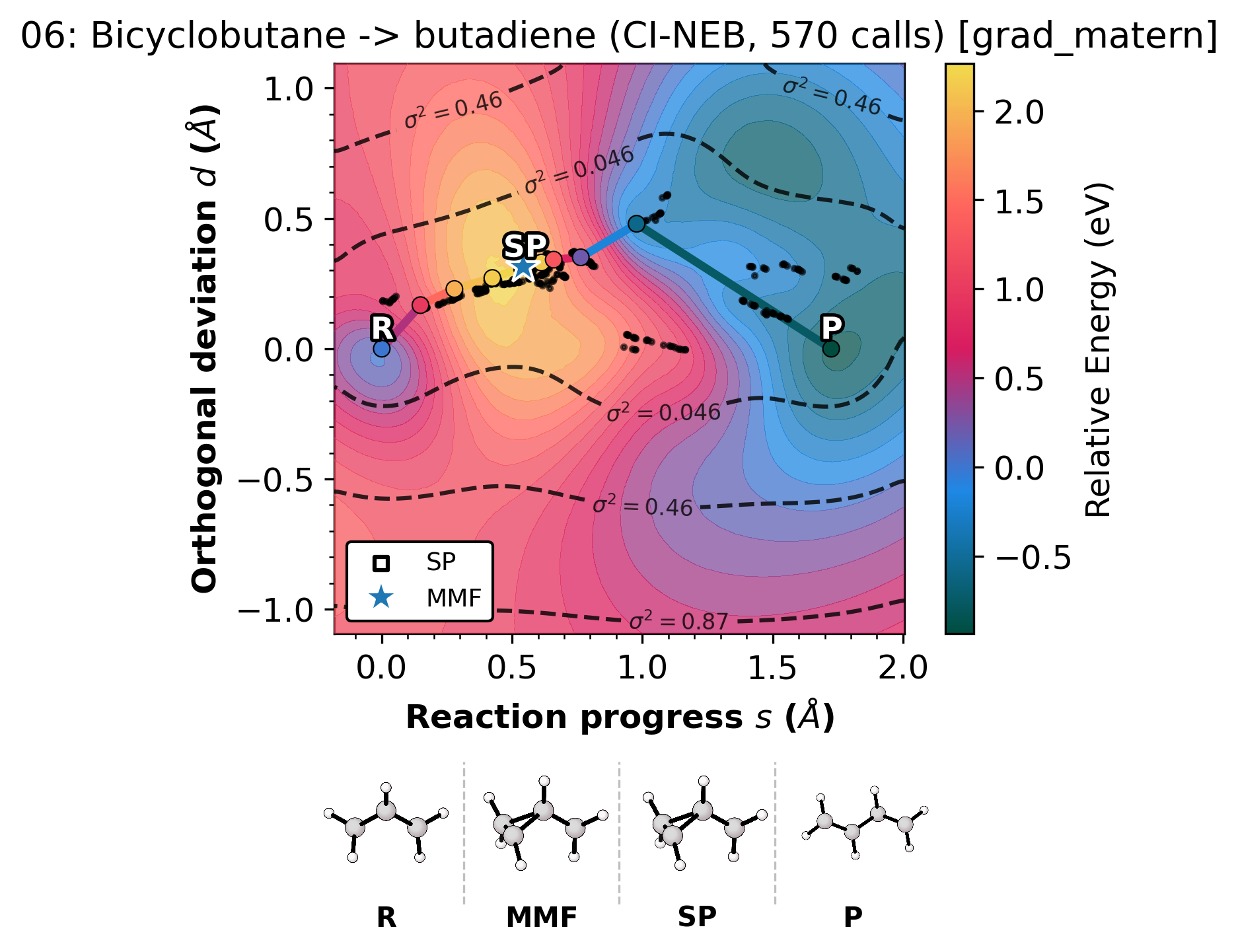}{Bicyclo[1.1.0]butane $\to$ \textit{trans}-butadiene}{06}
\bakerlandscape{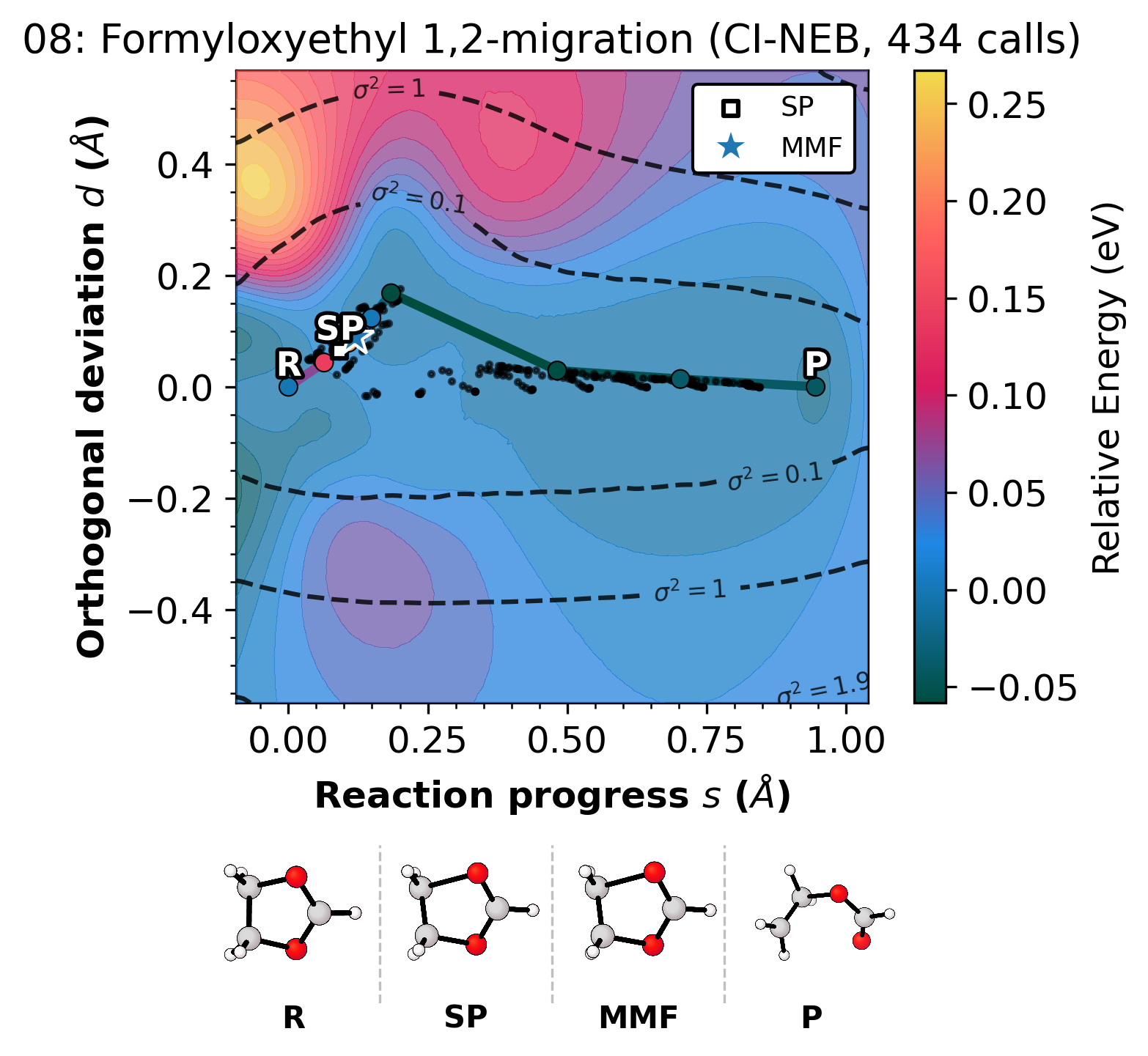}{Formyloxyethyl 1,2-migration}{08}
\bakerlandscape{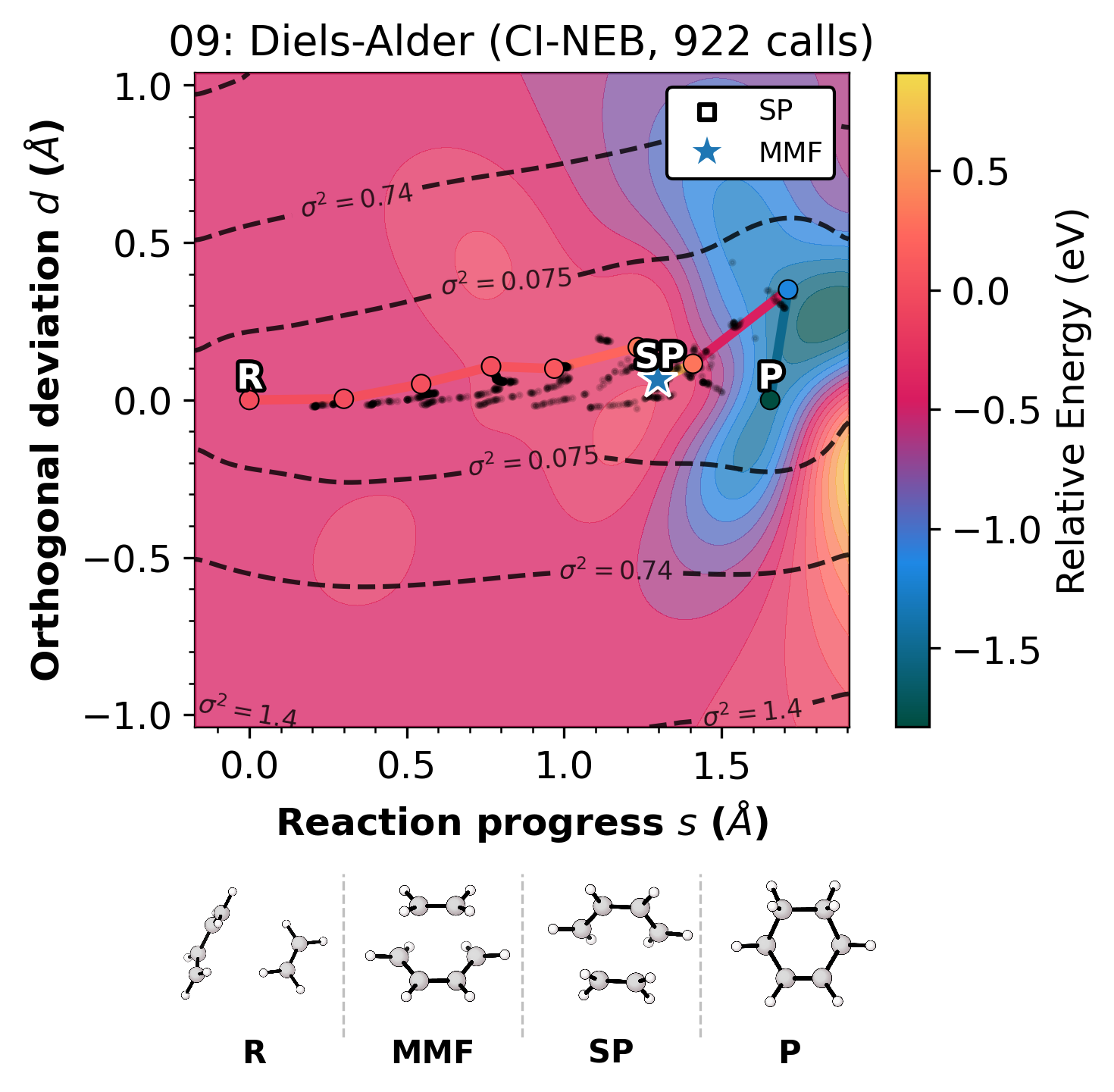}{Parent Diels--Alder cycloaddition}{09}
\bakerlandscape{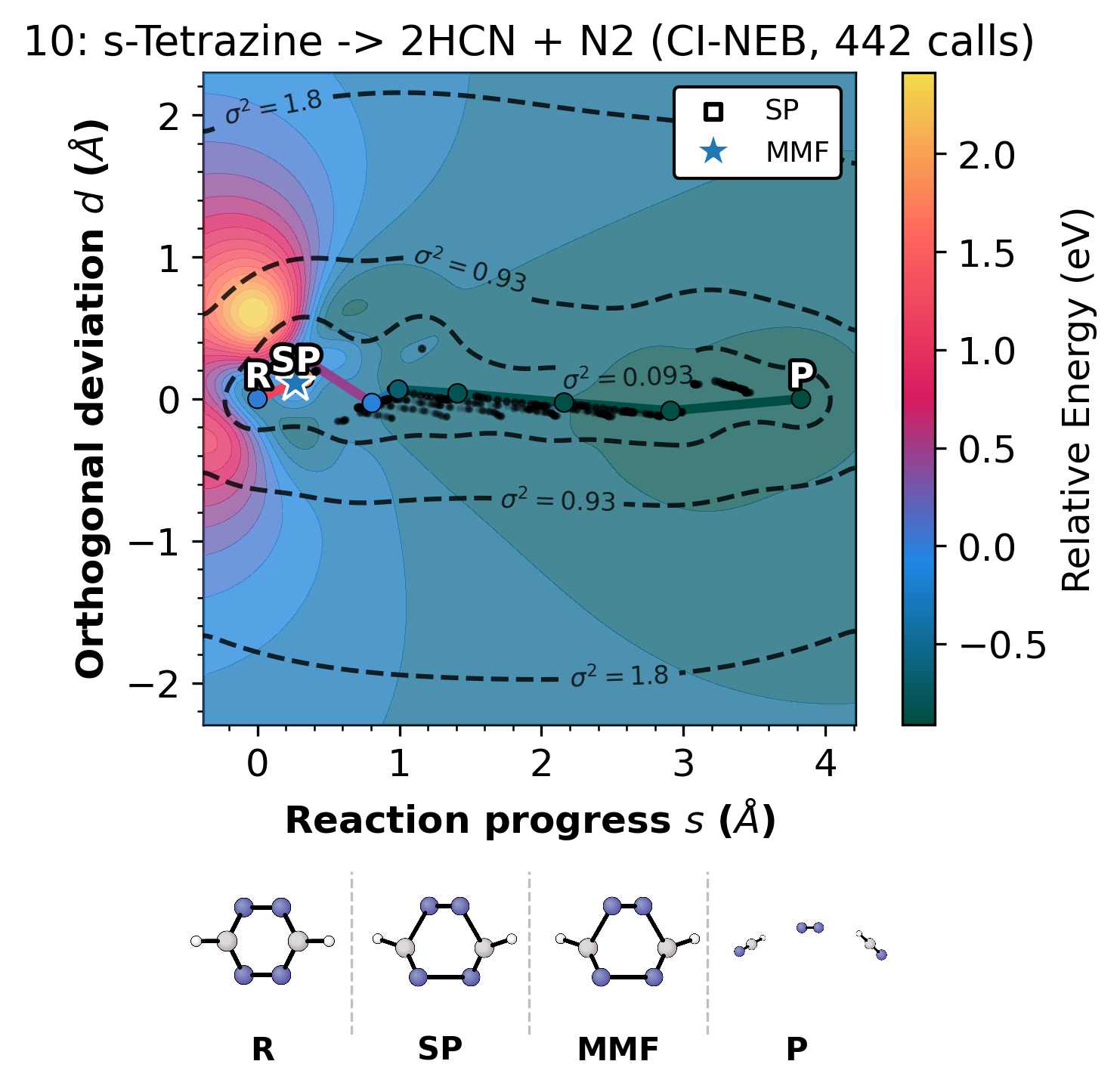}{\textit{s}-Tetrazine $\to$ \ce{2HCN + N2}}{10}
\bakerlandscape{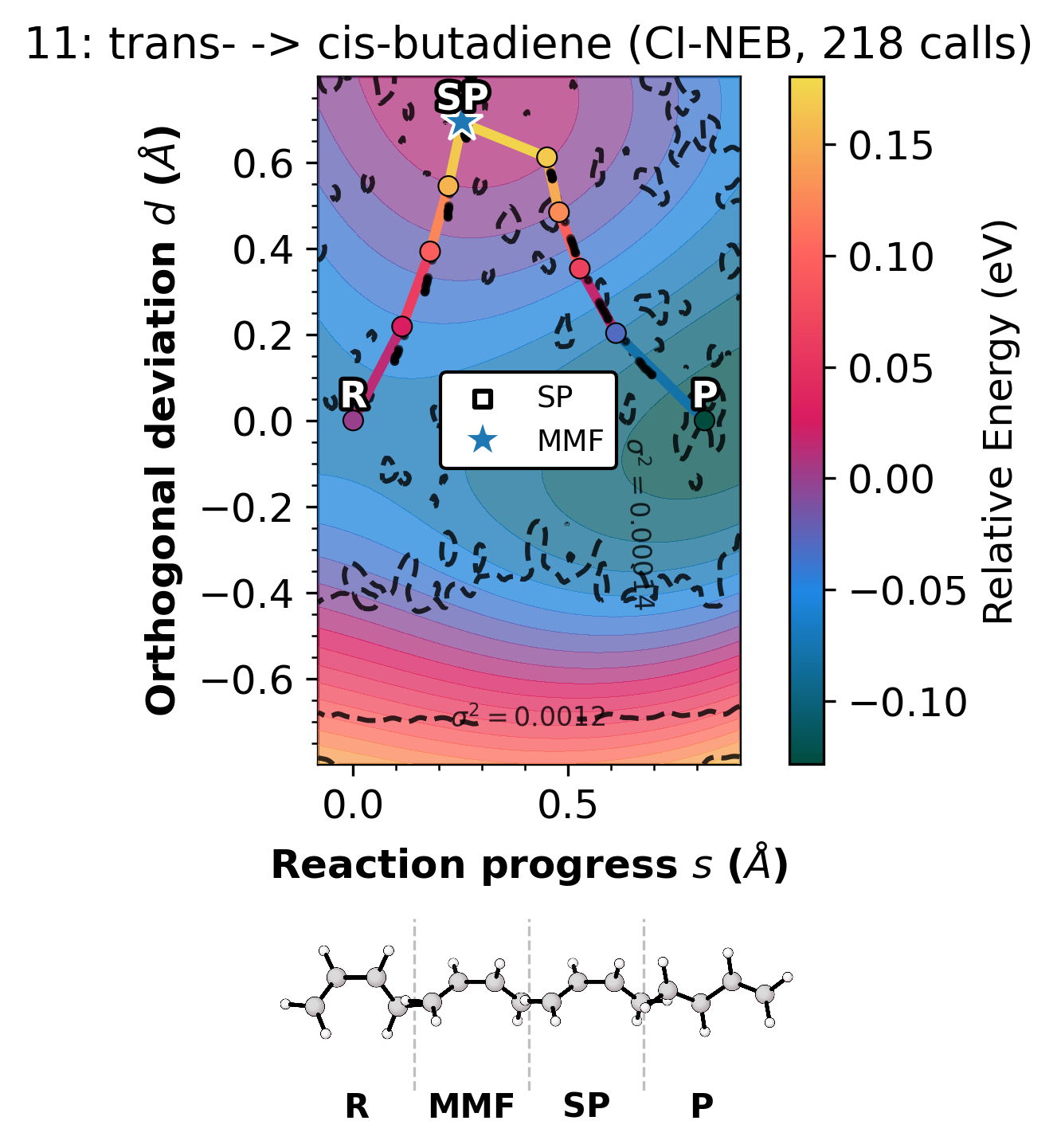}{\textit{trans}-Butadiene $\to$ \textit{cis}-butadiene}{11}
\bakerlandscape{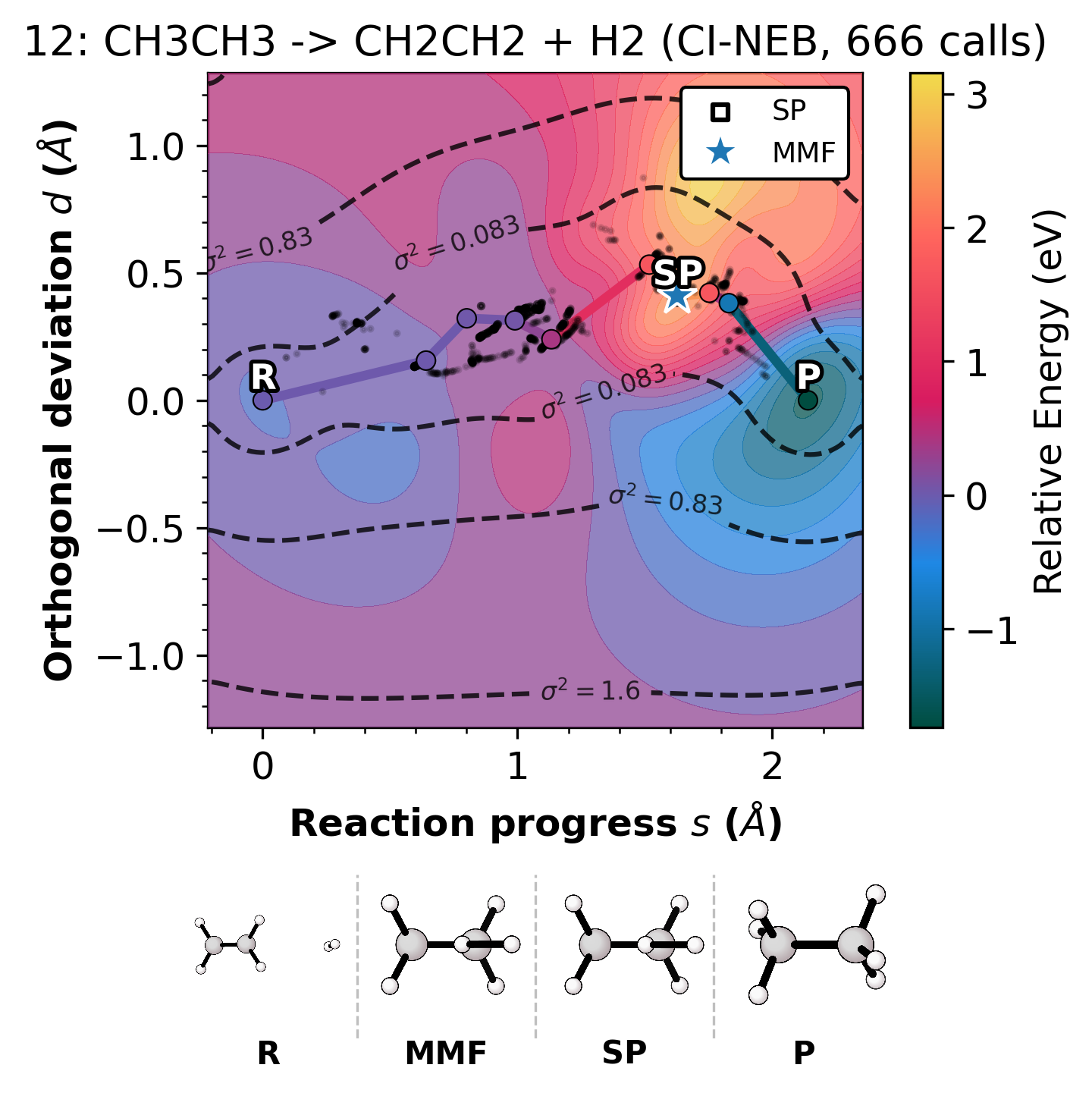}{\ce{CH3CH3 -> CH2CH2 + H2}}{12}
\bakerlandscape{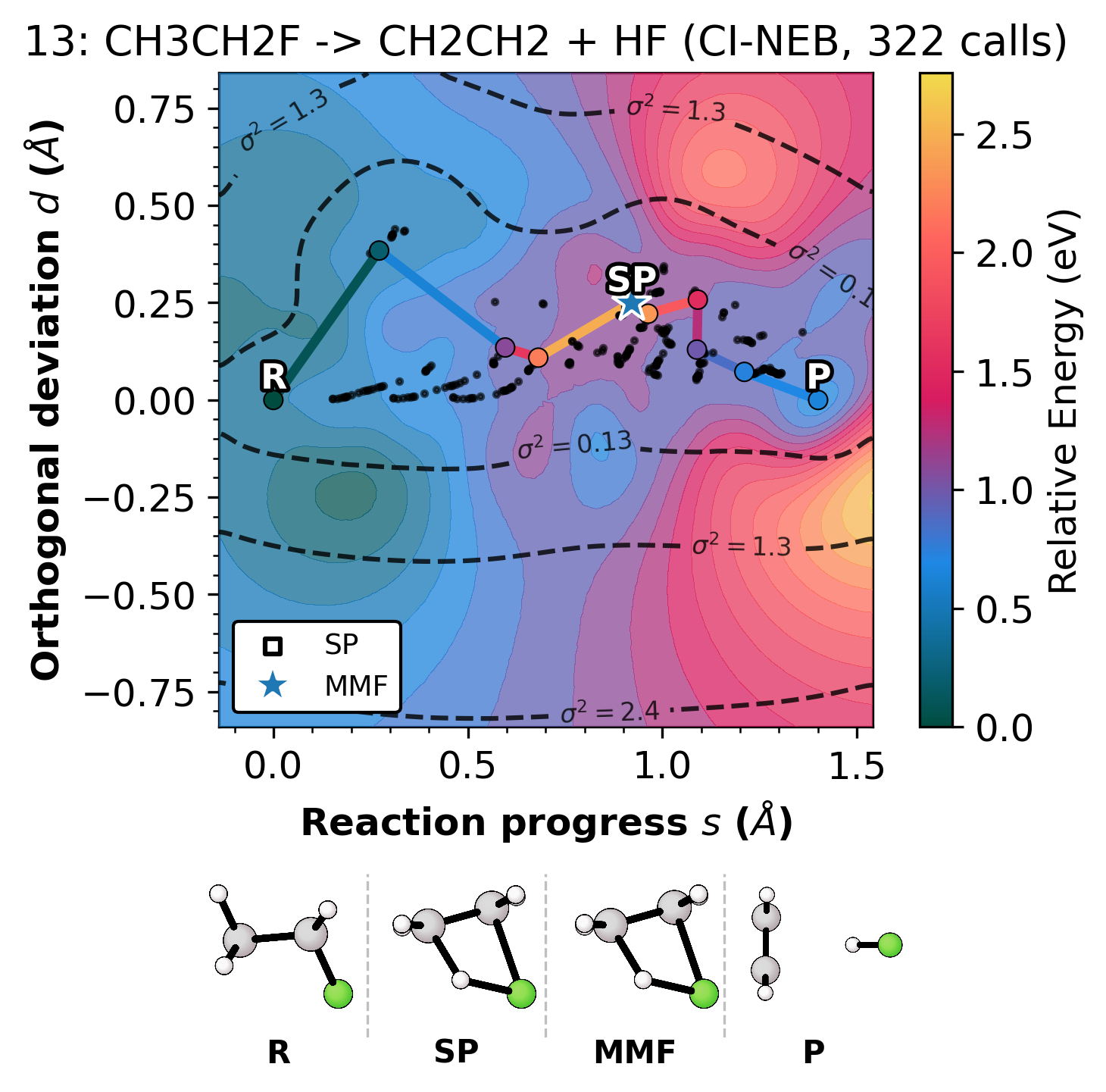}{\ce{CH3CH2F -> CH2CH2 + HF}}{13}
\bakerlandscape{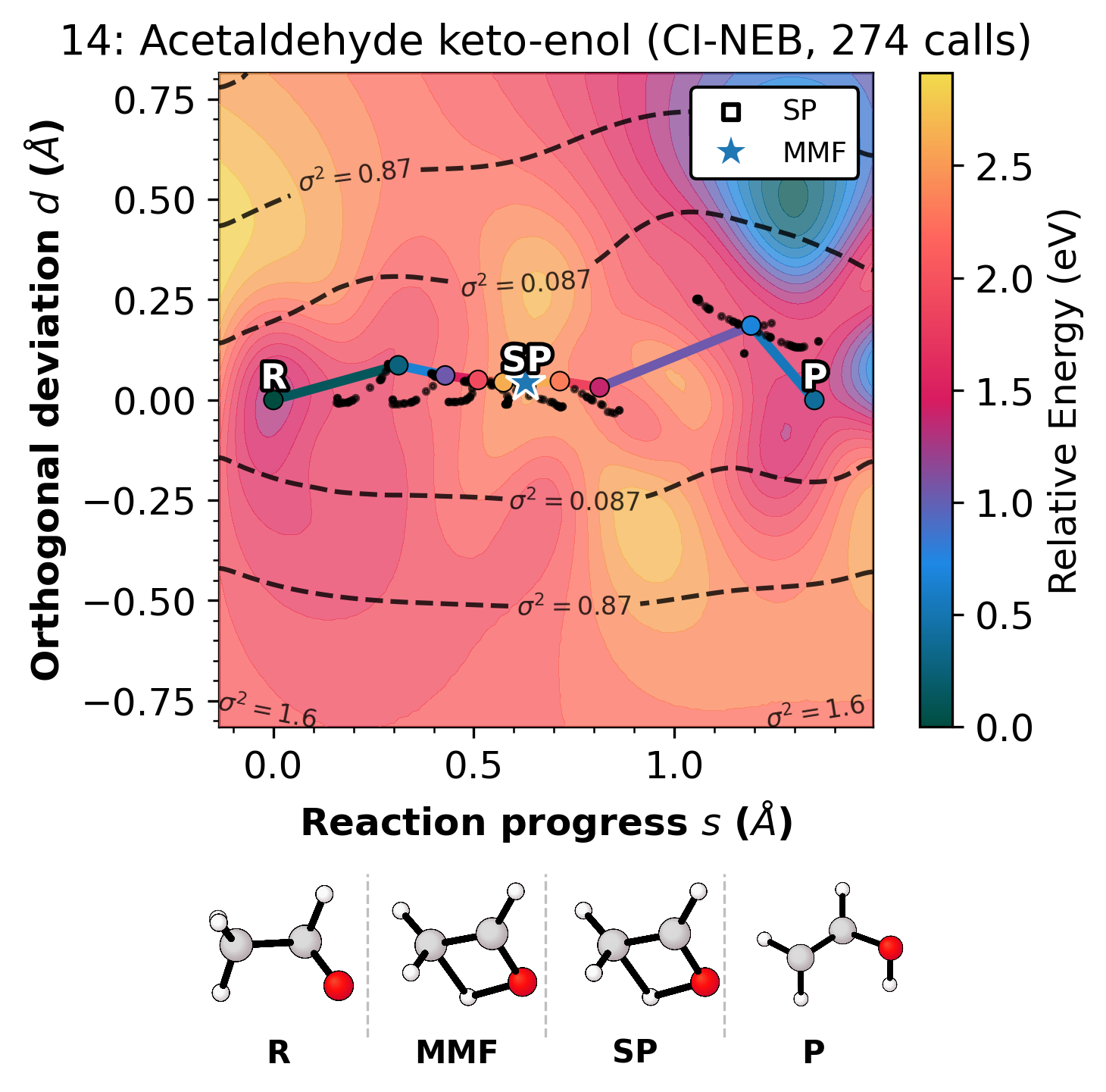}{Acetaldehyde keto--enol tautomerism}{14}
\bakerlandscape{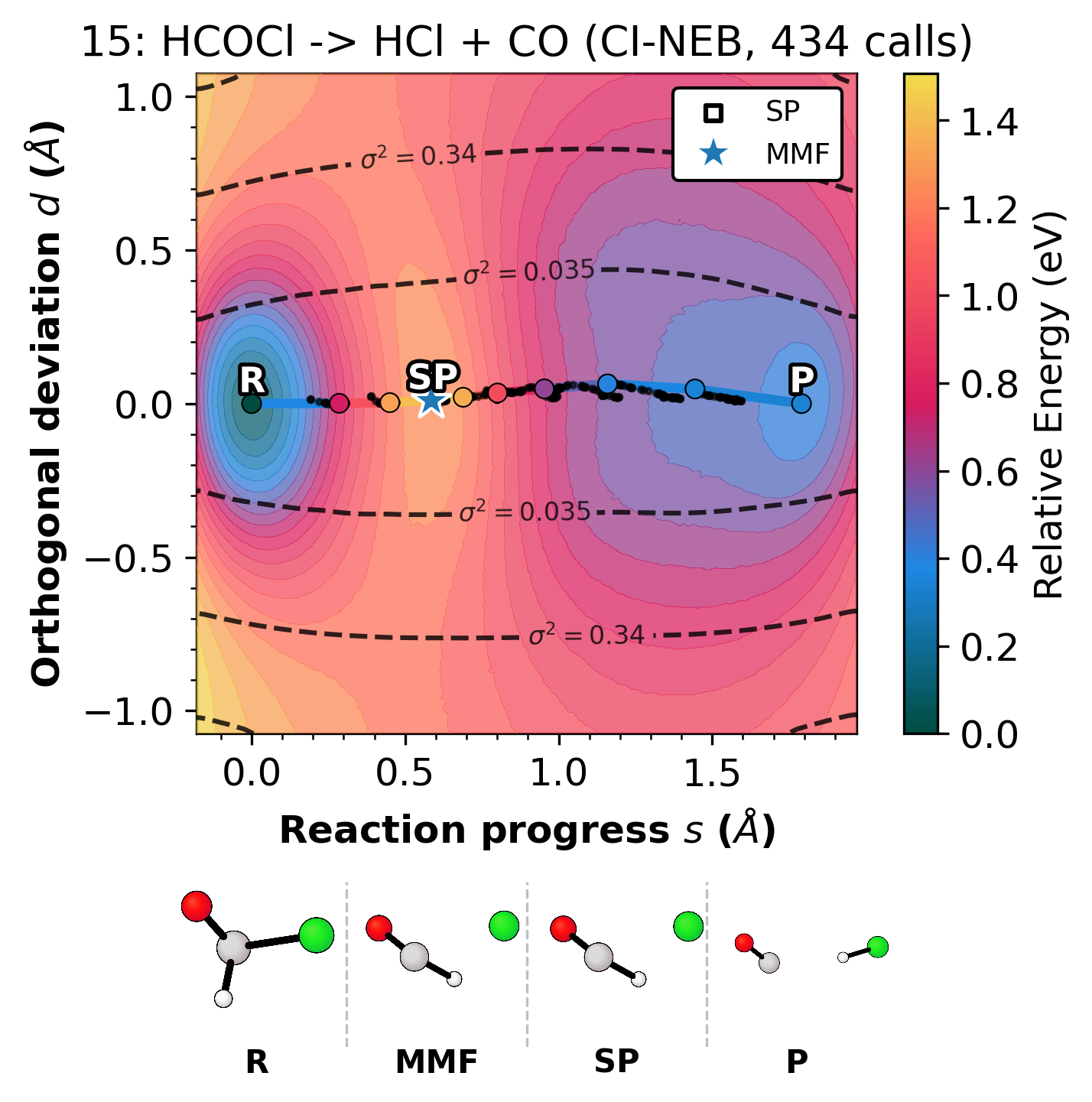}{\ce{HCOCl -> HCl + CO}}{15}
\bakerlandscape{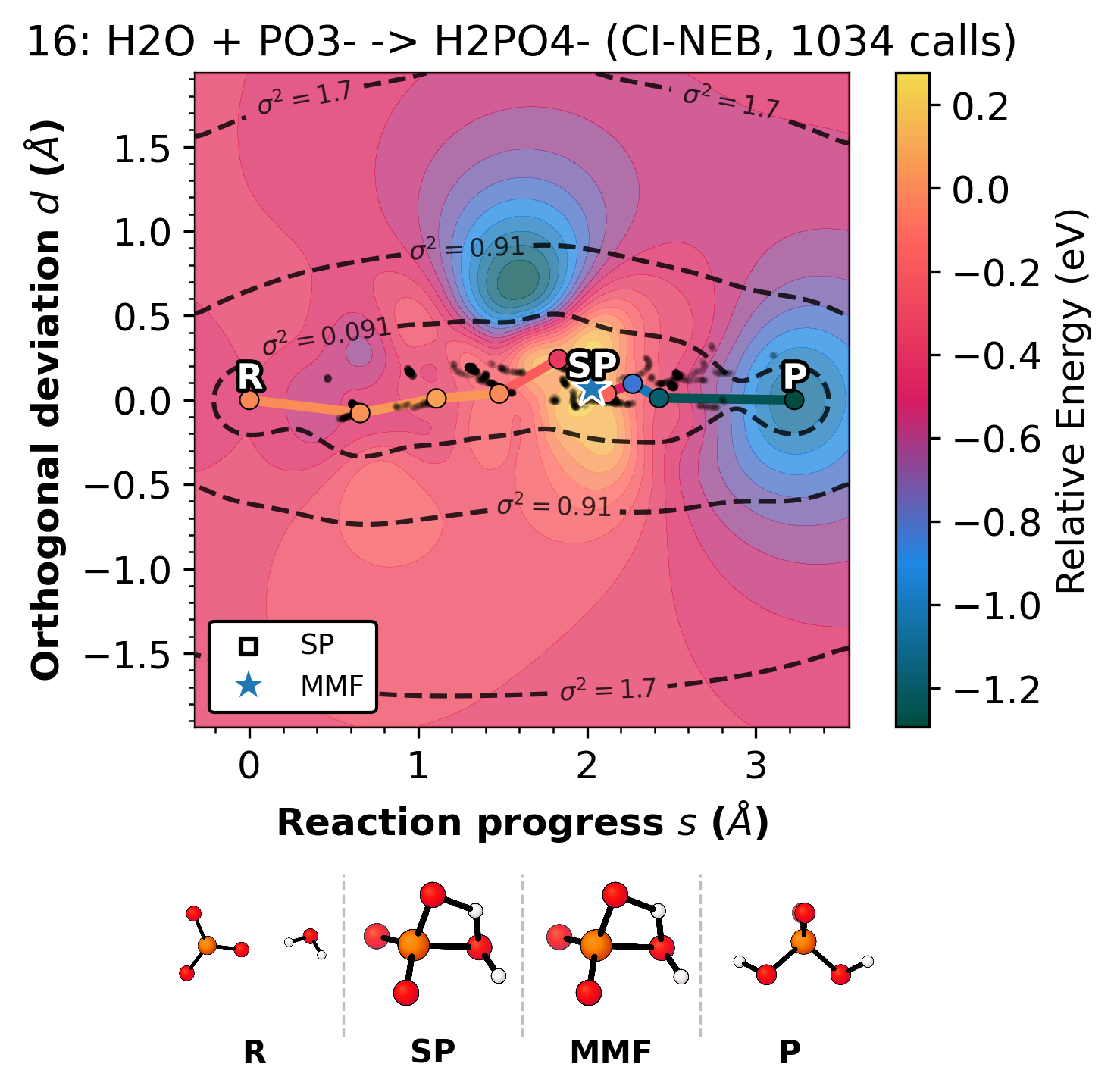}{\ce{H2O + PO3- -> H2PO4-}}{16}
\bakerlandscape{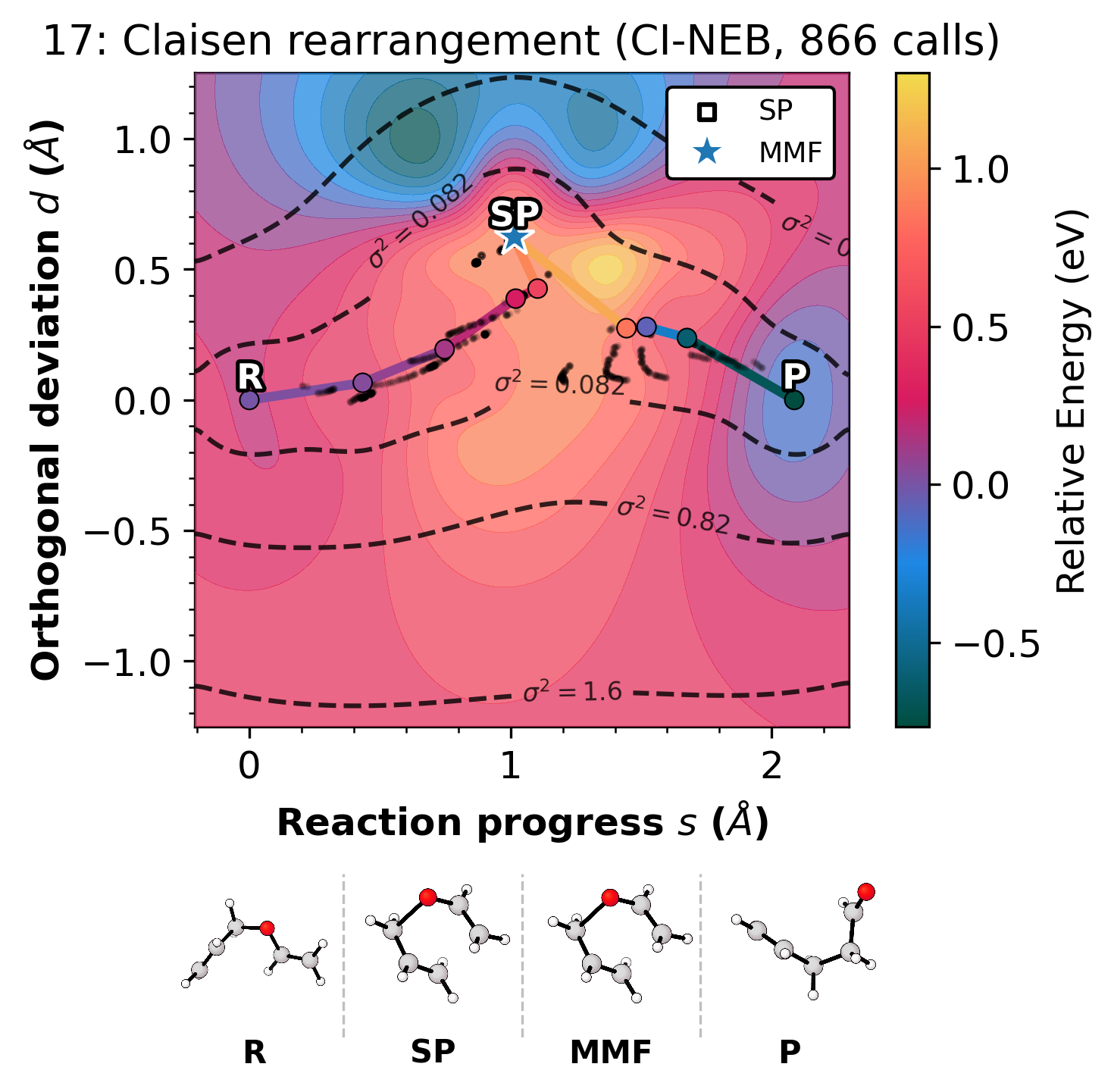}{\ce{CH2CHCH2CH2CHO} Claisen rearrangement}{17}
\bakerlandscape{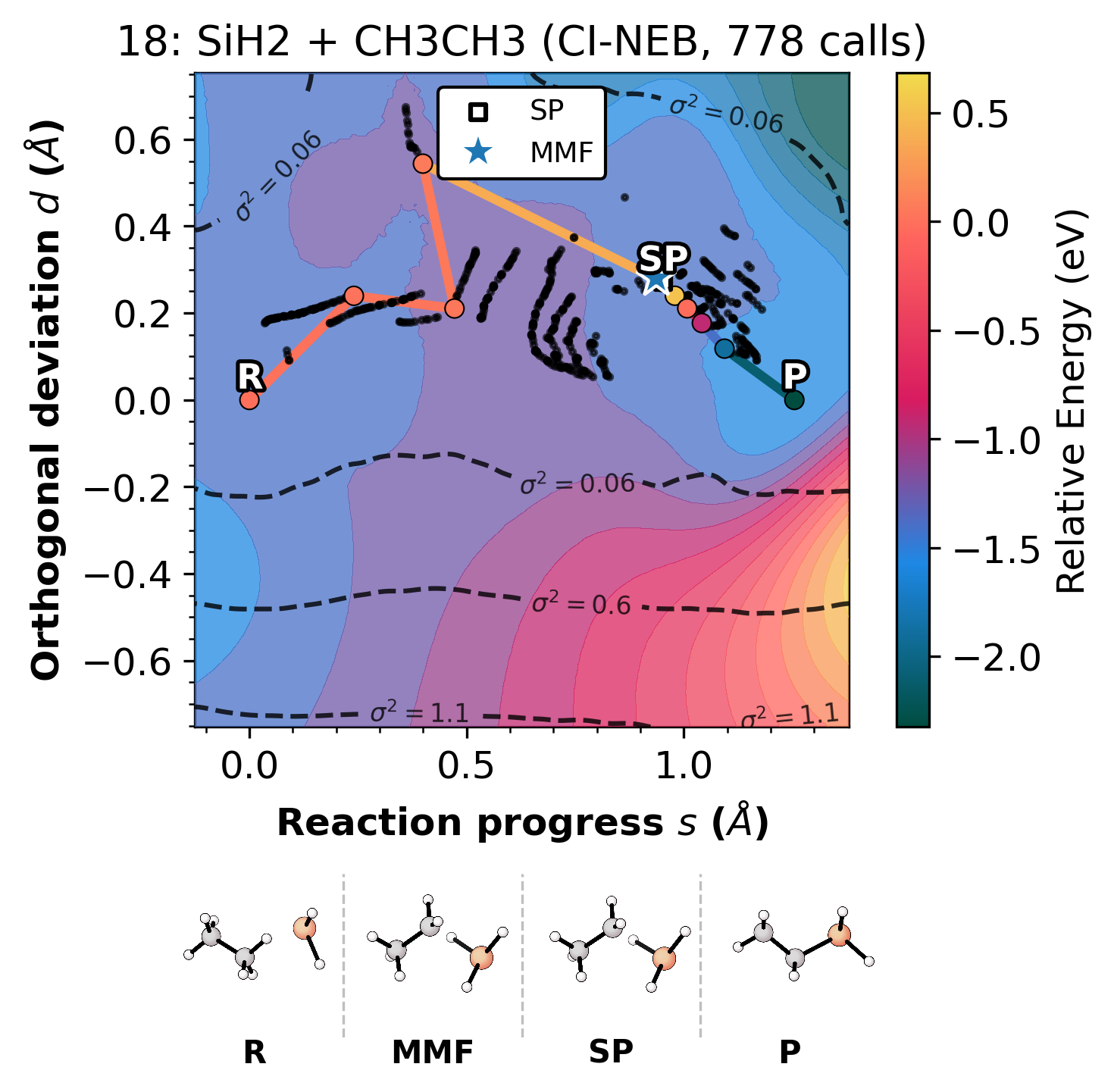}{\ce{SiH2 + CH3CH3 -> SiH3CH2CH3}}{18}
\bakerlandscape{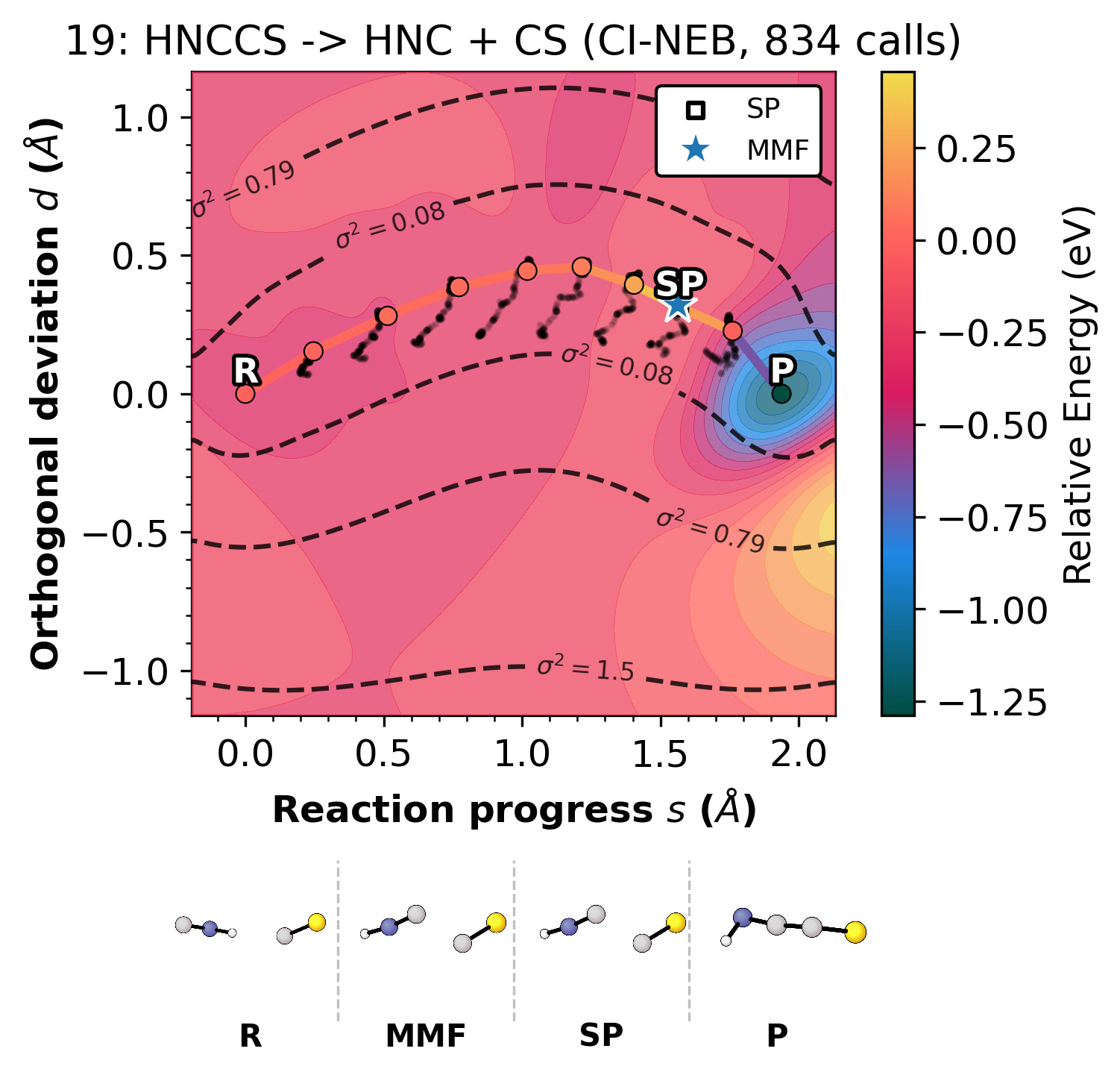}{\ce{HNCCS -> HNC + CS}}{19}
\bakerlandscape{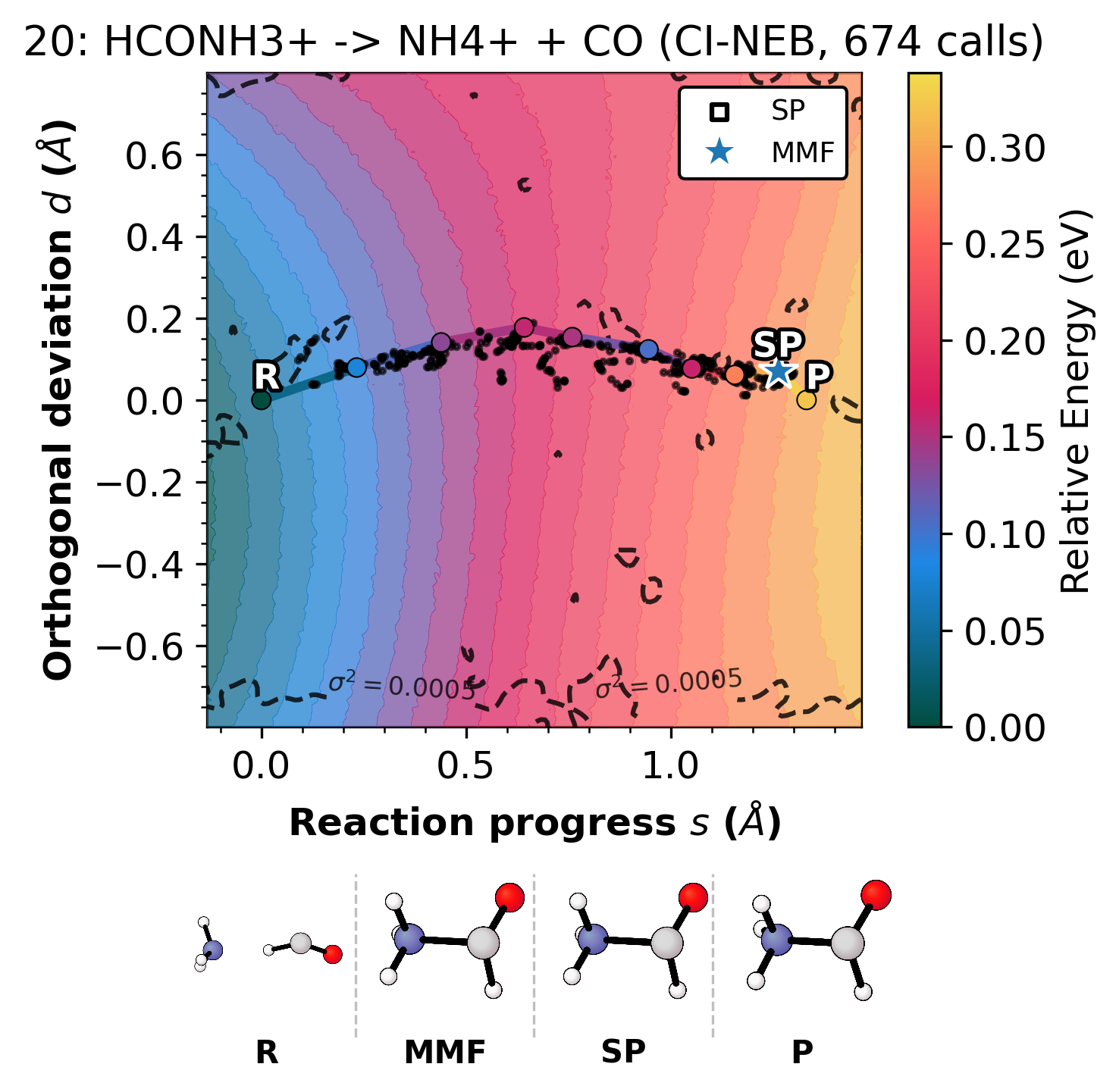}{\ce{HCONH3+ -> NH4+ + CO}}{20}
\bakerlandscape{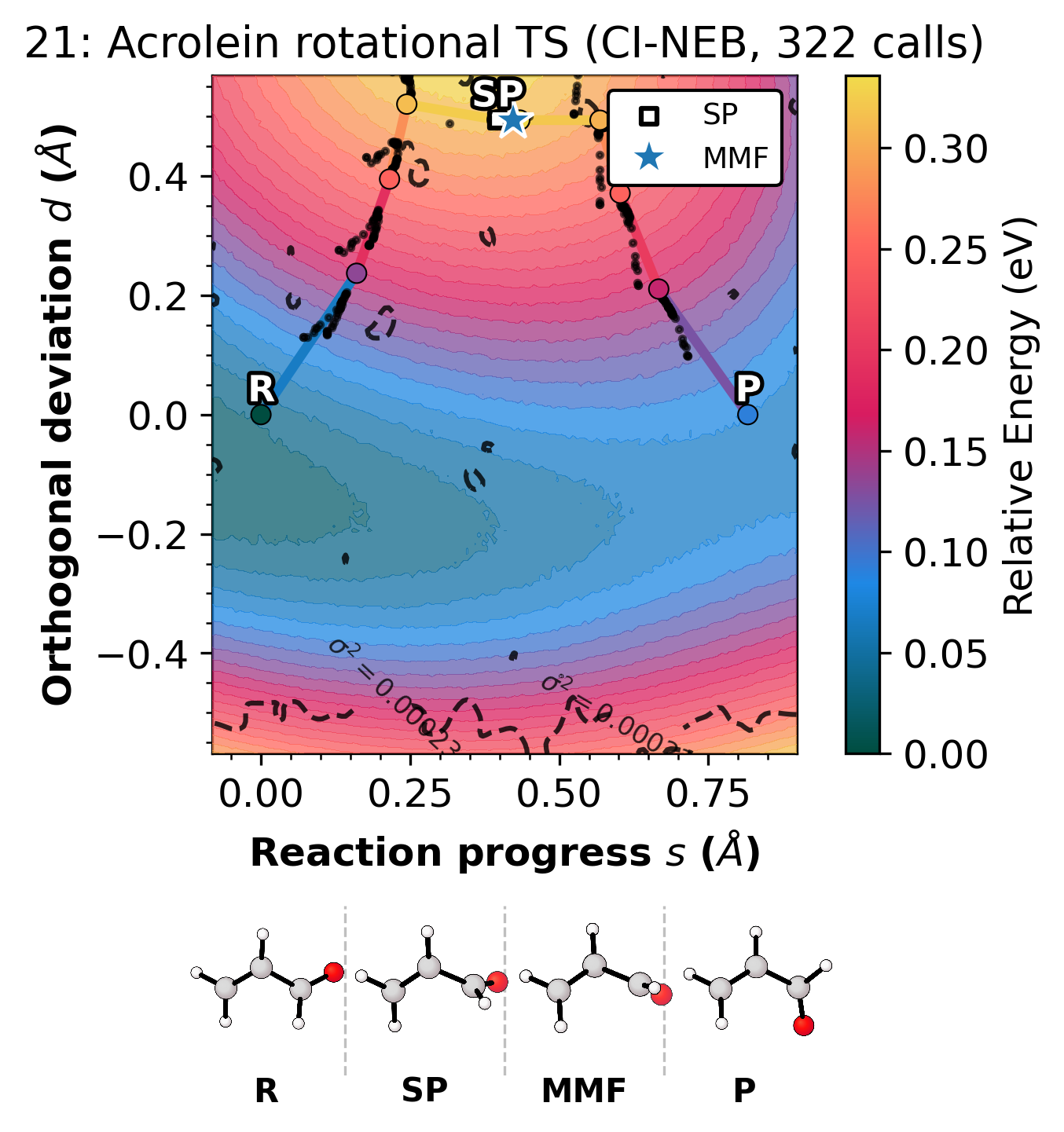}{Acrolein rotational TS}{21}
\bakerlandscape{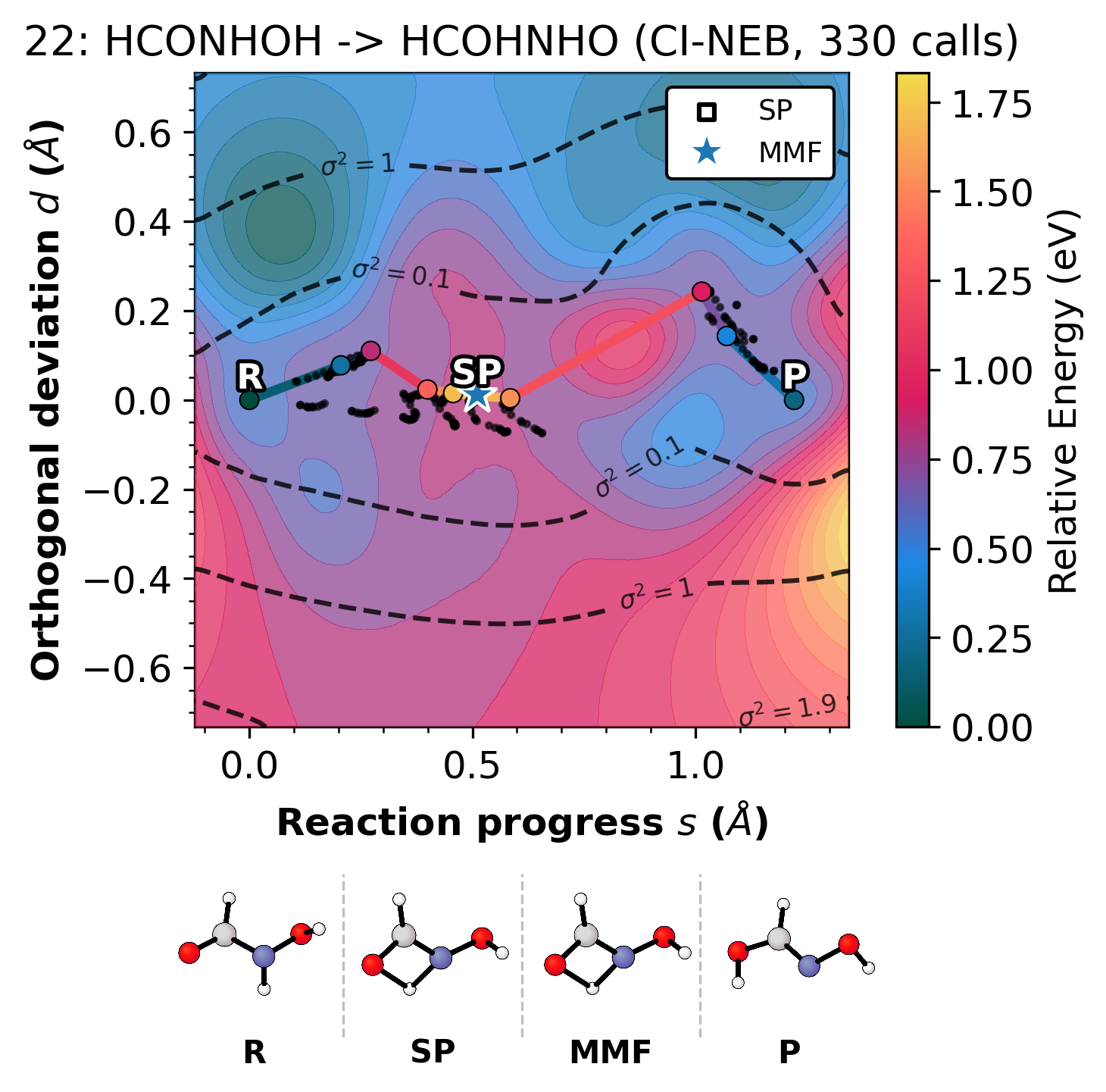}{\ce{HCONHOH -> HCOHNHO}}{22}
\bakerlandscape{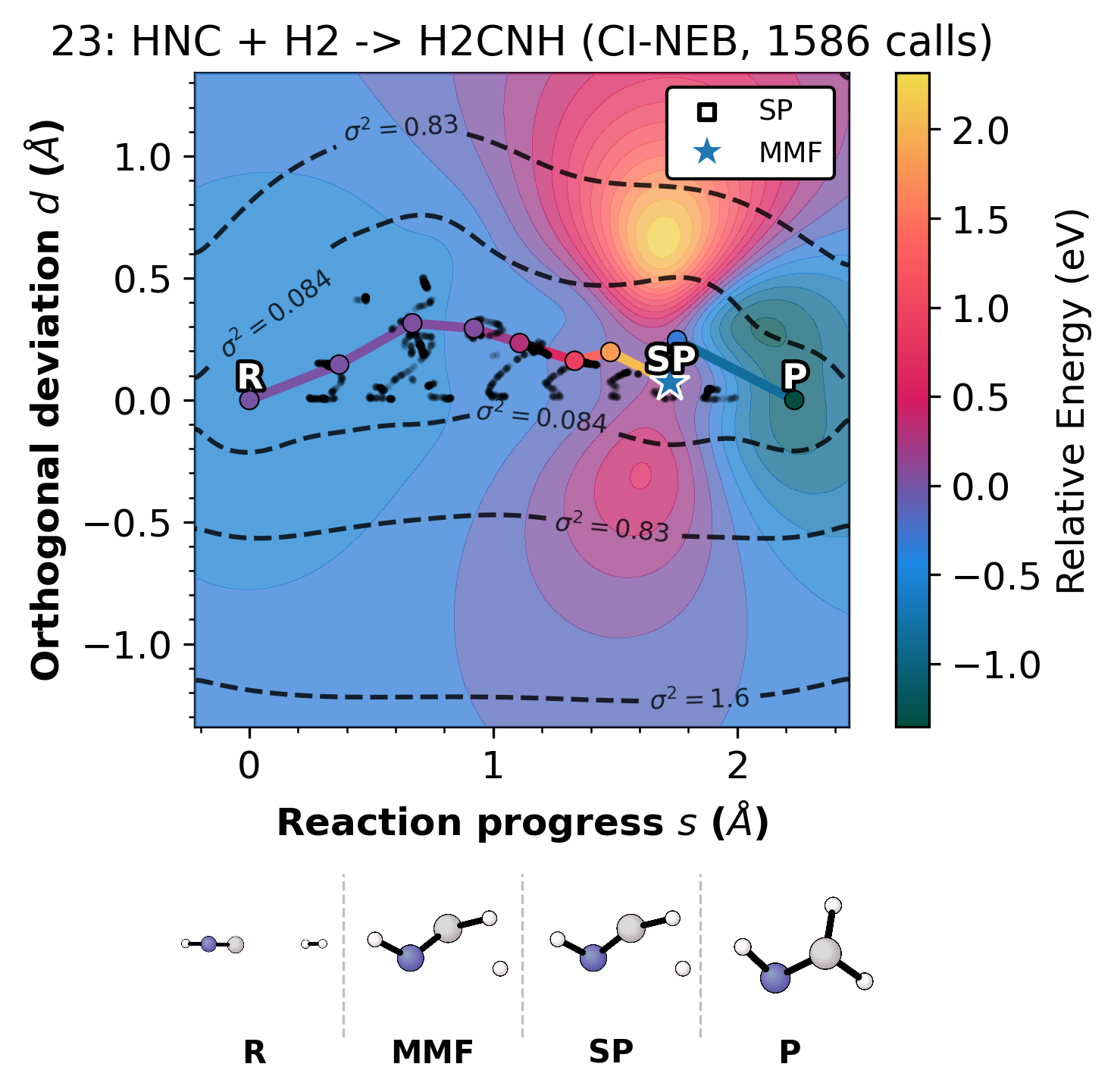}{\ce{HNC + H2 -> H2CNH}}{23}
\bakerlandscape{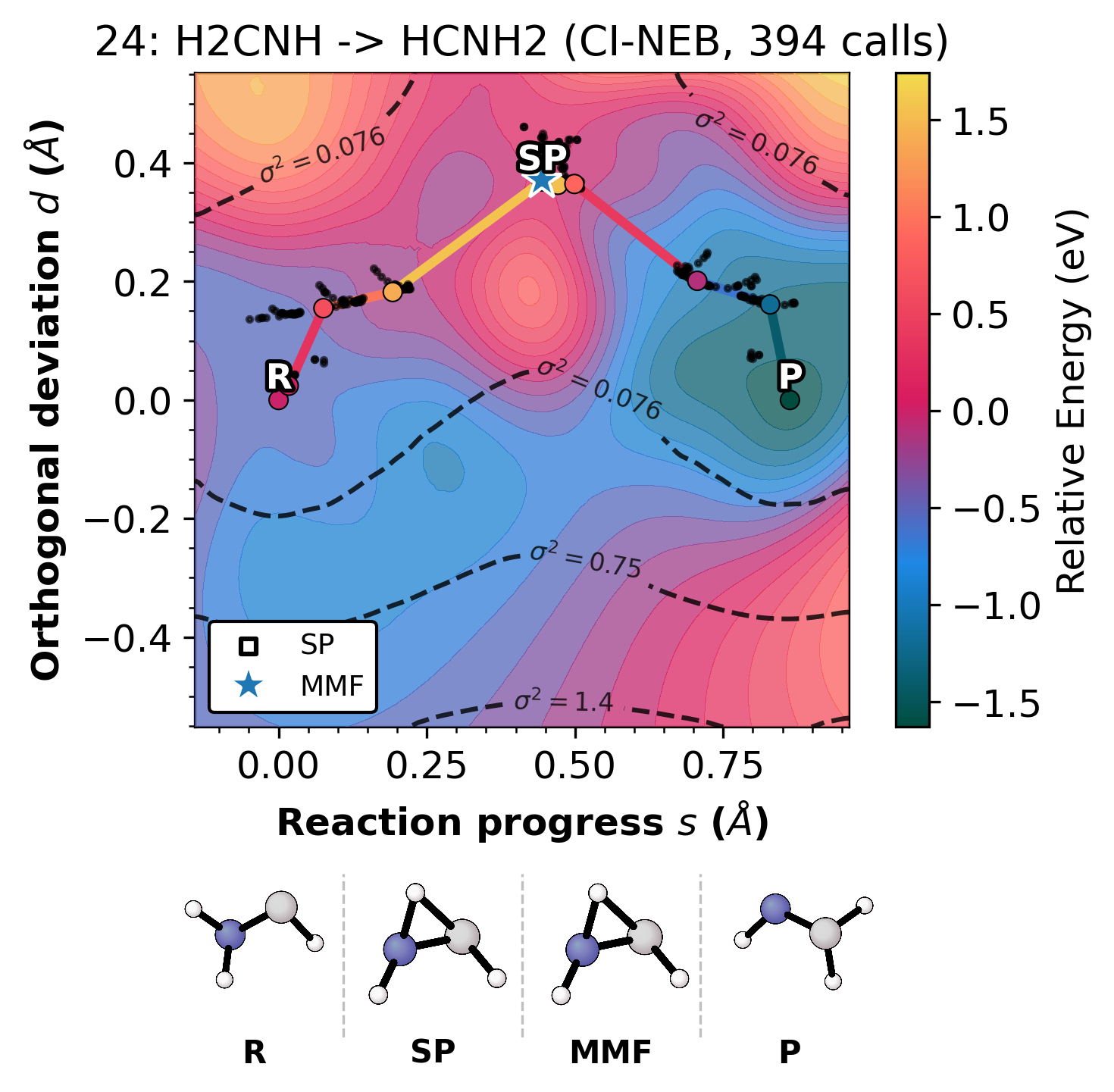}{\ce{H2CNH -> HCNH2}}{24}
\bakerlandscape{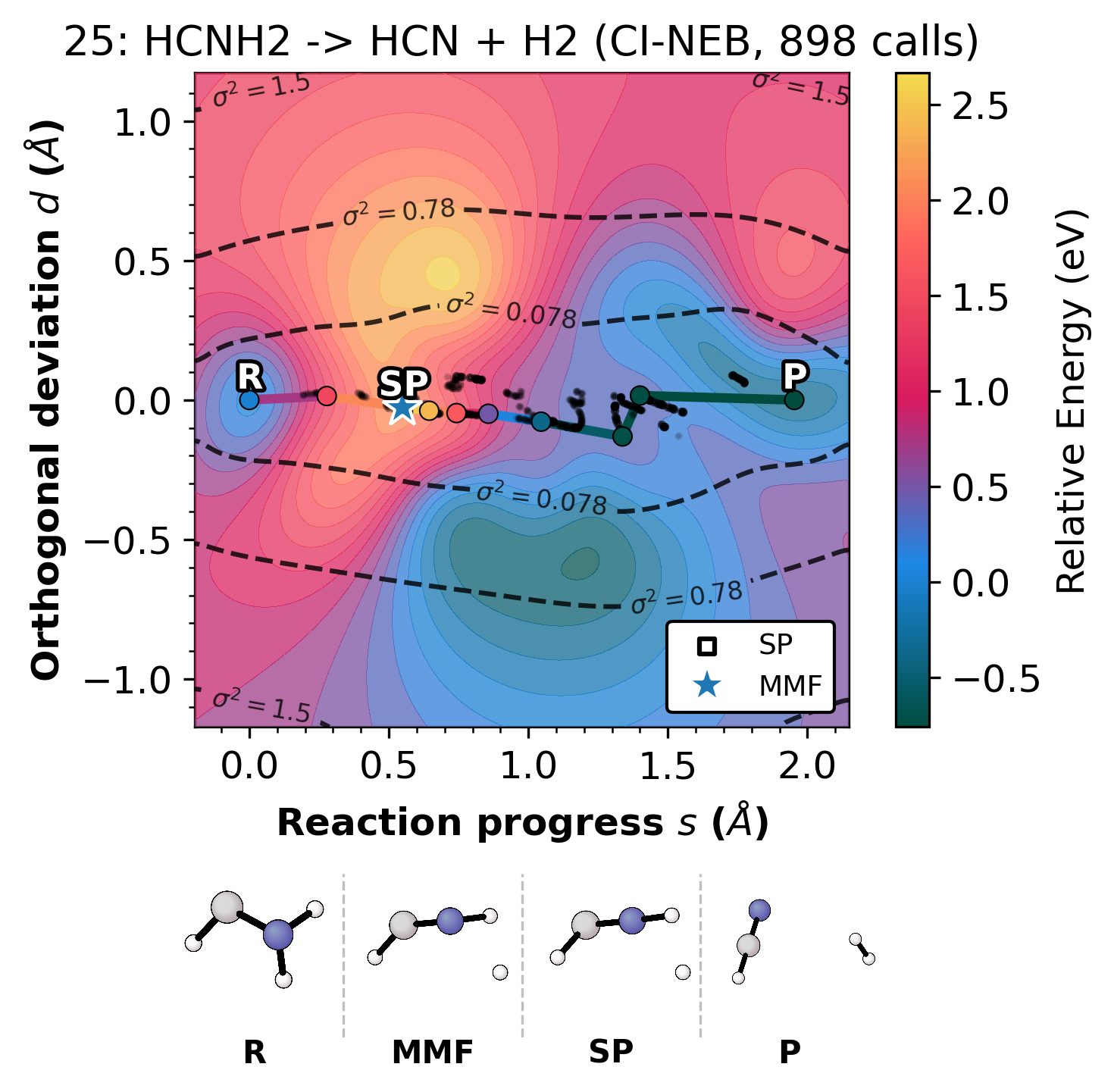}{\ce{HCNH2 -> HCN + H2}}{25}

\end{appendix}
\section{References}
\label{sec:org4ad90dd}
\bibliography{roneb}
\end{document}

%% file: authors_arxiv.tex
\author{ \href{https://orcid.org/0000-0002-2393-8056}{\includegraphics[scale=0.06]{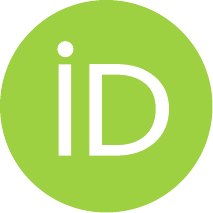}\hspace{1mm}Rohit Goswami}\thanks{Corresponding Author} \\
  Institute IMX and Lab-COSMO \\
  \'Ecole polytechnique f\'ed\'erale de Lausanne (EPFL) \\
  Station 12, CH-1015 Lausanne, Switzerland \\
  and \\
  Science Institute \& Faculty of Physical Sciences\\
  University of Iceland, 107 Reykjav\'ik, Iceland\\[1ex]
  \texttt{rgoswami@ieee.org}
  \And
  \href{https://orcid.org/0000-0003-0894-3233}{\includegraphics[scale=0.06]{orcid.pdf}\hspace{1mm}Miha Gunde} \\
  Institute Ruder Boskovic \\
  Bijenicka 54, 10000 Zagreb \\
  Croatia \\
  and \\
  Science Institute \& Faculty of Physical Sciences\\
  University of Iceland\\
  107 Reykjav\'ik, Iceland\\
  \texttt{mgunde@irb.hr}
  \And
  \href{https://orcid.org/0000-0001-8285-5421}{\includegraphics[scale=0.06]{orcid.pdf}\hspace{1mm}Hannes J\'onsson} \\
  Science Institute \& Faculty of Physical Sciences\\
  University of Iceland\\
  107 Reykjav\'ik, Iceland\\
  \texttt{hj@hi.is}
}

%% file: arxiv.bbl
\begin{thebibliography}{40}
\providecommand{\natexlab}[1]{#1}
\providecommand{\url}[1]{\texttt{#1}}
\expandafter\ifx\csname urlstyle\endcsname\relax
  \providecommand{\doi}[1]{doi: #1}\else
  \providecommand{\doi}{doi: \begingroup \urlstyle{rm}\Url}\fi

\bibitem[Jonsson et~al.(1998)Jonsson, Mills, and
  Jacobsen]{jonssonNudgedElasticBand1998}
Hannes Jonsson, Greg Mills, and Karsten~W. Jacobsen.
\newblock Nudged elastic band method for finding minimum energy paths of
  transitions.
\newblock In \emph{Classical and {{Quantum Dynamics}} in {{Condensed Phase
  Simulations}}}, pages 385--404. World Scientific, June 1998.
\newblock ISBN 978-981-02-3498-0.
\newblock \doi{10.1142/9789812839664_0016}.

\bibitem[Henkelman et~al.(2000)Henkelman, Uberuaga, and
  J{\'o}nsson]{henkelmanClimbingImageNudged2000}
Graeme Henkelman, Blas~P. Uberuaga, and Hannes J{\'o}nsson.
\newblock A climbing image nudged elastic band method for finding saddle points
  and minimum energy paths.
\newblock \emph{The Journal of Chemical Physics}, 113\penalty0 (22):\penalty0
  9901--9904, November 2000.
\newblock ISSN 0021-9606.
\newblock \doi{10.1063/1.1329672}.

\bibitem[Henkelman and J{\'o}nsson(1999)]{henkelmanDimerMethodFinding1999}
Graeme Henkelman and Hannes J{\'o}nsson.
\newblock A dimer method for finding saddle points on high dimensional
  potential surfaces using only first derivatives.
\newblock \emph{The Journal of Chemical Physics}, 111\penalty0 (15):\penalty0
  7010--7022, October 1999.
\newblock ISSN 0021-9606, 1089-7690.
\newblock \doi{10.1063/1.480097}.

\bibitem[{\'A}sgeirsson et~al.(2021){\'A}sgeirsson, Birgisson, Bjornsson,
  Becker, Neese, Riplinger, and J{\'o}nsson]{asgeirssonNudgedElasticBand2021}
Vilhj{\'a}lmur {\'A}sgeirsson, Benedikt~Orri Birgisson, Ragnar Bjornsson, Ute
  Becker, Frank Neese, Christoph Riplinger, and Hannes J{\'o}nsson.
\newblock Nudged {{Elastic Band Method}} for {{Molecular Reactions Using
  Energy-Weighted Springs Combined}} with {{Eigenvector Following}}.
\newblock \emph{Journal of Chemical Theory and Computation}, 17\penalty0
  (8):\penalty0 4929--4945, August 2021.
\newblock ISSN 1549-9618.
\newblock \doi{10.1021/acs.jctc.1c00462}.

\bibitem[Park et~al.(2025{\natexlab{a}})Park, Pritchard, and
  Wang]{parkHighthroughputApproachMinimum2025}
Heejune Park, Benjamin~P. Pritchard, and Lee-Ping Wang.
\newblock High-throughput approach for minimum energy pathway search using the
  nudged elastic band method with efficient data handling and parallel
  computing.
\newblock \emph{Journal of Chemical Theory and Computation}, 21\penalty0
  (23):\penalty0 12048--12063, December 2025{\natexlab{a}}.
\newblock ISSN 1549-9618.
\newblock \doi{10.1021/acs.jctc.5c01540}.

\bibitem[Goswami(2025{\natexlab{a}})]{goswamiEfficientExplorationChemical2025}
Rohit Goswami.
\newblock Efficient exploration of chemical kinetics, October
  2025{\natexlab{a}}.

\bibitem[Mousseau et~al.(2012)Mousseau, B{\'e}land, Brommer, Joly,
  {El-Mellouhi}, {Machado-Charry}, Marinica, and
  Pochet]{mousseauActivationRelaxationTechniqueART2012}
Normand Mousseau, Laurent~Karim B{\'e}land, Peter Brommer, Jean-Fran{\c c}ois
  Joly, Fedwa {El-Mellouhi}, Eduardo {Machado-Charry}, Mihai-Cosmin Marinica,
  and Pascal Pochet.
\newblock The {{Activation-Relaxation Technique}}: {{ART Nouveau}} and
  {{Kinetic ART}}.
\newblock \emph{Journal of Atomic and Molecular Physics}, 2012\penalty0
  (1):\penalty0 925278, 2012.
\newblock ISSN 2314-8020.
\newblock \doi{10.1155/2012/925278}.

\bibitem[Munro and Wales(1999)]{munroDefectMigrationCrystalline1999}
Lindsey~J. Munro and David~J. Wales.
\newblock Defect migration in crystalline silicon.
\newblock \emph{Physical Review B}, 59\penalty0 (6):\penalty0 3969--3980,
  February 1999.
\newblock \doi{10.1103/PhysRevB.59.3969}.

\bibitem[Cerjan and Miller(1981)]{cerjanFindingTransitionStates1981}
Charles~J. Cerjan and William~H. Miller.
\newblock On finding transition states.
\newblock \emph{The Journal of Chemical Physics}, 75\penalty0 (6):\penalty0
  2800--2806, September 1981.
\newblock ISSN 0021-9606, 1089-7690.
\newblock \doi{10.1063/1.442352}.

\bibitem[Olsen et~al.(2004)Olsen, Kroes, Henkelman, Arnaldsson, and
  J{\'o}nsson]{olsenComparisonMethodsFinding2004}
R.~A. Olsen, G.~J. Kroes, G.~Henkelman, A.~Arnaldsson, and H.~J{\'o}nsson.
\newblock Comparison of methods for finding saddle points without knowledge of
  the final states.
\newblock \emph{The Journal of Chemical Physics}, 121\penalty0 (20):\penalty0
  9776--9792, November 2004.
\newblock ISSN 0021-9606.
\newblock \doi{10.1063/1.1809574}.

\bibitem[Henkelman and J{\'o}nsson(2000)]{henkelmanImprovedTangentEstimate2000}
Graeme Henkelman and Hannes J{\'o}nsson.
\newblock Improved tangent estimate in the nudged elastic band method for
  finding minimum energy paths and saddle points.
\newblock \emph{The Journal of Chemical Physics}, 113\penalty0 (22):\penalty0
  9978--9985, December 2000.
\newblock ISSN 0021-9606.
\newblock \doi{10.1063/1.1323224}.

\bibitem[Mandelli and Parrinello(2021)]{mandelliModifiedNudgedElastic2021}
D.~Mandelli and M.~Parrinello.
\newblock A modified nudged elastic band algorithm with adaptive spring
  lengths.
\newblock \emph{Journal of Chemical Physics}, 155\penalty0 (7):\penalty0 74103,
  August 2021.
\newblock ISSN 0021-9606.
\newblock \doi{10.1063/5.0059593}.

\bibitem[Schmerwitz et~al.(2024)Schmerwitz, {\'A}sgeirsson, and
  J{\'o}nsson]{schmerwitzImprovedInitializationOptimal2024}
Yorick L.~A. Schmerwitz, Vilhj{\'a}lmur {\'A}sgeirsson, and Hannes J{\'o}nsson.
\newblock Improved {{Initialization}} of {{Optimal Path Calculations Using
  Sequential Traversal}} over the {{Image-Dependent Pair Potential Surface}}.
\newblock \emph{Journal of Chemical Theory and Computation}, 20\penalty0
  (1):\penalty0 155--163, January 2024.
\newblock ISSN 1549-9618.
\newblock \doi{10.1021/acs.jctc.3c01111}.

\bibitem[Smidstrup et~al.(2014)Smidstrup, Pedersen, Stokbro, and
  J{\'o}nsson]{smidstrupImprovedInitialGuess2014}
S{\o}ren Smidstrup, Andreas Pedersen, Kurt Stokbro, and Hannes J{\'o}nsson.
\newblock Improved initial guess for minimum energy path calculations.
\newblock \emph{The Journal of Chemical Physics}, 140\penalty0 (21):\penalty0
  214106, June 2014.
\newblock ISSN 0021-9606.
\newblock \doi{10.1063/1.4878664}.

\bibitem[Heyden et~al.(2005)Heyden, Bell, and
  Keil]{heydenEfficientMethodsFinding2005}
Andreas Heyden, Alexis~T. Bell, and Frerich~J. Keil.
\newblock Efficient methods for finding transition states in chemical
  reactions: {{Comparison}} of improved dimer method and partitioned rational
  function optimization method.
\newblock \emph{The Journal of Chemical Physics}, 123\penalty0 (22):\penalty0
  224101, December 2005.
\newblock ISSN 0021-9606.
\newblock \doi{10.1063/1.2104507}.

\bibitem[Peterson(2016)]{petersonAccelerationSaddlepointSearches2016}
Andrew~A. Peterson.
\newblock Acceleration of saddle-point searches with machine learning.
\newblock \emph{The Journal of Chemical Physics}, 145\penalty0 (7):\penalty0
  074106, August 2016.
\newblock ISSN 0021-9606.
\newblock \doi{10.1063/1.4960708}.

\bibitem[Koistinen et~al.(2019)Koistinen, {\'A}sgeirsson, Vehtari, and
  J{\'o}nsson]{koistinenNudgedElasticBand2019}
Olli-Pekka Koistinen, Vilhj{\'a}lmur {\'A}sgeirsson, Aki Vehtari, and Hannes
  J{\'o}nsson.
\newblock Nudged {{Elastic Band Calculations Accelerated}} with {{Gaussian
  Process Regression Based}} on {{Inverse Interatomic Distances}}.
\newblock \emph{Journal of Chemical Theory and Computation}, 15\penalty0
  (12):\penalty0 6738--6751, December 2019.
\newblock ISSN 1549-9618.
\newblock \doi{10.1021/acs.jctc.9b00692}.

\bibitem[Koistinen et~al.(2020)Koistinen, {\'A}sgeirsson, Vehtari, and
  J{\'o}nsson]{koistinenMinimumModeSaddle2020}
Olli-Pekka Koistinen, Vilhj{\'a}lmur {\'A}sgeirsson, Aki Vehtari, and Hannes
  J{\'o}nsson.
\newblock Minimum {{Mode Saddle Point Searches Using Gaussian Process
  Regression}} with {{Inverse-Distance Covariance Function}}.
\newblock \emph{Journal of Chemical Theory and Computation}, 16\penalty0
  (1):\penalty0 499--509, January 2020.
\newblock ISSN 1549-9618.
\newblock \doi{10.1021/acs.jctc.9b01038}.

\bibitem[Goswami and J{\'o}nsson(2025)]{goswamiAdaptivePruningIncreased2025b}
Rohit Goswami and Hannes J{\'o}nsson.
\newblock Adaptive {{Pruning}} for {{Increased Robustness}} and {{Reduced
  Computational Overhead}} in {{Gaussian Process Accelerated Saddle Point
  Searches}}.
\newblock \emph{ChemPhysChem}, November 2025.
\newblock ISSN 1439-7641.
\newblock \doi{10.1002/cphc.202500730}.

\bibitem[Goswami et~al.(2025{\natexlab{a}})Goswami, Masterov, Kamath,
  {Pe{\~n}a-Torres}, and
  J{\'o}nsson]{goswamiEfficientImplementationGaussian2025a}
Rohit Goswami, Maxim Masterov, Satish Kamath, Alejandro {Pe{\~n}a-Torres}, and
  Hannes J{\'o}nsson.
\newblock Efficient implementation of gaussian process regression accelerated
  saddle point searches with application to molecular reactions, May
  2025{\natexlab{a}}.

\bibitem[E et~al.(2007)E, Ren, and
  Vanden-Eijnden]{eSimplifiedImprovedString2007}
Weinan E, Weiqing Ren, and Eric Vanden-Eijnden.
\newblock Simplified and improved string method for computing the minimum
  energy paths in barrier-crossing events.
\newblock \emph{The Journal of Chemical Physics}, 126\penalty0 (16):\penalty0
  164103, 2007.
\newblock \doi{10.1063/1.2720838}.

\bibitem[Goswami et~al.(2026)Goswami, Chill, Terrell, Henkelman, Welborn,
  Zhang, Pedersen, {T-Brink}, Edelmann, Claude, Alejandro, Jung, Ghasemi,
  Chemist29, Satishskamath, Maxim, and
  Via9A]{rohitgoswamiTheochemUIEOnV29012026}
Rohit Goswami, Sam Chill, Rye Terrell, Graeme Henkelman, Matthew Welborn, Liang
  Zhang, Andreas Pedersen, {T-Brink}, Erik Edelmann, Jean Claude, Alejandro,
  Sung~Hoon Jung, Seyed~Alireza Ghasemi, Chemist29, Satishskamath, Maxim, and
  Via9A.
\newblock {{TheochemUI}}/{{eOn}}: V2.9.0.1.
\newblock Zenodo, February 2026.

\bibitem[Bigi et~al.(2025)Bigi, Abbott, Loche, Mazitov, Tisi, Langer,
  Goscinski, Pegolo, Chong, Goswami, Chorna, Kellner, Ceriotti, and
  Fraux]{bigiMetatensorMetatomicFoundational2025}
Filippo Bigi, Joseph~W. Abbott, Philip Loche, Arslan Mazitov, Davide Tisi,
  Marcel~F. Langer, Alexander Goscinski, Paolo Pegolo, Sanggyu Chong, Rohit
  Goswami, Sofiia Chorna, Matthias Kellner, Michele Ceriotti, and Guillaume
  Fraux.
\newblock Metatensor and metatomic: Foundational libraries for interoperable
  atomistic machine learning, August 2025.

\bibitem[Mazitov et~al.(2025{\natexlab{a}})Mazitov, Chorna, Fraux, Bercx,
  Pizzi, De, and Ceriotti]{mazitovMassiveAtomicDiversity2025}
Arslan Mazitov, Sofiia Chorna, Guillaume Fraux, Marnik Bercx, Giovanni Pizzi,
  Sandip De, and Michele Ceriotti.
\newblock Massive atomic diversity: A compact universal dataset for atomistic
  machine learning.
\newblock \emph{Scientific Data}, 12\penalty0 (1):\penalty0 1857, November
  2025{\natexlab{a}}.
\newblock ISSN 2052-4463.
\newblock \doi{10.1038/s41597-025-06109-y}.

\bibitem[Mazitov et~al.(2025{\natexlab{b}})Mazitov, Bigi, Kellner, Pegolo,
  Tisi, Fraux, Pozdnyakov, Loche, and
  Ceriotti]{mazitovPETMADLightweightUniversal2025}
Arslan Mazitov, Filippo Bigi, Matthias Kellner, Paolo Pegolo, Davide Tisi,
  Guillaume Fraux, Sergey Pozdnyakov, Philip Loche, and Michele Ceriotti.
\newblock {{PET-MAD}} as a lightweight universal interatomic potential for
  advanced materials modeling.
\newblock \emph{Nature Communications}, 16\penalty0 (1):\penalty0 10653,
  November 2025{\natexlab{b}}.
\newblock ISSN 2041-1723.
\newblock \doi{10.1038/s41467-025-65662-7}.

\bibitem[M{\"o}lder et~al.(2021)M{\"o}lder, Jablonski, Letcher, Hall,
  {Tomkins-Tinch}, Sochat, Forster, Lee, Twardziok, Kanitz, Wilm, Holtgrewe,
  Rahmann, Nahnsen, and K{\"o}ster]{molderSustainableDataAnalysis2021}
Felix M{\"o}lder, Kim~Philipp Jablonski, Brice Letcher, Michael~B. Hall,
  Christopher~H. {Tomkins-Tinch}, Vanessa Sochat, Jan Forster, Soohyun Lee,
  Sven~O. Twardziok, Alexander Kanitz, Andreas Wilm, Manuel Holtgrewe, Sven
  Rahmann, Sven Nahnsen, and Johannes K{\"o}ster.
\newblock Sustainable data analysis with {{Snakemake}}, April 2021.

\bibitem[K{\"a}stner and
  Sherwood(2008)]{kastnerSuperlinearlyConvergingDimer2008}
Johannes K{\"a}stner and Paul Sherwood.
\newblock Superlinearly converging dimer method for transition state search.
\newblock \emph{The Journal of Chemical Physics}, 128\penalty0 (1):\penalty0
  014106, January 2008.
\newblock ISSN 0021-9606.
\newblock \doi{10.1063/1.2815812}.

\bibitem[Goswami(2025{\natexlab{b}})]{goswamiBayesianHierarchicalModels2025a}
Rohit Goswami.
\newblock Bayesian hierarchical models for quantitative estimates for
  performance metrics applied to saddle search algorithms.
\newblock \emph{AIP Advances}, 15\penalty0 (8):\penalty0 85210, August
  2025{\natexlab{b}}.
\newblock ISSN 2158-3226.
\newblock \doi{10.1063/5.0283639}.

\bibitem[Liu and Nocedal(1989)]{liuLimitedMemoryBFGS1989a}
Dong~C. Liu and Jorge Nocedal.
\newblock On the limited memory {{BFGS}} method for large scale optimization.
\newblock \emph{Mathematical Programming}, 45\penalty0 (1):\penalty0 503--528,
  August 1989.
\newblock ISSN 1436-4646.
\newblock \doi{10.1007/BF01589116}.

\bibitem[Chapra and Canale(2015)]{chapraNumericalMethodsEngineers2015}
Steven~C. Chapra and Raymond~P. Canale.
\newblock \emph{Numerical Methods for Engineers}.
\newblock McGraw-Hill Education, New York, NY, 7. ed edition, 2015.
\newblock ISBN 978-0-07-339792-4.

\bibitem[Baker and Chan(1996)]{bakerLocationTransitionStates1996}
Jon Baker and Fora Chan.
\newblock The location of transition states: {{A}} comparison of {{Cartesian}},
  {{Z-matrix}}, and natural internal coordinates.
\newblock \emph{Journal of Computational Chemistry}, 17\penalty0 (7):\penalty0
  888--904, 1996.
\newblock ISSN 1096-987X.
\newblock \doi{10.1002/(SICI)1096-987X(199605)17:7<888::AID-JCC12>3.0.CO;2-7}.

\bibitem[Goswami(2025{\natexlab{c}})]{goswamiTwodimensionalRMSDProjections2025}
Rohit Goswami.
\newblock Two-dimensional {{RMSD}} projections for reaction path visualization
  and validation, December 2025{\natexlab{c}}.

\bibitem[Gunde et~al.(2021)Gunde, Salles, H{\'e}meryck, and
  {Martin-Samos}]{gundeIRAShapeMatching2021}
Miha Gunde, Nicolas Salles, Anne H{\'e}meryck, and Layla {Martin-Samos}.
\newblock {{IRA}}: A shape matching approach for recognition and comparison of
  generic atomic patterns.
\newblock \emph{Journal of Chemical Information and Modeling}, 61\penalty0
  (11):\penalty0 5446--5457, November 2021.
\newblock ISSN 1549-9596.
\newblock \doi{10.1021/acs.jcim.1c00567}.

\bibitem[Chill et~al.(2014)Chill, Stevenson, Ruehle, Shang, Xiao, Farrell,
  Wales, and Henkelman]{chillBenchmarksCharacterizationMinima2014}
Samuel~T. Chill, Jacob Stevenson, Victor Ruehle, Cheng Shang, Penghao Xiao,
  James~D. Farrell, David~J. Wales, and Graeme Henkelman.
\newblock Benchmarks for {{Characterization}} of {{Minima}}, {{Transition
  States}}, and {{Pathways}} in {{Atomic}}, {{Molecular}}, and {{Condensed
  Matter Systems}}.
\newblock \emph{Journal of Chemical Theory and Computation}, 10\penalty0
  (12):\penalty0 5476--5482, December 2014.
\newblock ISSN 1549-9618.
\newblock \doi{10.1021/ct5008718}.

\bibitem[Bergstra et~al.(2011)Bergstra, Bardenet, Bengio, and
  K{\'e}gl]{bergstraAlgorithmsHyperParameterOptimization2011}
James Bergstra, R{\'e}mi Bardenet, Yoshua Bengio, and Bal{\'a}zs K{\'e}gl.
\newblock Algorithms for {{Hyper-Parameter Optimization}}.
\newblock In \emph{Advances in {{Neural Information Processing Systems}}},
  volume~24. Curran Associates, Inc., 2011.

\bibitem[Goswami et~al.(2025{\natexlab{b}})Goswami, Masterov, Kamath,
  {Pena-Torres}, and J{\'o}nsson]{goswamiEfficientImplementationGaussian2025}
Rohit Goswami, Maxim Masterov, Satish Kamath, Alejandro {Pena-Torres}, and
  Hannes J{\'o}nsson.
\newblock Efficient {{Implementation}} of {{Gaussian Process Regression
  Accelerated Saddle Point Searches}} with {{Application}} to {{Molecular
  Reactions}}.
\newblock \emph{Journal of Chemical Theory and Computation}, July
  2025{\natexlab{b}}.
\newblock \doi{10.1021/acs.jctc.5c00866}.

\bibitem[Gunde et~al.(2024)Gunde, Salles, Grisanti, {Martin-Samos}, and
  Hemeryck]{gundeSOFIFindingPoint2024}
M.~Gunde, N.~Salles, L.~Grisanti, L.~{Martin-Samos}, and A.~Hemeryck.
\newblock {{SOFI}}: {{Finding}} point group symmetries in atomic clusters as
  finding the set of degenerate solutions in a shape-matching problem.
\newblock \emph{The Journal of Chemical Physics}, 161\penalty0 (6):\penalty0
  062503, August 2024.
\newblock ISSN 0021-9606, 1089-7690.
\newblock \doi{10.1063/5.0215689}.

\bibitem[Akiba et~al.(2019)Akiba, Sano, Yanase, Ohta, and
  Koyama]{akibaOptunaNextgenerationHyperparameter2019}
Takuya Akiba, Shotaro Sano, Toshihiko Yanase, Takeru Ohta, and Masanori Koyama.
\newblock Optuna: {{A Next-generation Hyperparameter Optimization Framework}}.
\newblock In \emph{Proceedings of the 25th {{ACM SIGKDD International
  Conference}} on {{Knowledge Discovery}} \& {{Data Mining}}}, pages
  2623--2631. ACM, 2019.
\newblock \doi{10.1145/3292500.3330701}.

\bibitem[Hutter et~al.(2014)Hutter, Hoos, and
  Leyton-Brown]{hutterEfficientApproachAssessing2014}
Frank Hutter, Holger~H. Hoos, and Kevin Leyton-Brown.
\newblock An {{Efficient Approach}} for {{Assessing Hyperparameter
  Importance}}.
\newblock In \emph{Proceedings of the 31st {{International Conference}} on
  {{Machine Learning}}}, Proceedings of {{Machine Learning Research}}, pages
  754--762. PMLR, 2014.

\bibitem[Park et~al.(2025{\natexlab{b}})Park, Lee, Kim, and
  Choi]{parkAutomatedWorkflowTransition2025}
Joonho Park, Sangmin Lee, Hyo~Jin Kim, and Yoon~Sup Choi.
\newblock Automated workflow for transition state searches using machine
  learning potentials.
\newblock \emph{Journal of Chemical Theory and Computation}, 21\penalty0
  (5):\penalty0 2145--2158, March 2025{\natexlab{b}}.
\newblock ISSN 1549-9618.
\newblock \doi{10.1021/acs.jctc.4c01234}.

\end{thebibliography}
